\documentclass[aps,twocolumn,pra,superscriptaddress,amsmath,amssymb]{revtex4-2} 
\usepackage{dcolumn}
\usepackage{amsmath}
\usepackage{mathrsfs}
\usepackage{txfonts}
\usepackage{bm}
\usepackage[T1]{fontenc}
\usepackage{xspace}
\usepackage{ulem}
\usepackage{comment}
\usepackage{braket}
\usepackage{lipsum}
\usepackage{bbold}
\usepackage{mathtools}
\usepackage{physics}
\newcommand{\means}[1]{\langle#1\rangle}

\newcommand{\backvec}[1]{\:\reflectbox{$\vec{\reflectbox{\!$#1$}}$}}   
\newcommand{\stco}{\overset{\,\star\,}{\,,\,}}    
\newcommand{\mTr}{\mathrm{Tr}}
\newcommand{\llangle}{\langle\hspace{-2pt}\langle}    
\newcommand{\rrangle}{\rangle\hspace{-2pt}\rangle}    

\usepackage{pifont}
\newcommand{\cmark}{\ding{51}\xspace}
\newcommand{\xmark}{\ding{55}\xspace}

\ifx\pdfoutput\undefined
\usepackage[dvipdfmx]{graphicx}
\usepackage[dvipdfmx]{hyperref}
\usepackage[dvipdfmx]{color}
\usepackage[dvipdfmx]{xcolor}
\else
\usepackage{graphicx}
\usepackage{hyperref}
\usepackage{color}
\usepackage{xcolor}
\usepackage[version=3]{mhchem}
\fi
\hypersetup{
        colorlinks=true,
        citecolor=blue,
        urlcolor=blue,
        linkcolor=blue
}

\begin{document}
\let\emph\textit

\title{
    Formulation of intrinsic nonlinear thermal conductivity for bosonic systems 
    \\using quantum kinetic equation
}
\author{Aoi Kuwabara}
\author{Joji Nasu}
\affiliation{
  Department of Physics, Graduate School of Science, Tohoku University, Sendai, Miyagi 980-8578, Japan
}

\date{\today}
\begin{abstract}
Nonlinear responses in transport phenomena have attracted significant attention because they can arise even when linear responses are forbidden by symmetry, with the quantum geometry of Bloch wave functions playing an essential role.
While such effects have been extensively studied in electric transport, similar quantum-geometric mechanisms are also expected to govern nonlinear thermal transport.
In particular, thermal responses are crucial in bosonic systems such as magnons and phonons, which are charge-neutral quasiparticles.
However, a consistent theoretical description of nonlinear thermal transport remains challenging because of the difficulty in the treatment of energy magnetization in higher-order responses with Luttinger's gravitational potential method.
Here, we formulate the intrinsic nonlinear thermal conductivity of bosonic systems using a quantum kinetic equation approach that avoids Luttinger's method and naturally incorporates contributions from energy magnetization.
We identify three distinct contributions to the nonlinear thermal conductivity: two expressed in terms of quantum-geometric quantities, namely the quantum metric and the thermal Berry-connection polarizability (TBCP), and a third determined solely by the band dispersions.
Applying our formalism to a specific quantum spin model within linear spin-wave theory, we show that the TBCP term dominates the nonlinear thermal Hall effect in the absence of threefold symmetry.
Our results differ quantitatively from those obtained using semiclassical theory, thereby highlighting the importance of quantum corrections beyond the semiclassical picture.
These findings establish a general framework for intrinsic nonlinear thermal responses in bosonic systems and reveal quantum-geometric mechanisms underlying thermal transport beyond linear response theory.
\end{abstract}
\maketitle


\section{Introduction}
\label{introduction}

In condensed matter physics, the geometric properties of Bloch wave functions in Hilbert space have been recognized as essential elements for understanding various physical phenomena.
One of the most well-known examples is the quantum Hall effect~\cite{Thouless1982}, where the quantized Hall conductivity is directly related to the Chern number, a topological invariant defined as the integral of the Berry curvature of a Bloch state over the Brillouin zone.
The Berry curvature is introduced via the Berry connection, which is defined as a gauge field associated with the adiabatic evolution of quantum states in momentum space~\cite{Berry1984,Simon1983}.
In the context of the geometry intrinsic to quantum states, another important quantity is the quantum metric, which characterizes the distance between two quantum states in Hilbert space~\cite{Provost1980,Resta2011}.
The quantum metric and Berry curvature are derived together from the quantum geometric tensor~\cite{Provost1980,Resta2011,Yu2025,Lhachemi2026}, which plays a crucial role in various physical phenomena, such as superfluidity in flat bands~\cite{Peotta2015,Liang2017} and the fractional quantum Hall effect~\cite{Wang2021APS}.

In recent years, nonlinear responses have been recognized as phenomena in which quantum geometric effects play an essential role~\cite{Gao2014,Gao2015,Das2023,Gao2023,Wang2023,Fang2024}.
For example, in electronic systems, nonlinear Hall effects~\cite{Gao2014,Sodemann2015,Du2021,Bandyopadhyay2024,Ulrich2025} and nonlinear optical responses~\cite{Sipe2000} have been intensively studied as phenomena originating from the quantum geometry of Bloch states.
In particular, for second-order responses in nonlinear Hall effects, two contributions can be considered: one is extrinsic and depends on the relaxation time, whereas the other is intrinsic and independent of the relaxation time; both contributions originate from the quantum geometry of Bloch states~\cite{Gao2014,Sodemann2015}.
It has been pointed out that the Berry curvature dipole contributes to the extrinsic nonlinear Hall effect~\cite{Sodemann2015}.
Whereas the quantum Hall effect, as a linear response to electric fields, requires the breaking of time-reversal symmetry, the extrinsic nonlinear Hall effect can occur even in time-reversal-symmetric systems, provided that inversion symmetry is broken~\cite{Sodemann2015}.
As possible candidates exhibiting nonlinear Hall effects, Weyl semimetals such as $\mathrm{WTe_2}$ and $\mathrm{MoTe_2}$ have been proposed on the basis of first-principles calculations~\cite{Zhang2018B,Zhang20182D}, and subsequently, the effect has been confirmed experimentally~\cite{Ma2019,Kang2019,Tiwari2021}.

On the other hand, the intrinsic nonlinear Hall effect is suggested to arise from changes in the Bloch wave function induced by an applied electric field~\cite{Gao2014}.
This effect is formulated in terms of a quantum geometric quantity known as the Berry-connection polarizability (BCP)~\cite{Gao2014,Lai2021} or the band-normalized quantum metric~\cite{Wang2023,Kaplan2024}, indicating that the intrinsic nonlinear Hall effect is characterized by the quantum metric.
Although the Berry curvature vanishes in systems possessing symmetry under the composite operation of spatial inversion $\mathcal P$ and time reversal $\mathcal T$, the so-called $\mathcal{PT}$ symmetry~\cite{Xiao2010}, the BCP can take a nonzero value, suggesting that the intrinsic nonlinear Hall effect can occur even in $\mathcal{PT}$-symmetric systems, unlike the linear Hall effect and the extrinsic nonlinear Hall effect.
Using this property, it has been proposed that the intrinsic nonlinear Hall effect can be utilized to detect a N\'eel vector in antiferromagnets with $\mathcal{PT}$ symmetry~\cite{Liu2021,Wang2021}.
Nevertheless, the formulation of the nonlinear Hall effect has remained under debate because the approaches based on semiclassical theory using wave-packet dynamics~\cite{Gao2014,Kaplan2024} and quantum kinetic theory~\cite{Kraut1979,Aversa1995} yield different results, although they provide consistent results in many cases~\cite{Luttinger1958,Sinitsyn2007,Sinitsyn2008,Xiao2019,Matsyshyn2019,Watanabe2020,Xiang2024,Mandal2024,Kaplan2024}.
This discrepancy suggests that quantum corrections beyond the semiclassical picture play a crucial role in the intrinsic nonlinear Hall effect.

So far, we have introduced nonlinear responses that appear in the form of an electric current.
However, such effects cannot be observed in insulating materials.
Even in insulating systems, collective excitations described by bosonic quasiparticles, such as magnons and phonons, as well as their composite particles, can contribute to transport phenomena because they can carry heat and spin.
For example, the thermal Hall effect, which is the generation of a heat current transverse to the temperature gradient, has been extensively studied in various insulating systems with quasiparticle excitations~\cite{Katsura2010,Qin2011,Matsumoto2011L,Matsumoto2011B,Qin2012}.
Similar to the quantum Hall effect, the thermal Hall effect can be understood in terms of the Berry curvature of bosonic quasiparticles and becomes nonzero when $\mathcal{PT}$ symmetry is broken~\cite{Fujiwara2022}.
This effect has been observed in various insulating magnets, where it is carried by magnons~\cite{Onose2010,Ideue2012,Hirschberger2015L,Hirschberger2015Science,Mcclarty2018,Joshi2018,Akazawa2020,Koyama2021,Zhang2021L,Czajka2023}, as well as in Kitaev systems with charge-neutral Majorana fermions~\cite{Kitaev2006,Nasu2017,Kasahara2018L,Kasahara2018N,Ye2018,Yokoi2021,Hwang2022,Imamura2024}.
Moreover, phonon contributions to the thermal Hall effect have also been investigated theoretically and experimentally~\cite{Sheng2006,Zhang2010,Zhang2011,Qin2012,Vinkler-Aviv2018,Chen2022,Chen2024,Sharma2024,Oh2025}.

As a natural extension of the thermal Hall effect, one can consider nonlinear (in particular, second-order) thermal Hall effects, which correspond to heat current generations by second-order terms in temperature gradients.
By analogy with nonlinear electric transport, we expect that the nonlinear thermal Hall effect has both extrinsic and intrinsic contributions.
The extrinsic contribution has been formulated using the Boltzmann equation, and it has been shown to arise from the (extended) Berry curvature dipole~\cite{Mukherjee2023,Varshney2023,Ni2025}.
On the other hand, the intrinsic nonlinear thermal Hall effect remains controversial.
So far, this effect has been studied using two different approaches, namely quantum kinetic theory~\cite{Varshney2023} and semiclassical theory~\cite{Li2024}, which yield qualitatively different results: the former study suggests that the intrinsic nonlinear thermal Hall conductivity can be expressed in terms of the band-resolved quantum metric, whereas the latter study shows that it is characterized by the thermal Berry-connection polarizability (TBCP)~\cite{Li2024,Zhang2025,Barman2025}, which is the thermal counterpart of the BCP in electric current responses.
One possible reason for this discrepancy is the limited applicability of Luttinger's method~\cite{Luttinger1964,Moreno1996,Tatara2015L,Tatara2015B}, which introduces a fictitious gravitational potential to evaluate thermal transport coefficients.
Both of the previous works~\cite{Varshney2023,Li2024} rely on Luttinger's method to evaluate the nonlinear thermal Hall effect; however, it remains unclear whether this method is valid in the nonlinear response regime.
Moreover, when Luttinger's method is applied, special care must be taken in the treatment of energy magnetization, which is known to play a crucial role in the formulation of the thermal Hall effect~\cite{Matsumoto2011L,Matsumoto2011B,Qin2011}.
Higher-order contributions of energy magnetization are expected to be significant also in the nonlinear thermal Hall effect, but such contributions have not been properly considered in previous studies~\cite{Varshney2023,Li2024}, because formalisms that incorporate these contributions have not yet been established.
To overcome these issues, it is necessary to develop a theoretical framework that does not rely on Luttinger's method and that includes quantum corrections beyond the semiclassical picture.

Recently, a quantum kinetic equation approach has been developed to describe heat transport phenomena in bosonic systems~\cite{Mangeolle2024}, and its applications to nonlinear responses and extrinsic contributions to heat transport have been proposed~\cite{Park2025,Mangeolle2026}.
In particular, diagrammatic expansions within the quantum kinetic framework have been developed to incorporate impurity effects in thermal transport~\cite{Mangeolle2026}, although clarifying the precise correspondence between the present formulation and a standard diagrammatic approach remains an important future problem.
The advantage of the quantum kinetic equation approach is that it can naturally incorporate quantum corrections beyond the semiclassical picture and does not rely on Luttinger's method.
Moreover, the energy magnetization is naturally included and does not need to be treated explicitly in this formalism~\cite{Mangeolle2024}.
These features make the quantum kinetic equation approach suitable for formulating the intrinsic nonlinear thermal Hall effect.

In this paper, we formulate intrinsic nonlinear heat transport in bosonic systems using a quantum kinetic equation approach.
We start from a general bosonic Bogoliubov-de Gennes (BdG) Hamiltonian and express both the Hamiltonian and the density matrix in the Wigner representation.
By expanding the Moyal product in the Liouville-von Neumann equation in the Wigner representation, which is referred to as the quantum kinetic equation, we obtain the heat current density up to second order in spatial gradients.
Using this expression, we calculate the intrinsic nonlinear thermal Hall conductivity in two-dimensional systems with translational symmetry under a local equilibrium condition.
The obtained expression consists of three terms, two of which are expressed in terms of quantum geometric quantities, namely the TBCP and the quantum metric, while the remaining term is expressed solely in terms of the band dispersion.
We find that the previous result obtained from semiclassical theory is included in the TBCP contribution, but additional quantum correction terms appear even in this part.
We apply our formulation to a specific model.
Specifically, we consider a spin-$S$ localized spin model on a honeycomb lattice with ferromagnetic nearest-neighbor interactions and next-nearest-neighbor Dzyaloshinskii-Moriya interactions, which has been introduced in previous studies to calculate the intrinsic nonlinear thermal Hall conductivity.
We assume a fully spin-polarized ferromagnetic state and employ linear spin-wave theory to describe magnon excitations.
We numerically calculate the intrinsic nonlinear thermal Hall conductivity in both undistorted and distorted honeycomb systems, where the latter breaks the three-fold rotational symmetry.
In both cases, we find that the intrinsic nonlinear thermal Hall conductivity is nonzero, whereas the contribution from the TBCP vanishes in the undistorted case.
This behavior is consistent with previous studies based on semiclassical theory; hence, the other two contributions, including the quantum metric, play a crucial role in the undistorted system.
In distorted systems, all three contributions are nonzero, and the TBCP contribution becomes dominant at higher temperatures.
In both cases, we find that our results quantitatively differ from those obtained in previous studies, suggesting the importance of quantum corrections beyond the semiclassical picture or approaches that do not employ Luttinger's method.

This paper is organized as follows.
In the next section, we formulate the heat current in general bosonic systems using the quantum kinetic equation.
The heat current density is derived within the Wigner representation in Sec.~\ref{sec:heat_current_qke}.
In Sec.~\ref{diag_star}, we diagonalize the bosonic BdG Hamiltonian in the Wigner representation using the star-product formulation of a paraunitary matrix.
In Sec.~\ref{subsec:gauge}, we discuss the gauge transformation intrinsic to the diagonalization of the Hamiltonian within the Wigner representation.
Section~\ref{subsec:QGT} introduces the quantum geometric tensor for bosonic systems, which comprises the quantum metric and the Berry curvature.
In Sec.~\ref{subsec:h_and_J}, we derive a gauge-invariant expression for the heat current density.
In Sec.~\ref{grad_expansion}, we introduce the gradient expansion of the Hamiltonian and the Berry connection.
In Sec.~\ref{sec:2_THE}, we derive an expression for the intrinsic nonlinear thermal conductivity tensor in translationally invariant systems.
First, in Sec.~\ref{subsec:thermal_current}, we expand the heat current density with respect to the spatial gradient and obtain its expression up to second order in the gradient.
Next, in Sec.~\ref{subsec:NLTHcoefficient}, we derive an expression for the second-order thermal conductivity tensor.
In Sec.~\ref{sec:model_calc}, we apply our formulation to a specific quantum spin model on a honeycomb lattice and numerically calculate the intrinsic nonlinear thermal Hall conductivity.
In Sec.~\ref{sec:model}, we introduce the quantum spin model and derive the spin-wave Hamiltonian, expressed as a bilinear form of bosonic operators, by assuming a fully spin-polarized ferromagnetic state.
In Sec.~\ref{subsec:result}, we present numerical results for the intrinsic nonlinear thermal Hall conductivity in both undistorted and distorted honeycomb lattices.
Finally, Sec.~\ref{sec:summary} is devoted to the summary and future perspectives.

\section{Heat current using quantum kinetic equation for bosonic systems}
\label{sec:general_NLTHE}

In this section, we formulate the heat current in general bosonic systems using the quantum kinetic equation, based on Refs.~\cite{Mangeolle2024,Park2025}.

\subsection{Heat current density in phase space}
\label{sec:heat_current_qke}

We start from a general $d$-dimensional bosonic BdG Hamiltonian and derive an expression for the heat current in the Wigner representation.
In general, the bosonic BdG Hamiltonian under the continuum approximation is given by
\begin{align}\label{eq:hamiltonian}
  \mathscr H = \frac{1}{2}\int_{\bm r_1,\bm r_2}\Psi^\dagger(\bm r_1)H(\bm r_1,\bm r_2)\Psi(\bm r_2).
\end{align}
Here, $H(\bm r_1,\bm r_2)$ is a $2N\times 2N$ matrix, and $\Psi^\dagger(\bm r)=[\beta^\dagger_1(\bm r),\dots,\beta^\dagger_N(\bm r),\beta_1(\bm r),\dots,\beta_N(\bm r)]$ is defined as a $2N$-component vector composed of the field operators $\beta_a(\bm r)$ and $\beta_a^\dagger(\bm r)$, which represent the creation and annihilation operators of bosons with index $a$ at position $\bm r$.
These bosonic operators satisfy the following commutation relations: $[\beta_a(\bm r_1),\beta_b^\dagger(\bm r_2)]=\delta_{ab}\delta(\bm r_1-\bm r_2)$ and $[\beta_a(\bm r_1),\beta_b(\bm r_2)]=[\beta_a^\dagger(\bm r_1),\beta_b^\dagger(\bm r_2)]=0$.
For simplicity, we introduce the notation $\int_{\bm r}=\int_{\mathbb{R}^{d}}d\bm r$ and $\int_{\bm p}=\int_{\mathbb{R}^{d}}d\bm p/(2\pi\hbar)^{d}$ to denote integrals over position and momentum, respectively.
Since the Hamiltonian is Hermitian, the condition $H_{ab}(\bm r_1,\bm r_2)=H_{ba}^*(\bm r_2,\bm r_1)$ must be satisfied.
In addition, we assume that the Hamiltonian possesses particle-hole symmetry, which imposes the constraint $\sigma_1 H(\bm r_1,\bm r_2)\sigma_1 = H^*(\bm r_1,\bm r_2)$, where $\sigma_1=
\begin{psmallmatrix}
0 & I_N\\
I_N &0
\end{psmallmatrix}$
is a $2N\times 2N$ matrix with $I_N$ denoting the $N\times N$ identity matrix.

Next, we introduce the density matrix as
\begin{align}\label{def_density_matrix}
  F_{ab}(\bm r_1,\bm r_2; t)\equiv \means{\Psi_b^\dagger(\bm r_1,t)\Psi_a(\bm r_2,t)},
\end{align}
where $\Psi_a(\bm r,t)=e^{i\mathscr H t/\hbar}\Psi_a(\bm r)e^{-i\mathscr H t/\hbar}$ and $\Psi_a^\dagger(\bm r,t)=e^{i\mathscr H t/\hbar}\Psi_a^\dagger(\bm r)e^{-i\mathscr H t/\hbar}$ are the bosonic field operators in the Heisenberg picture.
Using this definition, the time evolution of the density matrix is governed by the Liouville-von Neumann equation:
\begin{align}\label{L-vN_eq}
  &i\hbar\partial_t F_{ab}(\bm r_1,\bm r_2; t)\notag \\
  &=\sum_c\int_{\bm r}\Bigl\{\bar{H}_{ac}(\bm r_1,\bm r)F_{cb}(\bm r,\bm r_2; t) - F_{ac}(\bm r_1,\bm r; t)\bar{H}^{\dagger}_{cb}(\bm r,\bm r_2)\Bigr\},
\end{align}
where $\bar{H}(\bm r_1,\bm r_2)=\sigma_3 H(\bm r_1,\bm r_2)$ with 
$\sigma_3=
\begin{psmallmatrix}
   I_N & 0 \\
   0 & -I_N
\end{psmallmatrix}$,
and we introduce abbreviated notation for partial derivatives, such as $\partial_t=\frac{\partial}{\partial t}$.
The appearance of $\sigma_3$ in Eq.~\eqref{L-vN_eq} originates from bosonic statistics, which differ from fermionic statistics.
To proceed further, we employ the Wigner transformation to express the Hamiltonian and the density matrix as functions in phase space.

Here, we define the Wigner transform of a function $A(\bm r_1,\bm r_2)$ with $\bm r_1,\bm r_2\in \mathbb R^d$, which is given by
\begin{align}
  \mathsf A(\bm X,\bm p)=\int_{\bm x}e^{-i\frac{\bm p\cdot\bm x}{\hbar}}A\left(\bm X+\frac{\bm x}{2},\bm X-\frac{\bm x}{2}\right),
\end{align}
where $\bm p\in \mathbb R^d$, and we introduce the center-of-mass coordinate $\bm X=(\bm r_{1}+\bm r_{2})/2$ and the relative coordinate $\bm x=\bm r_{1}-\bm r_{2}$.
This transformation can be regarded as a Fourier transform with respect to the relative coordinate $\bm x$, and $\bm p$ represents the corresponding relative momentum.
Moreover, we introduce the star product (Moyal product) to express the convolution operation in the Wigner representation as follows:
\begin{align}
\label{def_star}
  \mathsf A \star \mathsf B
  &=\mathsf A \exp\left(\frac{i\hbar}{2}\sum_{\alpha\beta}\epsilon_{\alpha\beta}\backvec{\partial}_\alpha\vec{\partial}_{\beta}\right) \mathsf B\notag\\
  &=\mathsf A \mathsf B + \frac{i\hbar}{2}\sum_{j}\Bigl\{(\partial_{X_j}\mathsf A)(\partial_{p_j}\mathsf B) - (\partial_{p_j}\mathsf A)(\partial_{X_j}\mathsf B)\Bigr\} + \cdots,
\end{align}
where the indices $\alpha$ and $\beta$ run over the $2d$ variables $X_1,\dots ,X_d,p_1,\dots ,p_d$, and the set of partial differential operators is defined as
$\{\partial_\alpha\}_{\alpha=1}^{2d}=\{\partial_{X_1},\dots,\partial_{X_d},\partial_{p_1},\dots,\partial_{p_d}\}$.
In addition, $\epsilon_{\alpha\beta}$ is an antisymmetric tensor satisfying $\epsilon_{X_i p_j}=-\epsilon_{p_i X_j}=\delta_{ij}$ and $\epsilon_{X_i X_j}=\epsilon_{p_i p_j}=0$.
In the second line of Eq.~\eqref{def_star}, the arrows on the partial differential operators indicate whether the operator acts on the function to its left or right; namely, $f\backvec{\partial}_\alpha g=(\partial_\alpha f) g$ and $f\vec{\partial}_\alpha g=f\partial_\alpha g$ for functions $f$ and $g$.
When $\mathsf A$ and $\mathsf B$ are matrices, the star product $\mathsf A\star\mathsf B$ is defined to include both the star product given in Eq.~\eqref{def_star} and matrix multiplication.
Note that the star product satisfies the associative law, $(\mathsf A\star\mathsf B)\star\mathsf C=\mathsf A\star(\mathsf B\star\mathsf C)$; therefore, the star product of three elements can be written unambiguously as $\mathsf A\star\mathsf B\star\mathsf C$.
The proof is provided in Appendix~\ref{conv_star}.
A convolution integral, for example, the one appearing in Eq.~\eqref{L-vN_eq}, is replaced by the star product in the Wigner representation (see Appendix~\ref{conv_star}); hence, the star product plays a crucial role in this representation.
In Eq.~\eqref{def_star}, the spatial derivatives and $\hbar$ appear at the same order, indicating that the order of spatial gradients can be systematically controlled by the order of $\hbar$.

Using the above properties of the star product, we rewrite the Liouville-von Neumann equation~\eqref{L-vN_eq} in the Wigner representation as
\begin{align}\label{qke}
  i\hbar \partial_t \mathsf F = \bar{\mathsf H} \star \mathsf F - \mathsf F \star \bar{\mathsf H}^{\dagger},
\end{align}
where the $2N \times 2N$ matrices $\mathsf H = \mathsf H(\bm X, \bm p)$ and $\mathsf F = \mathsf F(\bm X, \bm p; t)$ denote the Wigner representations of $H(\bm r_1, \bm r_2)$ and $F(\bm r_1, \bm r_2; t)$, respectively, and $\bar{\mathsf H} = \sigma_3 \mathsf H$.
Here, we assume that the density matrix vanishes, i.e., $\mathsf F \to 0$, in the limit $|X_i|, |p_i| \to \infty$ so as to eliminate surface terms in phase-space integrals.
Equation~\eqref{qke} is referred to as the quantum kinetic equation~\cite{Mangeolle2024,Park2025}.

In this study, we focus on bosonic systems with zero chemical potential.
In such systems, the heat current is equivalent to the energy current; therefore, we first derive the energy density and then define the heat current as the energy current using the equation of continuity.
The expectation value of the total energy is given by
\begin{align}
  \means{\mathscr H}=\frac{1}{2}\sum_{ab}\int_{\bm r_1,\bm r_2}F_{ab}(\bm r_1,\bm r_2; t)H_{ba}(\bm r_2,\bm r_1),
\end{align}
which is independent of time due to energy conservation.
Using the Wigner representations $\mathsf F(\bm X,\bm p; t)$ and $\mathsf H(\bm X,\bm p)$, this expression can be rewritten as
\begin{align}
  \means{\mathscr H}=\frac{1}{2}\int_{\bm X,\bm p}\mTr(\mathsf F\star\mathsf H).
\end{align}
From this expression, we define the energy density $\mathcal H(\bm X;t)$ in phase space so that it satisfies $\means{\mathscr H}=\int_{\bm X}\mathcal H(\bm X;t)$.
Note that whereas the total energy $\means{\mathscr H}$ is time-independent due to energy conservation, the energy density $\mathcal H(\bm X;t)$ may depend on time.
Although the cyclic property of the trace holds under integration over the entire phase space, i.e., $\int_{\bm X,\bm p}\mTr(\mathsf A\star\mathsf B)=\int_{\bm X,\bm p}\mTr(\mathsf B\star\mathsf A)$ (see Appendix~\ref{conv_star} for a proof), the star product itself does not, in general, preserve the cyclic property of the trace, i.e., $\mTr(\mathsf A\star\mathsf B)\neq\mTr(\mathsf B\star\mathsf A)$.
Hence, the expression for the energy density is not uniquely determined~\cite{Hardy1963,Kapustin2020}.
Here, we adopt the symmetrized expression of $\mathsf F$ and $\mathsf H$ as the energy density~\footnote{The expression given in Eq.~\eqref{def_energy_density} does not completely eliminate the ambiguity in the definition of the energy density, as an uncertainty remains that is related to derivatives with respect to the position $\bm X$. 
The energy density is defined as a function $\mathcal H(\bm X)=\mathcal H^*(\bm X)$ that satisfies $\means{\mathscr H}=\int_{\bm X}\mathcal H(\bm X)$.
Under this definition, the energy density remains ambiguous in the sense that it can be modified to $\mathcal H(\bm X)+\partial_{\bm X}\cdot\bm F(\bm X)$ by adding any real function $\bm F(\bm X)$ that vanishes as $|X_i|\to\infty$.}:
\begin{align}\label{def_energy_density}
  \mathcal H=\frac{1}{4}\int_{\bm p}\mTr(\mathsf F\star\mathsf H+\mathsf H\star\mathsf F)=\frac{1}{2}\int_{\bm p}\Re\mTr(\mathsf F\star\mathsf H).
\end{align}

Since $\mathsf H$ is time independent, taking the time derivative of the energy density yields
\begin{align}\label{partial_tH}
  \partial_t\mathcal H = \frac{1}{2\hbar}\int_{\bm p}\Im\mTr([\bar{\mathsf H}\stco\mathsf F\star \mathsf H]).
\end{align}
Here, we define the commutator under the star product as $[\mathsf A\stco\mathsf B]=\mathsf A\star\mathsf B-\mathsf B\star\mathsf A$.
Because of total energy conservation, $\int_{\bm X}\partial_t\mathcal H=0$, the time derivative of the energy density can always be expressed in the form of a divergence.
In the collisionless limit, we define the energy current density $\mathcal J_i(\bm X)$ through the continuity equation:
\begin{align}\label{continuity_eq}
  \partial_t\mathcal H + \sum_{i}\partial_{X_i}\mathcal J_i = 0.
\end{align}
By expanding the star product in Eq.~\eqref{partial_tH} with respect to the spatial gradient, we derive an explicit expression for the energy current density up to second order in the spatial gradient, namely, second order in $\hbar$:
\begin{align}\label{energy_current}
  \mathcal J_i
  =&\frac{1}{2}\int_{\bm p}\Re\mTr\left[\partial_{p_i}\bar{\mathsf H}(\mathsf F\star\mathsf H)\right]
  \notag\\
  &-\frac{\hbar^2}{48}\sum_{jk}\partial^2_{X_jX_k}\int_{\bm p}\Re\mTr\left[\partial^3_{p_ip_jp_k}\bar{\mathsf H}\,\mathsf F\mathsf H\right]
  +\mathcal O(\hbar^3).
\end{align}
In this expression, we use abbreviated notations for partial derivatives such as $\partial_{\alpha\beta}^2=\partial_{\alpha}\partial_{\beta}$, and these derivatives act only on the immediately following symbol for simplicity.
For example, $\partial_{p_ip_jp_k}^3\bar{\mathsf H}\,\mathsf F\mathsf H$ indicates $(\partial_{p_i}\partial_{p_j}\partial_{p_k}\bar{\mathsf H})\mathsf F\mathsf H$.
These notations are also used in the following sections.

\subsection{Diagonalization under star product}
\label{diag_star}

In Eq.~\eqref{energy_current}, the heat current density is expressed in terms of the Wigner representations of the Hamiltonian and the density matrix; therefore, to evaluate the heat current, we need to diagonalize the Hamiltonian within the Wigner representation rather than treat it as a conventional matrix diagonalization problem.
Conventionally, the bosonic BdG Hamiltonian in momentum space is diagonalized using a paraunitary matrix that preserves the bosonic commutation relations~\cite{Colpa1978,Xiao2009}, where a paraunitary matrix $T$ satisfies $T\sigma_3 T^\dagger = T^\dagger \sigma_3 T = \sigma_3$.
In a similar manner, we introduce a star-paraunitary matrix ${\mathsf T}(\bm X,\bm p)$ in the Wigner representation, which satisfies the following condition:
\begin{align}\label{para_condition}
  {\mathsf T}\star\sigma_3\star {\mathsf T}^\dagger={\mathsf T}^\dagger\star\sigma_3\star {\mathsf T}=\sigma_3.
\end{align}
Note that since $\sigma_3$ is independent of $\bm X$ and $\bm p$, the above equation can be written as ${\mathsf T}\star\sigma_3\star {\mathsf T}^\dagger=({\mathsf T}\sigma_3)\star {\mathsf T}^\dagger={\mathsf T}\star(\sigma_3 {\mathsf T}^\dagger)$.
Using this star-paraunitary matrix, we can diagonalize the Hamiltonian $\mathsf H$ in the Wigner representation while preserving the bosonic commutation relations as follows:
\begin{align}
  {\mathsf T}^{\dagger}\star\mathsf H\star {\mathsf T}\equiv \tilde h^{\prime}=\mathrm{diag}(\tilde h^{\prime}_1,\dots,\tilde h^{\prime}_{N},\tilde h^{\prime}_{N+1},\dots,\tilde h^{\prime}_{2N}),
\end{align}
where we assume that all eigenvalues $\tilde h^{\prime}_n$ are positive.
From this equation, we obtain
\begin{align}\label{diag_H}
  {\mathsf T}^{-1}\star\bar{\mathsf H}\star {\mathsf T}=\tilde h=\mathrm{diag}(\tilde h^{\prime}_1,\dots,\tilde h^{\prime}_{N},-\tilde h^{\prime}_{N+1},\dots,-\tilde h^{\prime}_{2N}),
\end{align}
where $\sigma_3\tilde h^{\prime}=\tilde h$, and we define the inverse matrix of ${\mathsf T}$ such that ${\mathsf T}^{-1}\star {\mathsf T}={\mathsf T}\star {\mathsf T}^{-1}=I_{2N}$.
In particular, when the Hamiltonian does not depend on the center-of-mass coordinate $\bm X$, namely, when $\mathsf H$ is written as $\mathsf H(\bm p)$, the star product reduces to the ordinary matrix product; hence, the diagonalization problem in Eq.~\eqref{diag_H} becomes equivalent to the conventional matrix diagonalization problem, leading to the equivalence between the energy eigenvalues $\tilde h(\bm p)$ and the dispersion relations of bosonic particle bands.

To obtain the eigenvalues $\tilde h$ and the transformation matrix ${\mathsf T}$ that satisfy Eq.~\eqref{diag_H}, we expand the star product in Eq.~\eqref{diag_H} in powers of $\hbar$.
In particular, at the zeroth order in $\hbar$, Eq.~\eqref{diag_H} reduces to the conventional matrix diagonalization problem as
\begin{align}\label{diag_H0}
  ({\mathsf T}^{(0)})^{-1}\bar{\mathsf H} {\mathsf T}^{(0)}=\tilde h^{(0)},
\end{align}
where ${\mathsf T}^{(0)}$ and $\tilde{h}^{(0)}$ denote the zeroth-order terms of ${\mathsf T}$ and $\tilde{h}$ in the $\hbar$ expansion, respectively.
The higher-order terms of ${\mathsf T}$ and $\tilde{h}$ can be systematically obtained from the flow equation introduced in Sec.~\ref{grad_expansion}~\cite{Park2025}.

In Eq.~\eqref{energy_current}, we find that the density matrix $\mathsf F$ always appears in the form $\mathsf F\star\mathsf H$ in the expression for the heat current density.
Thus, under the transformation that diagonalizes the Hamiltonian, $\tilde h^{\prime}={\mathsf T}^{-1}\star\bar{\mathsf H}\star{\mathsf T}$, the density matrix transforms as
\begin{align}\label{eq:tilde_F}
  \tilde F={\mathsf T}^{-1}\star\mathsf F\star\sigma_3\star{\mathsf T}.
\end{align}
Note that, since ${\mathsf T}$ is merely the transformation matrix that diagonalizes $\mathsf H$, $\tilde F$ is not necessarily diagonal.
For later convenience, we decompose $\tilde F$ into its diagonal and off-diagonal components as $\tilde F=\tilde f+\tilde{\mathcal F}$, where $\tilde f$ and $\tilde{\mathcal F}$ denote the diagonal and off-diagonal parts, respectively.

\subsection{Gauge transformation}
\label{subsec:gauge}

In the diagonalization of the bosonic BdG Hamiltonian using the star paraunitary matrix ${\mathsf T}$, it is important to note that there exists a gauge redundancy in the choice of the transformation matrix.
Here, we introduce a gauge transformation for the star paraunitary matrix ${\mathsf T}$ as
\begin{align}\label{eq:gauge_T}
  {\mathsf T}\to {\mathsf T}\star \mathrm E^{i\theta},
\end{align}
which does not change the star paraunitary condition given in Eq.~\eqref{para_condition} and retains the diagonalized form in Eq.~\eqref{diag_H} as long as $\theta=\theta(\bm X,\bm p)$ is a diagonal matrix written as $\theta\equiv \mathrm{diag}(\theta_1,\dots,\theta_{2N})$, where $\mathrm E^{i\theta}$ is defined as $\mathrm E^{\lambda}\equiv 1+\lambda+\frac{1}{2!}\lambda\star\lambda+\cdots$, and satisfies $\mathrm E^{\lambda}\star\mathrm E^{-\lambda}=\mathrm E^{-\lambda}\star\mathrm E^{\lambda}=I_{2N}$.
Under this gauge transformation, the diagonalized Hamiltonian transforms as $\tilde{h}\to \mathrm E^{-i\theta}\star\tilde{h}\star\mathrm E^{i\theta}$ and is not gauge invariant, although it remains diagonal.
In a similar manner, one can show that the diagonal part of the density matrix, $\tilde f$, is not gauge invariant.
By expanding the star product, we find that under the gauge transformation a diagonal element $\tilde d$ (where $\tilde d=\tilde h,\tilde f$) transforms as
\begin{align}\label{tilde_d}
    \tilde{d}
  \to\tilde{d}-&\hbar\sum_{\alpha\beta}\epsilon_{\alpha\beta}\partial_{\alpha}\tilde{d}\left(\partial_{\beta}\theta+\frac{\hbar}{2}\sum_{\gamma\sigma}\epsilon_{\gamma\sigma}\partial_{\beta\sigma}^2\theta\,\partial_{\gamma}\theta\right)
  \notag\\
  &\qquad
  +\frac{\hbar^2}{2}\sum_{\alpha\beta\gamma\sigma}\epsilon_{\alpha\beta}\epsilon_{\gamma\sigma}\partial_{\alpha\gamma}^2\tilde{d}\,\partial_{\beta}\theta\,\partial_{\sigma}\theta+\mathcal O(\hbar^3)
\end{align}
To describe physical observables such as the total energy and the heat current density, we need to use gauge-invariant quantities.
The gauge-invariant expressions for the Hamiltonian and the density matrix are given by
\begin{align}
  \label{diag_h}
  h\equiv&
  \tilde{h}+\hbar\sum_{\alpha\beta}\epsilon_{\alpha\beta}\partial_{\alpha}\tilde h
  \left(A_\beta+\hbar\sum_{\gamma\sigma}\epsilon_{\gamma\sigma}A_{\sigma}\,\partial_\gamma A_\beta\right)\notag\\
  &-\frac{\hbar^2}{4}\sum_{\alpha\beta\gamma\sigma}\epsilon_{\alpha\beta}\epsilon_{\gamma\sigma}
  \partial_{\alpha\gamma}^2\tilde h\left(\mathrm{diag}(\{\Lambda_\beta,\Lambda_\sigma\})-4A_\beta A_\sigma\right)
  +\mathcal O(\hbar^3)
\end{align}
and
\begin{align}
  \label{diag_f}
  f\equiv
  &\tilde{f}+\hbar\sum_{\alpha\beta}\epsilon_{\alpha\beta}\partial_{\alpha}(\tilde f A_\beta)\notag\\
  &+\frac{\hbar^2}{4}\sum_{\alpha\beta\gamma\sigma}\epsilon_{\alpha\beta}\epsilon_{\gamma\sigma}\partial_{\alpha\gamma}^2\left(\tilde f \,\mathrm{diag}(\{\Lambda_\beta,\Lambda_\sigma\})\right)
  +\mathcal O(\hbar^3),
\end{align}
respectively, up to second order in $\hbar$~\cite{Park2025}.
Here, we introduce $\Lambda_\alpha=-i{\mathsf T}^{-1}\star\partial_\alpha {\mathsf T}$ and $A_\alpha=\mathrm{diag}(\Lambda_\alpha)$, where $(\Lambda_\alpha)_{nm}$ with $n\neq m$ corresponds to the inter-band Berry connection, and $(\Lambda_\alpha)_{nn}=A_{n,\alpha}$ corresponds to the intra-band Berry connection.

Moreover, from the star-paraunitarity condition for ${\mathsf T}$ in Eq.~\eqref{para_condition}, we find that the relation $A_\alpha^*=\mathrm{diag}(i\sigma_3\star\partial_\alpha {\mathsf T}^{-1}\star {\mathsf T}\star\sigma_3)=A_\alpha$ holds, which indicates that $A_\alpha$ is a diagonal matrix with real components.
Under the gauge transformation, $\Lambda_\alpha$ and $A_\alpha$ transform as
\begin{align}\label{eq:gauge_Lambda}
  \Lambda_\alpha
  \to e^{-i\theta}\left(\Lambda_\alpha+\frac{\hbar}{2}\sum_{\gamma\sigma}\epsilon_{\gamma\sigma}(i\partial_\gamma\theta\,\Lambda_\alpha\,\partial_\sigma\theta-\{\partial_\gamma\Lambda_\alpha,\partial_\sigma\theta\})\right)e^{i\theta}\notag\\
  +\partial_\alpha\theta+\frac{\hbar}{2}\sum_{\gamma\sigma}\epsilon_{\gamma\sigma}\partial_\gamma\theta\,\partial_{\alpha\sigma}^2\theta+\mathcal O(\hbar^2)
\end{align}
and
\begin{align}\label{eq:gauge_A}
  A_\alpha
  \to A_\alpha+\partial_\alpha\theta
  -\hbar\sum_{\gamma\sigma}\epsilon_{\gamma\sigma}\left(\partial_\gamma A_\alpha\,\partial_\sigma\theta+\frac{1}{2}\partial_\gamma\theta\,\partial_{\alpha\sigma}^2\theta\right)+\mathcal O(\hbar^2),
\end{align}
respectively.
Using these transformation rules, we can verify that $h$ and $f$, introduced in Eqs.~\eqref{diag_h} and \eqref{diag_f}, are gauge-invariant up to second order in $\hbar$.

\subsection{Quantum geometric tensor in phase space}
\label{subsec:QGT}

As mentioned at the beginning of Sec.~\ref{diag_star}, conventionally, the bosonic BdG Hamiltonian written as $H_{\bm k}$ in momentum space is diagonalized using a $2N\times 2N$ paraunitary matrix $T_{\bm k}$.
Here, the projection operator onto the $n$-th band is introduced as~\cite{Shindou2013}
\begin{align}
  P_{n\bm k}\equiv T_{\bm k}p_n T_{\bm k}^{-1},
\end{align}
where $p_n$ is defined as $(p_n)_{ab}\equiv \delta_{na}\delta_{nb}$, which is a $2N\times 2N$ matrix whose only nonzero element is the $(n,n)$ component, equal to $1$.
Using this projection operator, the quantum geometric tensor in momentum space for bosonic systems is defined as~\cite{Koyama2025,Tesfaye2025}
\begin{align}\label{eq:QGT1}
  \mathcal T_{n\bm k,ij}\equiv\mTr\left[\partial_{k_i} P_{n\bm k}(I_{2N}-P_{n\bm k})\partial_{k_j} P_{n\bm k}\right].
\end{align}

Based on the conventional definition of the quantum geometric tensor in momentum space, we introduce the quantum geometric tensor in phase space, which is defined so as to be gauge invariant under the transformation given in Eq.~\eqref{eq:gauge_T} and to coincide with $\mathcal T_{n\bm k,ij}$ given in Eq.~\eqref{eq:QGT1} when the Hamiltonian $\mathsf H(\bm X,\bm p)$ depends only on the momentum $\bm p$.
To this end, we first introduce the projection operator in phase space as
\begin{align}\label{def_P_n}
  P_n(\bm X,\bm p)\equiv {\mathsf T}(\bm X,\bm p)\star p_n\star {\mathsf T}^{-1}(\bm X,\bm p),
\end{align}
where $(p_n)_{ab}\equiv \delta_{na}\delta_{nb}$.
Since $P_n$ is not Hermitian due to the relation $P_n^\dagger=\sigma_3P_n\sigma_3$, it is not an orthogonal projection.
Nevertheless, from its definition, we can easily verify that $P_n\star P_m=\delta_{nm}P_n$ and $\sum_{n=1}^{2N}P_n=I_{2N}$ hold.
Moreover, under the gauge transformation given in Eq.~\eqref{eq:gauge_T}, we find that
  $P_n \to {\mathsf T}\star E^{i\theta}\star p_n \star E^{-i\theta}\star {\mathsf T}^{-1}
  =P_n$,
indicating that $P_n$ is gauge invariant.

Using the projection operator $P_n$ defined in Eq.~\eqref{def_P_n}, we introduce the quantum geometric tensor in phase space as follows.
\begin{align}\label{eq:def_QGT}
  \mathcal T_{n,\alpha\beta}\equiv\mTr(\partial_\alpha P_n\star(I_{2N}-P_n)\star\partial_\beta P_n).
\end{align}
Because $P_n$ is gauge invariant, the quantum geometric tensor defined above is also gauge invariant.
Moreover, when the Hamiltonian $\mathsf H(\bm X,\bm p)$ does not depend on the center-of-mass coordinate $\bm X$, the transformation matrix $\mathsf T$ depends only on $\bm p$, and the star product reduces to the ordinary matrix product.
This reduction leads to the equivalence between the quantum geometric tensor defined in Eq.~\eqref{eq:def_QGT} and that defined in Eq.~\eqref{eq:QGT1} in momentum space.

In the quantum geometric tensor $\mathcal T_{n,\alpha\beta}$ defined in Eq.~\eqref{eq:def_QGT}, the real part corresponds to the quantum metric $g_{n,\alpha\beta}=\Re(\mathcal T_{n,\alpha\beta})$, whereas the imaginary part corresponds to the Berry curvature, $\Omega_{n,\alpha\beta}=2\Im(\mathcal T_{n,\alpha\beta})$.
Note that, from the relation $\mathcal T_{n,\alpha\beta}^*=\mathcal T_{n,\beta\alpha}$, it follows that the quantum metric $g_{n,\alpha\beta}$ is symmetric with respect to the exchange of the indices $\alpha$ and $\beta$, whereas the Berry curvature $\Omega_{n,\alpha\beta}$ is antisymmetric with respect to the exchange of the indices $\alpha$ and $\beta$.
From Eq.~\eqref{eq:def_QGT}, explicit expressions for the Berry curvature and the quantum metric can be derived as follows:
\begin{align}\label{eq:def_Curv}
  \Omega_{\alpha\beta}&=\partial_\alpha A_\beta-\partial_\beta A_\alpha\notag\\
  &-\hbar\sum_{\gamma\sigma}\epsilon_{\gamma\sigma}\left(\partial_\gamma A_\alpha\partial_\sigma A_\beta+\partial_\gamma\left((\partial_\alpha A_\beta-\partial_\beta A_\alpha)A_\sigma\right)\right)+\mathcal O(\hbar^2),\\
\label{eq:def_QM}
  g_{\alpha\beta}&=\frac12\mathrm{diag}(\Lambda_\alpha\Lambda_\beta+\Lambda_\beta\Lambda_\alpha)-A_\alpha A_\beta+\mathcal O(\hbar).
\end{align}
As shown later in Eq.~\eqref{diag_J}, $\Omega_{\alpha\beta}$ and $g_{\alpha\beta}$ appear in the heat current density at the first and second orders in $\hbar$, respectively; therefore, their expressions are derived up to the first and zeroth orders in $\hbar$, respectively, as presented above.

\subsection{Gauge-invariant heat current density}
\label{subsec:h_and_J}

In this section, we derive gauge-invariant expressions for the energy density and heat current density using the gauge-invariant forms of the diagonalized Hamiltonian and density matrix given in Eqs.~\eqref{diag_h} and \eqref{diag_f}, respectively.
The star product $\mathsf F\star \mathsf H$ appearing in Eq.~\eqref{def_energy_density} is expressed as
\begin{align}\label{FstarH}
  \mathsf F\star\mathsf H=\sum_{n=1}^{2N}P_n\star{d}_n\star P_n+{\mathsf T}\star\tilde{\mathcal F}\star\tilde{h}\star {\mathsf T}^{-1}.
\end{align}
Here, $P_n$ is the projection operator defined in Eq.~\eqref{def_P_n}, and the diagonal matrix $d$ is given by
\begin{widetext}
\begin{align}\label{eq:def_uline_d}
  {d}=\underline{f} h-\hbar^2\sum_{\alpha\beta\gamma\sigma}\epsilon_{\alpha\beta}\epsilon_{\gamma\sigma}\left(\partial_\alpha\underline{f}\,\partial_\gamma h\, g_{\beta\sigma}+\frac{1}{8}\partial_{\alpha\gamma}^2\underline{f} \,\partial_{\beta\sigma}^2 h\right)
  +i\frac{\hbar}{2}\sum_{\alpha\beta}\epsilon_{\alpha\beta}\left(\partial_\alpha\underline{f}\,\partial_\beta h-\hbar\sum_{\gamma\sigma}\epsilon_{\gamma\sigma}\partial_\alpha\underline{f}\,\partial_\gamma h \,\Omega_{\beta\sigma}\right)+\mathcal O(\hbar^3),
\end{align}
where $\underline{f}$ is defined as
\begin{align}\label{eq:uline_f}
  \underline{f}
  &=\left(1-\frac{\hbar}{2}\sum_{\alpha\beta}\epsilon_{\alpha\beta}\Omega_{\alpha\beta}
  -\frac{\hbar^2}{2}\sum_{\alpha\beta\gamma\sigma}\epsilon_{\alpha\beta}\epsilon_{\gamma\sigma}\left(\frac{1}{2}\Omega_{\alpha\gamma}\Omega_{\beta\sigma}-\frac{3}{4}\Omega_{\alpha\beta}\Omega_{\gamma\sigma}+\partial_{\alpha\gamma}^2 g_{\beta\sigma}\right)\right)f
-\hbar^2\sum_{\alpha\beta\gamma\sigma}\epsilon_{\alpha\beta}\epsilon_{\gamma\sigma}\partial_\gamma(\partial_\alpha f g_{\beta\sigma}).
\end{align}
The Hamiltonian $\bar{\mathsf H}$ can be expressed using the projection operator $P_n$ as
\begin{align}\label{proj_h}
  \bar{\mathsf H}=\sum_{n=1}^{2N}P_n\star h_n\star P_n.
\end{align}
Details of the derivations of the expressions for $\mathsf F\star\mathsf H$ and $\bar{\mathsf H}$ are provided in Appendix~\ref{energy_current_calc}.
Here, we assume that $\tilde{\mathcal F}$ does not contribute to the heat current, which is always satisfied when considering a local equilibrium state.
Under this assumption, to calculate Eq.~\eqref{energy_current}, we use Eq.~\eqref{proj_h} and only need to consider the first term in Eq.~\eqref{FstarH} for $\mathsf F\star \mathsf H$.
Although the detailed derivation is provided in Appendix~\ref{energy_current_calc}, we present the final result for the gauge-invariant heat current density as follows:
\begin{align}\label{diag_J}
  \mathcal{J}_i
  &=\frac{1}{2}\int_{\bm p}\mTr\left[f\left(h\partial_{p_i} h-\hbar\sum_{\gamma\sigma}\epsilon_{\gamma\sigma}\partial_\gamma h\left(h\Omega_{p_i \sigma}(1-\hbar\Omega_{X_j p_j})+\hbar\sum_{\mu\nu}\epsilon_{\mu\nu}\partial_{p_i \mu}^2 h g_{\sigma\nu}\right)\vphantom{\frac{\hbar^2}{2}}
  +\frac{\hbar^2}{2}\sum_{\gamma\sigma\mu\nu}\epsilon_{\gamma\sigma}\epsilon_{\mu\nu}\partial_{\gamma\mu}^2 h\left(h\partial_{p_i}g_{\sigma\nu}-\frac{1}{4}\partial_{p_i\sigma\nu}^3 h\right)\right)\right]\notag\\
  &\qquad +\frac{\hbar}{2}\sum_j\partial_{X_j}\int_{\bm p}\mTr[f(b_{ij}^s +b_{ij}^a)]
  +\frac{\hbar^2}{24}\sum_{jk}\partial_{X_j X_k}^2\int_{\bm p}\mTr[f(b_{ijk}^s + 2b_{ij;k}^a)]+\mathcal O(\hbar^3)
\end{align}
where $b_{ij}^s$ and $b_{ij}^a$ are symmetric and antisymmetric tensors with respect to the exchange of the indices $i$ and $j$, respectively, which are defined as
\begin{align}
  b_{ij}^s\equiv\frac{\hbar}{2}\sum_{\mu\nu}\epsilon_{\mu\nu}\left(h\partial_\mu h \,\partial_\nu g_{p_i p_j}+(h\,\partial_{p_i\mu}^2h+\partial_{p_i} h\,\partial_\mu h)g_{p_j\nu}+(h\,\partial_{p_j \mu}^2h+\partial_{p_j} h\,\partial_\mu h)g_{p_i\nu}\right),
\end{align}
and
\begin{align}
  b_{ij}^a\equiv&-h m_{ij}\left(1-\hbar\sum_{k}\Omega_{X_k p_k}\right)
  -\frac{\hbar}{2}\sum_{\mu\nu}\epsilon_{\mu\nu}\left(\partial_{p_j}\left(h\,\partial_\mu h \,g_{p_i\nu}+\frac{1}{4}\partial_\mu h\,\partial_{p_i\nu}^2 h\right)-\partial_{p_i}\left(h\,\partial_\mu h\, g_{p_j\nu}+\frac{1}{4}\partial_\mu h \,\partial_{p_j\nu}^2 h\right)\right).
\end{align}
We also introduce $b_{ijk}^s$ as a symmetric tensor with respect to the exchange of the indices $i$, $j$, and $k$, which is defined as
\begin{align}
  b_{ijk}^s\equiv-\frac{1}{4}\partial_{p_i p_j p_k}^3 h^2,
\end{align}
and $b_{ij;k}^a$ as
\begin{align}
  b_{ij;k}^a\equiv h\partial_{p_i} c_{jk}-h\partial_{p_j} c_{ik}+2hc_{jk;i}-2hc_{ik;j}-\frac{1}{2}\partial_{p_i} h\,\partial_{p_j p_k}^2 h+\frac{1}{2}\partial_{p_j} h\,\partial_{p_i p_k}^2 h,
\end{align}
\end{widetext}
where $b_{ij;k}^a$ is antisymmetric with respect to the exchange of the first two indices $i$ and $j$, namely, $b_{ij;k}^a=-b_{ji;k}^a$.
Moreover, $m_{ij}$, $c_{ij}$, and $c_{ij;k}$ are diagonal matrices whose $n$-th diagonal components are defined as~\cite{Park2025}
\begin{align}
  \label{eq:def_m}
  m_{n,ij}&\equiv\Im\mTr[\partial_{p_i}P_n\star(\bar{\mathsf H}-h_n)\star \partial_{p_j}P_n]\\
  \label{eq:def_c}
  c_{n,ij}&\equiv\Re\mTr[\partial_{p_i}P_n\star(\bar{\mathsf H}-h_n)\star \partial_{p_j}P_n]\\
  \label{eq:def_c2}
  c_{n,ij;k}&\equiv\Re\mTr[\partial_{p_ip_j}^2P_n\star(\bar{\mathsf H}-h_n)\star \partial_{p_k}P_n],
\end{align}
Note that $m_{ij}$ corresponds to the energy magnetization matrix introduced in Refs.~\cite{Cooper1997,Qin2011}; in particular, it exactly coincides with the energy magnetization expressed in terms of Bloch wave functions when the system possesses translational symmetry.
Moreover, as shown in Sec.~\ref{sec:2_THE}, the last two terms in Eq.~\eqref{diag_J}, which are expressed in the form of spatial derivatives, do not contribute to the nonlinear thermal conductivity in this study.

\subsection{Gradient expansion of Hamiltonian}
\label{grad_expansion}

As discussed in Sec.~\ref{diag_star}, the diagonalization of the Hamiltonian in the Wigner representation requires expanding the star product in Eq.~\eqref{diag_H} and solving the resulting equations order by order in $\hbar$.
At zeroth order in $\hbar$, as seen from Eq.~\eqref{diag_H0}, we only need to perform ordinary matrix diagonalization, which yields $\tilde{h}^{(0)}$ as the zeroth-order term of the energy eigenvalue and $T^{(0)}$ as the zeroth-order term of the corresponding eigenvector.
Furthermore, from Eq.~\eqref{diag_h}, since $h=\tilde{h}+\mathcal O(\hbar)$ holds, we obtain $h^{(0)}=\tilde{h}^{(0)}$ as the zeroth-order term of the gauge-invariant Hamiltonian.
In this section, we introduce a systematic framework for computing higher-order energy eigenvalues and eigenvectors that satisfy Eq.~\eqref{diag_H}, taking the zeroth-order terms as the starting point.
This procedure enables us to obtain correction terms to $\tilde{h}^{(0)}$ and ${\mathsf T}^{(0)}$ that arise from the spatial modulation of the Hamiltonian in phase space~\cite{Park2025}.

We focus on the Hamiltonian $\tilde{h}$ and the Berry connection $\Lambda_\alpha$, which are particularly important for calculating correction terms, as can be seen from the expression for the heat current density in Eq.~\eqref{diag_J}.
For these quantities, we introduce the following notation with respect to the order of spatial gradients (or, equivalently, the order in $\hbar$):
\begin{align}
  \tilde{h}&=\tilde{h}^{(0)}+\tilde{h}^{(1)}+\cdots,\\
  \Lambda_\alpha&=\Lambda_\alpha^{(0)}+\Lambda_\alpha^{(1)}+\cdots,
\end{align}
where $\tilde{h}^{(n)}$ and $\Lambda_\alpha^{(n)}$ denote the $n$th-order terms in $\hbar$.
With this notation, the zeroth-order term of the Berry connection is given by $\Lambda_\alpha^{(0)}=-i({\mathsf T}^{(0)})^{-1}\partial_\alpha {\mathsf T}^{(0)}$.

Next, we derive the equation connecting quantities at different orders, referred to as the flow equation~\cite{Park2025}.
By differentiating both sides of the relation ${\mathsf T}^{-1}\star {\mathsf T}=I_{2N}$ with respect to $\hbar$, we obtain
\begin{align}
  \partial_\hbar {\mathsf T}^{-1}\star {\mathsf T}+\frac{i}{2}\sum_{\alpha\beta}\epsilon_{\alpha\beta}\partial_\alpha {\mathsf T}^{-1}\star \partial_\beta {\mathsf T} + {\mathsf T}^{-1}\star\partial_\hbar {\mathsf T}=0.
\end{align}
From this equation and the star-paraunitarity condition for ${\mathsf T}$ given in Eq.~\eqref{para_condition}, the following relation is obtained:
\begin{align}
  Q+\sigma_3 Q^{\dagger}\sigma_3=-\frac{i\hbar}{2}\sum_{\alpha\beta}\epsilon_{\alpha\beta}\Lambda_\alpha\star\Lambda_\beta,
\end{align}
where $Q\equiv \hbar {\mathsf T}^{-1}\star\partial_\hbar {\mathsf T}$.
This $2N\times 2N$ matrix $Q$ can be decomposed into a pseudo-Hermitian component and an anti-pseudo-Hermitian component $Y$ as~\footnote{$2N\times 2N$ matrices $A$ and $B$ are called pseudo-Hermitian and anti-pseudo-Hermitian, respectively, if they satisfy the relations $A^{\dagger}=\sigma_3 A\sigma_3$ and $B^{\dagger}=-\sigma_3 B\sigma_3$.
An arbitrary $2N\times 2N$ matrix $C$ can be decomposed into pseudo-Hermitian and anti-pseudo-Hermitian components as $C=\frac{1}{2}(C+\sigma_3 C^{\dagger}\sigma_3)+\frac{1}{2}(C-\sigma_3 C^{\dagger}\sigma_3)$.}
\begin{align}
  Q=Y-\frac{i\hbar}{4}\sum_{\alpha\beta}\epsilon_{\alpha\beta}\Lambda_\alpha\star\Lambda_\beta.
\end{align}
With these preparations, differentiating both sides of Eq.~\eqref{diag_H} with respect to $\hbar$ yields the flow equation for $\tilde{h}$:
\begin{align}
  \hbar\partial_\hbar \tilde{h}
  &=[\tilde{h}\stco Y]+\frac{\hbar}{2}\sum_{\alpha\beta}\epsilon_{\alpha\beta}\{\Lambda_\alpha\stco\partial_\beta\tilde{h}\}\notag\\
  &\hphantom{{}={}}+\frac{i\hbar}{4}\sum_{\alpha\beta}\epsilon_{\alpha\beta}\{\Lambda_\alpha\stco[\Lambda_\beta\stco\tilde{h}]\},
\end{align}
where we define the anticommutator under the star product as $\{\mathsf A\stco\mathsf B\}=\mathsf A\star\mathsf B+\mathsf B\star\mathsf A$.
Using this equation, we obtain the expression for $\tilde{h}^{(1)}$ as follows:
\begin{align}\label{flow_h1}
  \tilde{h}^{(1)}
  &=[\tilde{h}^{(0)},Y^{(1)}]+\frac{\hbar}{2}\sum_{\alpha\beta}\epsilon_{\alpha\beta}\{\Lambda_\alpha^{(0)},\partial_\beta\tilde{h}^{(0)}\}\notag\\
  &\hphantom{{}={}}-\frac{i\hbar}{4}\sum_{\alpha\beta}\epsilon_{\alpha\beta}\{\Lambda_\alpha^{(0)},[\Lambda_\beta^{(0)},\tilde{h}^{(0)}]\},
\end{align}
where $Y^{(1)}$ denotes the first-order term of $Y$ with respect to $\hbar$.
Because the left-hand side is a diagonal matrix, the off-diagonal part of the anti-pseudo-Hermitian matrix $Y^{(1)}$ is given by
\begin{align}\label{Y1}
  &(Y^{(1)})_{nm}=\notag\\
  &-\frac{\hbar}{2(\tilde{h}^{(0)}_n-\tilde{h}^{(0)}_m)}\sum_{\alpha\beta}\epsilon_{\alpha\beta}\left(\{\Lambda_\alpha^{(0)},\partial_\beta\tilde{h}^{(0)}\}+\frac{i}{2}\{\Lambda_\alpha^{(0)},[\Lambda_\beta^{(0)},\tilde{h}^{(0)}]\}\right)_{nm},
\end{align}
which leads to the following expression for $\tilde{h}^{(1)}$:
\begin{align}
  \tilde{h}^{(1)}=-\hbar\sum_{\alpha\beta}\epsilon_{\alpha\beta}\partial_\alpha\tilde{h}^{(0)}A_\beta^{(0)}+\frac{i\hbar}{4}\sum_{\alpha\beta}\epsilon_{\alpha\beta}\,\mathrm{diag}(\{\Lambda_\alpha^{(0)},[\Lambda_\beta^{(0)},\tilde{h}^{(0)}]\}).
\end{align}
Finally, from Eq.~\eqref{diag_h}, the first-order correction to the gauge-invariant Hamiltonian $h$ is obtained as
\begin{align}\label{eq:h1}
  h^{(1)}=\frac{i\hbar}{4}\sum_{\alpha\beta}\epsilon_{\alpha\beta}\,\mathrm{diag}(\{\Lambda_\alpha^{(0)},[\Lambda_\beta^{(0)},h^{(0)}]\}).
\end{align}
By applying the flow equation to higher-order terms, correction terms of arbitrary order can be derived systematically.

In a similar manner, by differentiating $\Lambda_\alpha=-i{\mathsf T}^{-1}\star\partial_\alpha {\mathsf T}$ with respect to $\hbar$, we obtain the flow equation for $\Lambda_\alpha$ as follows:
\begin{align}
  \hbar\partial_\hbar\Lambda_\alpha=[\Lambda_\alpha\stco Y]-i\partial_\alpha Y
  +\frac{\hbar}{4}\sum_{\gamma\sigma}\epsilon_{\gamma\sigma}\{\Lambda_\gamma\stco\partial_\sigma\Lambda_\alpha\}.
\end{align}
From this equation, the first-order correction to the Berry connection is derived as
\begin{align}\label{A1}
  A_\alpha^{(1)}=\mathrm{diag}([\Lambda_\alpha^{(0)},Y^{(1)}])
  +\frac{\hbar}{4}\sum_{\gamma\sigma}\epsilon_{\gamma\sigma}\mathrm{diag}(\{\Lambda_\gamma^{(0)},\partial_\sigma\Lambda_\alpha^{(0)}\}).
\end{align}
In the above expression, we choose ${\mathsf T}$ such that the diagonal part of $Y^{(1)}$ vanishes, because it does not affect the diagonalization of the Hamiltonian.
This can be understood from the fact that the diagonal part of $Y^{(1)}$ does not contribute to $\tilde{h}^{(1)}$, as shown in Eq.~\eqref{flow_h1}.

\section{Nonlinear thermal conductivity tensor in systems with translational symmetry}
\label{sec:2_THE}

In this section, we calculate the second-order nonlinear thermal Hall conductivity in two-dimensional systems with translational symmetry using the results obtained in Sec.~\ref{sec:general_NLTHE}.
Instead of adopting the conventional approach based on the Kubo formula with Luttinger's fictitious gravitational potential~\cite{Luttinger1964}, we compute the thermal Hall conductivity by directly evaluating the heat current density in the presence of a temperature gradient, assuming that a system with edges is in local equilibrium~\cite{Matsumoto2011B,Matsumoto2011L,Mangeolle2024}.
Hereafter, because we consider two-dimensional systems, the phase-space variables are written as $\bm X=(x,y)$ and $\bm{p}=(p_x,p_y)$.

\subsection{Heat current up to second order in spatial gradients}
\label{subsec:thermal_current}

\begin{figure}[t]
\centering
    \includegraphics[width=0.9\columnwidth,clip]{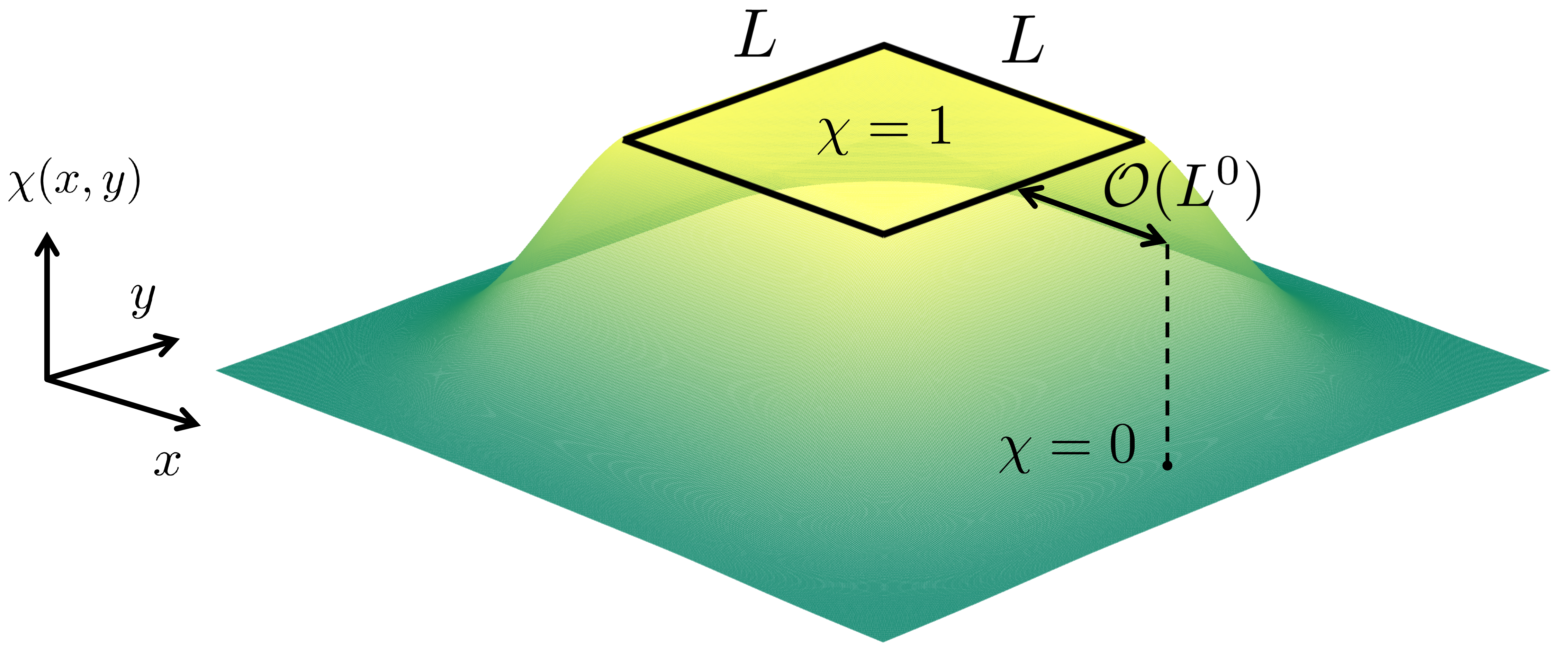}
    \caption{
Schematic illustration of $\chi(x,y)$ introduced in Sec.~\ref{subsec:thermal_current}.
The $L\times L$ square enclosed by the black solid line denotes region $D$, where $\chi(x,y)=1$.
Outside this region, $\chi(x,y)$ decays to zero over a length scale of $\mathcal O(L^0)$.
Here, $\chi(x,y)$ is plotted over its support $D'$.
  }
    \label{fig:chi_xy}
\end{figure}

As in Ref.~\cite{Mangeolle2024}, we assume that the Hamiltonian is given by $\mathsf H(\bm X,\bm p)=\chi(\bm X)\mathsf H_0(\bm p)$, where $\mathsf H_0(\bm p)$ is a $2N\times 2N$ matrix describing a system with translational symmetry, and $\chi(\bm X)=\chi(x,y)$ is a smooth function satisfying $\chi=1$ within a finite region $D=\left\{(x,y)\in\mathbb{R}^2\,\middle|\,-\frac{L}{2}\leq x\leq \frac{L}{2},-\frac{L}{2}\leq y\leq \frac{L}{2}\right\}$ and that decays to zero outside $D$ with a decay length of order $\mathcal{O}(L^0)$~(see Fig.~\ref{fig:chi_xy}).
Here we introduce the region $D'=\left\{(x,y)\in\mathbb{R}^2\,\middle|\,-\frac{L'}{2}\leq x\leq \frac{L'}{2},-\frac{L'}{2}\leq y\leq \frac{L'}{2}\right\}$, where $L'>L$ and $L'-L=\mathcal O(L^0)$, and we assume that $D'$ is the support of $\chi$, i.e., $\chi(x,y)=0$ for $(x,y)\notin D'$.
Note that $L$ is kept finite to define a local-equilibrium region with a temperature gradient, and $\chi(\bm X)$ smoothly connects this region to the outside.

Since $\chi(x,y)$ is a scalar function, we only need to diagonalize $\mathsf H_0$ when diagonalizing the Hamiltonian.
Because $\mathsf H_0$ depends only on $\bm p$, as discussed in Sec.~\ref{diag_star}, we can perform an ordinary matrix diagonalization as ${\mathsf T}_0^{-1}\bar{\mathsf H}_0{\mathsf T}_0\equiv\mathrm{diag}(\varepsilon_1,\dots,\varepsilon_{2N})$, where the transformation matrix is given by ${\mathsf T}^{(0)}(\bm X,\bm p)={\mathsf T}_0(\bm p)$.
Hence, the zeroth-order term of the gauge-invariant Hamiltonian is written as
\begin{align}
  h^{(0)}(\bm{X},\bm p)=\chi(\bm X)\mqty(\dmat{\varepsilon_1(\bm p),\ddots,\varepsilon_{2N}(\bm p)}).
\end{align}
In a similar manner, the zeroth-order term of the Berry connection is given by
\begin{align}
  \label{Conn0}
  \Lambda_{p_i}^{(0)}=-i{\mathsf T}_0^{-1}\partial_{p_i}{\mathsf T}_0,\quad
  \Lambda_{X_i}^{(0)}=0.
\end{align}
From Eq.~\eqref{Y1}, the off-diagonal part of $Y^{(1)}$ is calculated as
\begin{align}
  (Y^{(1)})_{nm}=\hbar\sum_i \frac{\partial_{X_i}\chi}{2\chi}\frac{\varepsilon_n+\varepsilon_m}{\varepsilon_n-\varepsilon_m}(\Lambda_{p_i}^{(0)})_{nm}.
\end{align}
Moreover, from Eq.~\eqref{A1}, the first-order correction term of the Berry connection is obtained as
\begin{align}
  A_{n,p_i}^{(1)}=\hbar \sum_j G^t_{n,ij}\psi_j,
\end{align}
where $\psi_i\equiv \partial_{X_i}\chi/\chi$, and $G^t_{n,ij}$ is the thermal Berry-connection polarizability (TBCP), defined as follows~\cite{Li2024}:
\begin{align}\label{eq:def_TBCP}
  G^t_{n,ij}
  &=-\sum_{m\neq n}^{2N}(\varepsilon_n+\varepsilon_m)\frac{\Re\left[(\Lambda_{p_i}^{(0)})_{nm}(\Lambda_{p_j}^{(0)})_{mn}\right]}{\varepsilon_n-\varepsilon_m},
\end{align}
where we use the relation $\Re\left[(\Lambda_{p_i}^{(0)})_{nm}(\Lambda_{p_j}^{(0)})_{mn}\right]=\frac{1}{2}\left[(\Lambda_{p_i}^{(0)})_{nm}(\Lambda_{p_j}^{(0)})_{mn}+(\Lambda_{p_j}^{(0)})_{nm}(\Lambda_{p_i}^{(0)})_{mn}\right]$.
This TBCP is a symmetric tensor with respect to the exchange of indices $i$ and $j$, namely, $G_{n,ij}^t=G_{n,ji}^t$, and is a thermal analog of the Berry-connection polarizability (BCP)~\cite{Gao2014,Liu2021}, which appears in intrinsic second-order nonlinear conductivity.
Using the above results, the Berry curvature and quantum metric given in Eqs.~\eqref{eq:def_Curv} and \eqref{eq:def_QM} are calculated as
\begin{align}
  \label{Omega_pp}
  \Omega_{p_i p_j}&=\partial_{p_i}A_{p_j}^{(0)}-\partial_{p_j}A_{p_i}^{(0)}\notag\\
  &\hphantom{{}={}}+\hbar\sum_k (\partial_{p_i}G^t_{jk}-\partial_{p_j}G^t_{ik})\psi_k+\mathcal O(\hbar^2),\\
  \label{Omega_Xp}
  \Omega_{X_i p_j}&=\hbar\sum_k G^t_{jk}\partial_{X_i}\psi_k+\mathcal O(\hbar^2),\\
  \label{g_pp}
  g_{n,p_i p_j}&=
  \sum_{m\neq n}^{2N}
  \Re\left[(\Lambda_{p_i}^{(0)})_{nm}(\Lambda_{p_j}^{(0)})_{mn}\right]
  +\mathcal O(\hbar),\\
  \label{Omega_XX}
  \Omega_{X_i X_j}&=0+\mathcal O(\hbar^2),\quad g_{X_i p_j}=g_{X_i X_j}=0+\mathcal O(\hbar).
\end{align}
The matrices $G^t_{ij}$ and $g_{p_i p_j}$ are diagonal matrices with diagonal entries $G^t_{n,ij}$ and $g_{n,p_i p_j}$, respectively.
We also find that the first-order correction of the Hamiltonian given in Eq.~\eqref{eq:h1} vanishes because of Eq.~\eqref{Conn0}, i.e., $h^{(1)}=0$.

As mentioned at the beginning of this section, we consider a local equilibrium state under a spatially varying temperature $T(\bm X)$, which is constructed from the global equilibrium state.
Thus, we first need to determine the gauge-invariant distribution function as a function of temperature under global equilibrium.
The statistical properties of bosonic systems originate from the density operator $F$ given in Eq.~\eqref{def_density_matrix}, which is expressed in terms of the expectation values of the bosonic operators.
Here, by applying the Matsubara formulation to Eq.~\eqref{def_density_matrix}, we introduce the gauge-invariant density matrix $f_{\mathrm{eq}}$ from the Wigner representation of the density operator $\mathsf F$ under global equilibrium.
The details of this procedure are provided in Appendix~\ref{subsec:f_eq}, where we derive the expression for $f_{\mathrm{eq}}$ up to second order in $\hbar$.

Assuming that the system is in local equilibrium under a spatially dependent temperature $T(\bm X)$, the gauge-invariant density matrix $f$ is obtained by replacing the temperature $T$ in $f_{\mathrm{eq}}$ with $T(\bm X)$.
From Eq.~\eqref{eq:f_eq_app} in Appendix~\ref{subsec:f_eq}, $f^{(n)}$, which represents the $n$-th order term of $f$ with respect to $\hbar$, is calculated as
\begin{align}
  f^{(0)}&=n_B(h^{(0)},T)|_{T(\bm X)}+\frac{1}{2},\\
  f^{(1)}&=0,
\end{align}
\begin{widetext}
\begin{align}
  \label{f^(2)}
  f^{(2)}&=h^{(2)}n_B^{\prime}(h^{(0)},T)|_{T(\bm X)}
  +\hbar^2\sum_{ij} G_{ij}^t\partial_{X_i}\psi_j n_B(h^{(0)},T)|_{T(\bm X)}
  +\hbar^2\sum_{ij} \mathcal{G}_{ij}\partial_{X_iX_j}^2 h^{(0)} n_B^{\prime}(h^{(0)},T)|_{T(\bm X)}
  \notag\\&\quad
  +\frac{\hbar^2}{2}\sum_i\partial_{X_i}h^{(0)}\sum_j\partial_{X_j}h^{(0)}\mathcal{G}_{ij}n_B^{\prime\prime}(h^{(0)},T)|_{T(\bm X)}
  -\frac{\hbar^2}{16}\sum_{\alpha\beta\gamma\sigma}\epsilon_{\alpha\beta}\epsilon_{\gamma\sigma}\partial_{\alpha\gamma}^2 h^{(0)}\partial_{\beta\sigma}^2 h^{(0)}n_B^{\prime\prime}(h^{(0)},T)|_{T(\bm X)}
  \notag\\&\quad
  -\frac{\hbar^2}{24}\sum_{\alpha\beta\gamma\sigma}\epsilon_{\alpha\beta}\epsilon_{\gamma\sigma}\partial_{\alpha\gamma}^2 h^{(0)}\partial_{\beta} h^{(0)}\partial_{\sigma} h^{(0)}n_B^{\prime\prime\prime}(h^{(0)},T)|_{T(\bm X)},
\end{align}
\end{widetext}
where $n_B(\varepsilon,T)=(e^{\varepsilon/(k_B T)}-1)^{-1}$ is the Bose-Einstein distribution function at temperature $T$, and $\mathcal{G}_{ij}\equiv g_{p_i p_j}^{(0)}$ is the quantum metric in momentum space defined in Eq.~\eqref{g_pp}.
In addition, $n_B^{\prime}(\varepsilon,T)=\partial_\varepsilon n_B(\varepsilon,T)$ denotes the derivative of $n_B(\varepsilon,T)$ with respect to $\varepsilon$, and $n_B^{\prime\prime}(\varepsilon,T)$ and $n_B^{\prime\prime\prime}(\varepsilon,T)$ are defined similarly.

Here, we calculate the heat current driven by the temperature gradient up to second order in spatial gradients.
The heat current density, defined by $J_i\equiv\frac{1}{V}\int_{\bm X}\mathcal J_i(\bm X)$ for the volume $V=L^2$, is obtained from Eq.~\eqref{diag_J} as
\begin{widetext}
\begin{align}
    \label{int_XJ}
  J_i
  &=\frac{1}{2V}\int_{\bm X,\bm p}\mTr\left[f\left(h\partial_{p_i} h-\hbar\sum_{\gamma\sigma}\epsilon_{\gamma\sigma}\partial_\gamma h\left(h\Omega_{p_i \sigma}(1-\hbar\Omega_{X_j p_j})+\hbar\sum_{\mu\nu}\epsilon_{\mu\nu}\partial_{p_i \mu}^2 h g_{\sigma\nu}\right)\vphantom{\frac{\hbar^2}{2}}
  \right.\right.
  \notag\\
  &\qquad\qquad\qquad\qquad
  \left.\left.
  +\frac{\hbar^2}{2}\sum_{\gamma\sigma\mu\nu}\epsilon_{\gamma\sigma}\epsilon_{\mu\nu}\partial_{\gamma\mu}^2 h\left(h\partial_{p_i}g_{\sigma\nu}-\frac{1}{4}\partial_{p_i\sigma\nu}^3 h\right)\right)\right]+\mathcal O(\hbar^3),
\end{align}
where we omit the second and third terms involving spatial derivatives in Eq.~\eqref{diag_J}, because they reduce to surface terms upon spatial integration and therefore do not contribute to the thermal conductivity.
Thus, by using Eqs.~\eqref{Omega_pp}--\eqref{f^(2)}, the second-order term $J_i^{(2)}$ in $\hbar$ can be expressed as
\begin{align}
  J_i^{(2)}
  =\frac{1}{2V}\int_{\bm X,\bm p}\mTr\Biggl[&h^{(2)}\partial_{p_i}h^{(0)}n_B+h^{(0)}\partial_{p_i}h^{(2)}n_B+h^{(0)}\partial_{p_i}h^{(0)}h^{(2)}n_B^{\prime}
  +\hbar^2 h^{(0)}\partial_{p_i}h^{(0)}\sum_{jk} G_{jk}^t\partial_{X_{j}}\psi_{k} n_B+\hbar^2 h^{(0)}\partial_{p_i}h^{(0)}\sum_{jk} \mathcal{G}_{{jk}}\partial_{X_{j}X_{k}}^2 h^{(0)} n_B^{\prime}
  \notag\\
  &
  +\frac{\hbar^2}{2}h^{(0)}\partial_{p_i}h^{(0)}\sum_{jk}\partial_{X_{j}}h^{(0)}\partial_{X_{k}}h^{(0)}\mathcal{G}_{jk}n_B^{\prime\prime}-\frac{\hbar^2}{16}\sum_{\alpha\beta\gamma\sigma}\epsilon_{\alpha\beta}\epsilon_{\gamma\sigma}h^{(0)}\partial_{p_i}h^{(0)}\partial_{\alpha\gamma}^2 h^{(0)}\partial_{\beta\sigma}^2 h^{(0)}n_B^{\prime\prime}
  \notag\\
  &
  -\frac{\hbar^2}{24}\sum_{\alpha\beta\gamma\sigma}\epsilon_{\alpha\beta}\epsilon_{\gamma\sigma}h^{(0)}\partial_{p_i}h^{(0)}\partial_{\alpha\gamma}^2 h^{(0)}\partial_{\beta} h^{(0)}\partial_{\sigma} h^{(0)}n_B^{\prime\prime\prime}
  -\hbar h^{(0)}\sum_j \partial_{X_j}h^{(0)}\Omega_{p_i p_j}^{(1)}n_B
  +\hbar h^{(0)}\sum_j \partial_{p_j}h^{(0)}\Omega_{p_i X_j}^{(1)}n_B
  \notag\\
  &
  -\hbar^2 \sum_{jk} \partial_{X_j}h^{(0)}\partial_{p_iX_k}^2 h^{(0)} \mathcal{G}_{jk}n_B
  +\frac{\hbar^2}{2} h^{(0)}\sum_{jk}\partial_{X_j X_k}^2h^{(0)}\partial_{p_i}\mathcal{G}_{jk}n_B
  -\frac{\hbar^2}{8}\sum_{\alpha\beta\gamma\sigma}\epsilon_{\alpha\beta}\epsilon_{\gamma\sigma}\partial_{\alpha\gamma}^2 h^{(0)}\partial_{p_i\beta\sigma}^3h^{(0)}n_B\Biggr],
\end{align}
where we introduce the following notation: $n_B=n_B(h^{(0)},T)|_{T(\bm X)}+\frac{1}{2}$, $n_B^{\prime}=n_B^{\prime}(h^{(0)},T)|_{T(\bm X)},n_B^{\prime\prime}=n_B^{\prime\prime}(h^{(0)},T)|_{T(\bm X)}$, and $n_B^{\prime\prime\prime}=n_B^{\prime\prime\prime}(h^{(0)},T)|_{T(\bm X)}$.
To proceed further with the calculations, we use the following relations:
\begin{align}
  \int_{\bm p} (h^{(2)}\partial_{p_i}h^{(0)}n_B+h^{(0)}\partial_{p_i}h^{(2)}n_B+h^{(2)}h^{(0)}\partial_{p_i}h^{(0)}n_B^{\prime})=\int_{\bm p} \partial_{p_i}(h^{(0)}h^{(2)}n_B)=0,
\end{align}
and
\begin{align}
  &\sum_{\alpha\beta\gamma\sigma}\int_{\bm X,\bm p}\left(-\frac{\hbar^2}{16}\epsilon_{\alpha\beta}\epsilon_{\gamma\sigma}h^{(0)}\partial_{p_i}h^{(0)}\partial_{\alpha\gamma}^2 h^{(0)}\partial_{\beta\sigma}^2 h^{(0)}n_B^{\prime\prime}
  -\frac{\hbar^2}{24}\epsilon_{\alpha\beta}\epsilon_{\gamma\sigma}h^{(0)}\partial_{p_i}h^{(0)}\partial_{\alpha\gamma}^2 h^{(0)}\partial_{\beta} h^{(0)}\partial_{\sigma} h^{(0)}n_B^{\prime\prime\prime}
  -\frac{\hbar^2}{8}\epsilon_{\alpha\beta}\epsilon_{\gamma\sigma}\partial_{\alpha\gamma}^2 h^{(0)}\partial_{p_i\beta\sigma}^3h^{(0)}n_B\right)\notag\\
  &=\frac{\hbar^2}{24}\sum_{\alpha\beta}\epsilon_{\alpha\beta}\sum_j \int_{\bm X,\bm p}\Bigl((h^{(0)}\partial_{p_i \alpha p_j}^3h^{(0)}\partial_\beta h^{(0)}+2h^{(0)}\partial_{\alpha p_j}^2 h^{(0)}\partial_{p_i \beta}^2 h^{(0)})\partial_T n_B^{\prime}\partial_{X_j}T
  -(2\partial_{p_i \alpha p_j}^3h^{(0)}\partial_\beta h^{(0)}+4\partial_{\alpha p_j}^2 h^{(0)}\partial_{p_i \beta}^2 h^{(0)})\partial_T n_B\partial_{X_j}T\Bigr).
\end{align}
Finally, using the above relations, we obtain the heat current up to second order in spatial gradients as
\begin{align}
  J_i^{(2)}&=J_i^{(2),\mathrm{QG}}+J_i^{(2),\mathrm{disp}}
\end{align}
where the first term $J_i^{(2),\mathrm{QG}}$, arising from quantum-geometric effects such as TBCP and the quantum metric, is given by
\begin{align}
    \label{J2QG1}
  J_i^{(2),\mathrm{QG}}
  &=\frac{\hbar^2}{2V}\sum_{jk}\int_{\bm X,\bm p}\mTr\Biggl[n_B \mathcal E (\partial_{p_i}\mathcal E G^t_{jk}-\partial_{p_j}\mathcal E G^t_{ik})\chi^2\partial_{X_j}\psi_k-n_B\mathcal{E}^2(\partial_{p_i}G^t_{jk}-\partial_{p_j}G^t_{ik})\chi^2\psi_j\psi_k\vphantom{\frac{\hbar^2}{2}}\notag\\
  &\qquad
  +n_B^{\prime}\mathcal E^2\partial_{p_i}\mathcal E \mathcal{G}_{jk}\chi^2\partial_{X_jX_k}^2\chi+\frac{1}{2}n_B^{\prime\prime}\mathcal E^3\partial_{p_i}\mathcal E \mathcal{G}_{jk}\chi^4\psi_j\psi_k-n_B \mathcal E\partial_{p_i}\mathcal E \mathcal{G}_{jk}\chi^2\psi_j\psi_k
  +\frac{1}{2}n_B \mathcal{E}^2 \partial_{p_i}\mathcal{G}_{jk}\chi\partial_{X_j X_k}^2\chi\Biggr].
\end{align}
The second term $J_i^{(2),\mathrm{disp}}$ represents the contributions that depend only on the band dispersion $\varepsilon_n(\bm p)$ and is given by
\begin{align}
  \label{J2disp}
  J_i^{(2),\mathrm{disp}}
  =\frac{\hbar^2}{48V}\sum_{jk}\int_{\bm X,\bm p}\mTr\left[\mathcal E^{-2}(-\mathcal E\partial_{p_i p_j p_k}^3\mathcal E - 2\partial_{p_i}\mathcal E\partial_{p_j p_k}^2\mathcal E + 3\partial_{p_i p_j}^2\mathcal E\partial_{p_k}\mathcal E)
  ((\chi\mathcal E)^2\partial_{X_k}(\chi \mathcal E)\partial_T n_B^{\prime}\partial_{X_j}T-2\chi\mathcal E \partial_{X_k}(\chi \mathcal E)\partial_T n_B\partial_{X_j}T)\right],
\end{align}
\end{widetext}
where $\mathcal{E}\equiv\mathrm{diag}(\varepsilon_1,\dots,\varepsilon_{2N})$ is a diagonal matrix consisting of the band energies.

\subsection{Nonlinear thermal Hall conductivity}
\label{subsec:NLTHcoefficient}

In this section, we derive the second-order nonlinear thermal Hall conductivity from the heat current obtained in the previous section.
The second-order nonlinear thermal Hall conductivity describes the transverse heat current generated by the second order of the temperature gradient and is defined as
\begin{align}
  J_i^{(2)}=\sum_{jk}\kappa_{i;jk}\,\partial_{X_j}T(\bm X)\partial_{X_k}T(\bm X),
\end{align}
in the thermodynamic limit with $L\to \infty$.
This limit is taken after extracting the bulk contribution.
Since Eqs.~\eqref{J2QG1} and \eqref{J2disp} contain second-order spatial derivatives of $\chi(\bm X)$, we need to perform integration by parts, which leads to an expression in terms of the second order of the temperature gradient~\cite{Mangeolle2024}.
The detailed derivation is provided in Appendix~\ref{app:kappa}, and here we present only the final result:
\begin{align}\label{eq:kappa_ijk_total}
\kappa_{i;jk}=\kappa_{i;jk}^{\mathrm{TBCP}}+\kappa_{i;jk}^{\mathrm{QM}}+\kappa_{i;jk}^{\mathrm{disp}}.
\end{align}
Here, the first two terms originate from $J_i^{(2),\mathrm{QG}}$ in Eq.~\eqref{J2QG1}, while $\kappa_{i;jk}^{\mathrm{disp}}$ originates from $J_i^{(2),\mathrm{disp}}$ in Eq.~\eqref{J2disp}.
Each term in Eq.~\eqref{eq:kappa_ijk_total} is given by
\begin{widetext}
\begin{align}
  \label{eq:kappa_ijk^TBCP}
  \kappa_{i;jk}^{\mathrm{TBCP}}
  &=\hbar^2 k_B^2\sum_{n=1}^{N}\int_{\bm p}\varepsilon_n^{-1}\left(\partial_{p_j}(\varepsilon_n G_{n,ik}^t)-\partial_{p_i}(\varepsilon_n G_{n,jk}^t)\right)c_2[n_B(\varepsilon_n,T)],\\
  \label{eq:kappa_ijk^QM}
  \kappa_{i;jk}^{\mathrm{QM}}
  &=\hbar^2 k_B^2\sum_{n=1}^{N}\int_{\bm p}\left[\left(2\varepsilon_n^{-1}\partial_{p_i}\varepsilon_n \mathcal{G}_{n,jk}+\frac{5}{2}\partial_{p_i}\mathcal{G}_{n,jk}\right)c_2[n_B(\varepsilon_n,T)]
  -\frac{\varepsilon_n^3}{2 k_B^2 T^2}\partial_{p_i}\mathcal{G}_{n,jk}n_B^{\prime}(\varepsilon_n,T)\right],\\
  \label{eq:kappa_ijk^disp}
  \kappa_{i;jk}^{\mathrm{disp}}
  &=\frac{\hbar^2 k_B^2}{6}\sum_{n=1}^{N}\int_{\bm p}\varepsilon_n^{-2}(\varepsilon_n\partial_{p_i p_j p_k}^3\varepsilon_n + 2\partial_{p_i}\varepsilon_n\partial_{p_j p_k}^2\varepsilon_n - 3\partial_{p_i p_j}^2\varepsilon_n\partial_{p_k}\varepsilon_n)\left(c_2[n_B(\varepsilon_n,T)]+\frac{\varepsilon_n^3}{4k_B^2T^2}(\varepsilon_n n_B^{\prime\prime}(\varepsilon_n,T)-2n_B^{\prime}(\varepsilon_n,T))\right)\notag\\
  &=\frac{\hbar^2}{24}\sum_{n=1}^{N}\int_{\bm p}\varepsilon_n^2\partial_{p_i p_j p_k}^3\varepsilon_n\frac{\varepsilon_n n_B^{\prime\prime}(\varepsilon_n,T)-n_B^{\prime}(\varepsilon_n,T)}{T^2},
\end{align}
\end{widetext}
where we introduce $c_2[n_B]\equiv\int_{0}^{n_B}\left(\ln \frac{1+t}{t}\right)^2dt$, which can be expressed as $c_2[n_B]=(1+n_B)\left(\ln \frac{1+n_B}{n_B}\right)^2-(\ln n_B)^2-2\mathrm{Li}_2(-n_B)$ using the dilogarithm function $\mathrm{Li}_2(z)=\sum_{n=1}^{\infty}\frac{z^n}{n^2}$.
This function also plays an important role in the linear thermal Hall conductivity~\cite{Matsumoto2011L,Matsumoto2011B,Matsumoto2014}.
Note that $c_2[n_B(\varepsilon,T)]$ is a monotonically decreasing function of $\varepsilon/T$, taking the value $\frac{\pi^2}{3}$ at $\varepsilon/T=0$ and approaching $0$ as $\varepsilon/T\to\infty$.
As seen in Eqs.~\eqref{eq:kappa_ijk^TBCP} and \eqref{eq:kappa_ijk^QM}, we separate the contributions from $J_i^{(2),\mathrm{QG}}$ into $\kappa_{i;jk}^{\mathrm{TBCP}}$ and $\kappa_{i;jk}^{\mathrm{QM}}$, where the former arises from the TBCP and the latter from the quantum metric.
Since $\kappa_{i;jk}^{\mathrm{disp}}$ originates from $J_i^{(2),\mathrm{disp}}$, it depends only on the band dispersion $\varepsilon_n(\bm p)$.

Here, we discuss the characteristic features of each term in Eq.~\eqref{eq:kappa_ijk_total}.
By expanding the momentum derivatives in Eq.~\eqref{eq:kappa_ijk^TBCP}, we find that two terms proportional to $\partial_p\varepsilon$ and $\partial_p G^t$ appear in $\kappa_{i;jk}^{\mathrm{TBCP}}$.
The latter term involving $\partial_p G^t$ is exactly equivalent to the thermal conductivity obtained from the semiclassical wave-packet approach introduced in Ref.~\cite{Li2024}, which is explicitly given by
\begin{align}\label{eq:kappa_SC}
\kappa^{\mathrm{SC}}_{i;jk}=\hbar^2 k_B^2 \sum_{n=1}^N\int_{\bm p}(\partial_{p_j}G^t_{n,ik}-\partial_{p_i}G^t_{n,jk})c_2[n_B(\varepsilon_n,T)].
\end{align}
The presence of additional terms beyond $\kappa^{\mathrm{SC}}_{i;jk}$ suggests that our result contains quantum corrections that cannot be captured by the semiclassical approximation.
Furthermore, the former term proportional to $\partial_{p}\varepsilon$ in Eq.~\eqref{eq:kappa_ijk^TBCP} becomes dominant, particularly at higher temperatures, leading to qualitatively different behavior of the thermal conductivity in the high-temperature limit compared with previous studies~\cite{Li2024,Varshney2023}, as will be discussed in Sec.~\ref{sec:model_calc}.

To further analyze the nature of the nonlinear thermal Hall conductivity, we decompose $\kappa_{i;jk}$ into the ``Ohmic'' part, $\kappa_{i;jk}^{\mathrm{O}}$, and the ``Hall'' part, $\kappa_{i;jk}^{\mathrm{H}}$, within the intrinsic contributions as follows~\cite{Stepan2022}.
\begin{align}
  &\kappa_{i;jk}^{\mathrm{O}}=\frac{1}{6}(\kappa_{i;jk}+\kappa_{i;kj}+\kappa_{j;ik}+\kappa_{j;ki}+\kappa_{k;ij}+\kappa_{k;ji}),\\
  &\kappa_{i;jk}^{\mathrm{H}}=\kappa_{i;jk}-\kappa_{i;jk}^{\mathrm{O}}.\label{eq:kappa_Hall-def}
\end{align}
From these definitions, it follows that the longitudinal component of the ``Hall'' conductivity always vanishes; namely, $\kappa_{i;ii}^{\mathrm{H}}=0$.
We find that the TBCP contribution always satisfies $\kappa_{i;ii}^{\mathrm{TBCP}}=0$, whereas the quantum-metric contribution and the contributions determined solely by the band dispersion can exhibit nonzero longitudinal components; namely, $\kappa_{i;ii}^{\mathrm{QM}}\neq0$ and $\kappa_{i;ii}^{\mathrm{disp}}\neq0$.
These features indicate that $\kappa_{i;jk}^{\mathrm{TBCP}}$ is a ``Hall''-type conductivity, whereas $\kappa_{i;jk}^{\mathrm{QM}}$ and $\kappa_{i;jk}^{\mathrm{disp}}$ contain ``Ohmic'' contributions.
Such ``Ohmic'' contributions also arise in the second-order intrinsic nonlinear Hall effect when quantum corrections are taken into account~\cite{Das2023,Kaplan2024}.
In this paper, because it is practically difficult to experimentally extract the ``Hall''-type contribution defined in Eq.~\eqref{eq:kappa_Hall-def}, we refer to $\kappa_{x;yy}$ and $\kappa_{y;xx}$ as the nonlinear thermal Hall conductivities~\cite{Li2024,Varshney2023}.

\begin{table}[t]
    \centering
    \begin{tabular}{ccccccc}
        \hline\hline
         & $\mathcal P$ & $\mathcal T$ & $\mathcal{PT}$ & $\mathcal M_x \mathcal T$ & $\mathcal M_y \mathcal T$ & $\mathcal C_3^z$ \\\hline
        $\kappa_{xy}^{\mathrm{H}}$ & \cmark & \xmark & \xmark & \cmark & \cmark & \cmark \\
        $\kappa_{x;yy}$ & \xmark & \xmark & \cmark & \cmark & \xmark & \cmark \\
        $\kappa_{y;xx}$ & \xmark & \xmark & \cmark & \xmark & \cmark & \cmark \\ 
        $\kappa_{i;jk}^{\mathrm{TBCP}}$, $\kappa_{i;jk}^{\mathrm{SC}}$ & -- & -- & -- & -- & -- & \xmark \\ \hline\hline
    \end{tabular}
    \caption{
Symmetry constraints on the linear and nonlinear thermal Hall conductivities.
Here, $\kappa_{xy}^{\mathrm{H}}$ denotes the linear thermal Hall conductivity.
In this table, $\mathcal P$ represents spatial inversion symmetry, $\mathcal T$ time-reversal symmetry, $\mathcal M_y$ ($\mathcal M_x$) mirror symmetry with respect to the $yz$ ($zx$) plane, and $\mathcal C_3^z$ threefold rotational symmetry about the $z$ axis present in the system.
The symbols \cmark and \xmark indicate symmetry-allowed and symmetry-forbidden responses, respectively.
The symbol ``--'' in $\kappa_{i;jk}^{\mathrm{TBCP}}$ and $\kappa_{i;jk}^{\mathrm{SC}}$ indicates the same symmetry constraints as those for $\kappa_{i;jk}$, where the indices $i,j,k$ are restricted to $x$ and $y$.
    }    \label{tab:NLTHE_sym}
  \end{table}

Finally, we discuss symmetry-imposed restrictions on the nonlinear thermal Hall conductivity.
Table~\ref{tab:NLTHE_sym} summarizes whether the linear and nonlinear thermal Hall conductivities are allowed or forbidden under several symmetries~\cite{Zhang2023}.
The linear thermal Hall conductivity is allowed when time-reversal symmetry $\mathcal T$ is broken, whereas it is forbidden when either spatial inversion symmetry $\mathcal P$ or the combined symmetry $\mathcal{PT}$ is present.
On the other hand, the nonlinear thermal Hall conductivity $\kappa_{i;jk}$ can be nonzero even in the presence of $\mathcal{PT}$ symmetry because of contributions from the quantum metric.
We also consider the combined symmetries of mirror reflection and time reversal, $\mathcal M_x \mathcal T$ and $\mathcal M_y \mathcal T$, where $\mathcal M_x$ ($\mathcal M_y$) denotes the mirror reflection with respect to the $yz$ ($xz$) plane.
In the presence of $\mathcal M_x \mathcal T$ ($\mathcal M_y \mathcal T$) symmetry, $\kappa_{y;xx}$ ($\kappa_{x;yy}$) must vanish.
Furthermore, when the system has threefold rotational symmetry around the $z$ axis, $\kappa_{i;jk}$ with $i,j,k=x$ or $y$ can be nonzero, but $\kappa_{i;jk}^{\mathrm{TBCP}}$, including $\kappa_{i;jk}^{\mathrm{SC}}$, cannot contribute to $\kappa_{i;jk}$ because they are antisymmetric with respect to the exchange of the first two indices $i$ and $j$.
These properties will be used in Sec.~\ref{sec:model_calc} to analyze the nonlinear thermal Hall conductivity in a specific spin model.

\section{Application to a spin model}
\label{sec:model_calc}

\subsection{Model}
\label{sec:model}

\begin{figure}[t]
\centering
    \includegraphics[width=\columnwidth,clip]{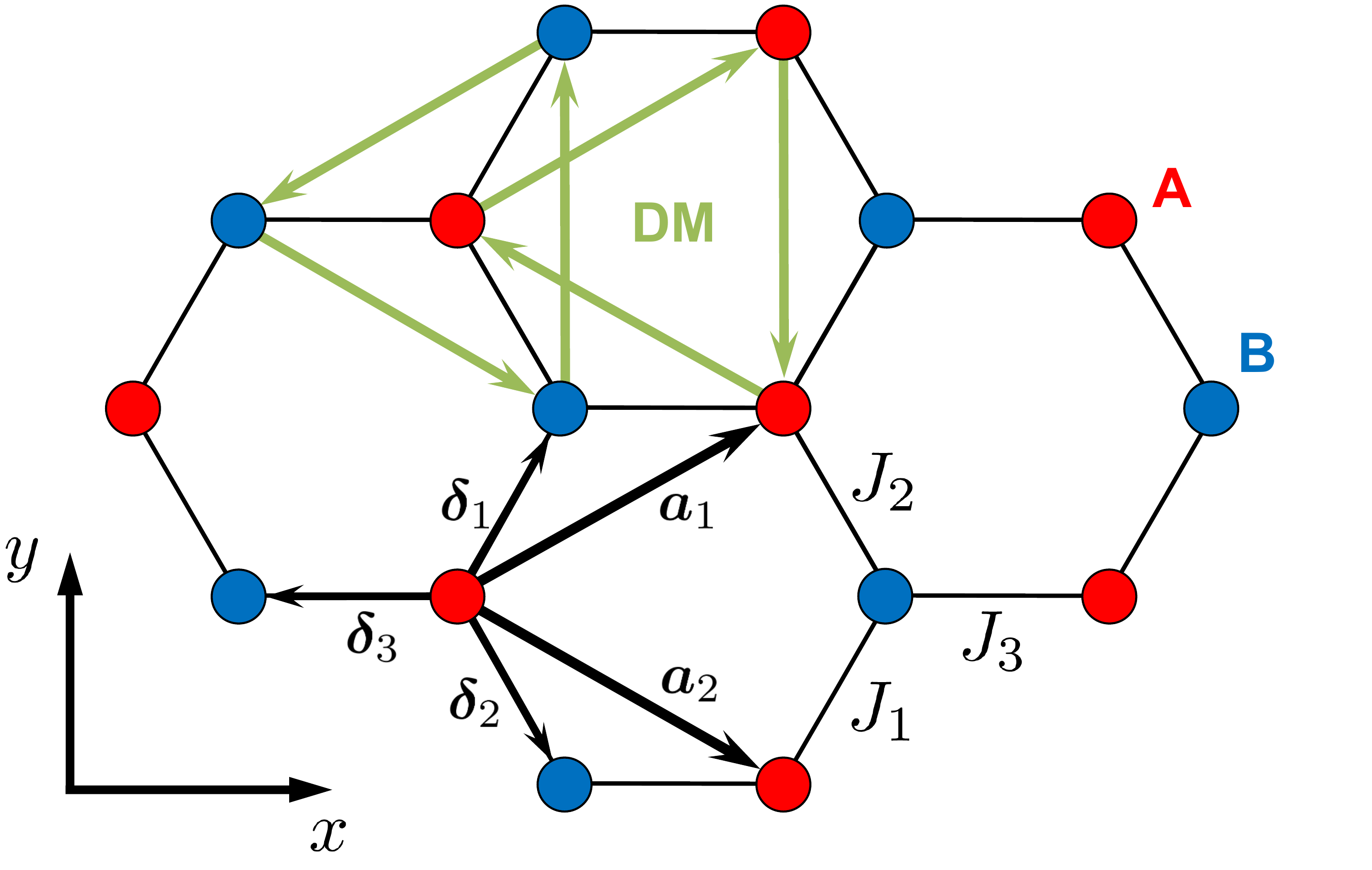}
    \caption{
Schematic illustration of a spin model on a honeycomb lattice, where $\bm a_1$ and $\bm a_2$ are primitive translation vectors, and $\bm \delta_1$, $\bm \delta_2$, and $\bm \delta_3$ are vectors connecting nearest-neighbor sites.
The honeycomb lattice consists of two sublattices, A and B, shown by red and blue circles, respectively.
The green arrows indicate the directions associated with the DM interaction; we define $\nu_{ij}=-\nu_{ji}=+1$ when the direction of the vector from site $i$ to site $j$ is the same as that of the corresponding green arrow.
  }
    \label{fig:lattice}
\end{figure}

We apply the formula derived in the previous section to a specific model with bosonic excitations.
Here, we consider a spin model defined on a honeycomb lattice, whose Hamiltonian is given by
\begin{align}\label{eq:spin_hamiltonian}
  \mathscr H &=  -\sum_{\gamma=1}^{3}\sum_{\langle ij\rangle_\gamma}J_\gamma\bm S_i\cdot\bm S_j
  + D \sum_{\llangle ij\rrangle}\nu_{ij}[\bm S_i\times\bm S_j]_z\notag\\
   &\qquad - B\sum_{i\in \mathrm{A}}S_i^z + B\sum_{i\in \mathrm{B}}S_i^z - \mathcal K \sum_i (S_i^z)^2,
\end{align}
where $\bm S_i=(S_i^x,S_i^y,S_i^z)$ denotes the spin operator with magnitude $S$ at site $i$ on the honeycomb lattice, which consists of two sublattices, A and B, as shown in Fig.~\ref{fig:lattice}.
In this Hamiltonian, $J_\gamma$ with $\gamma=1,2,3$ represents the exchange coupling between nearest-neighbor sites $\langle ij\rangle_\gamma$ in the direction of $\bm \delta_\gamma$, $D$ denotes the strength of the Dzyaloshinskii-Moriya (DM) interaction between next-nearest-neighbor sites $\llangle ij\rrangle$, where $\nu_{ij}=\pm 1$ depending on the direction of the DM interaction as shown in Fig.~\ref{fig:lattice}, $B$ is a staggered magnetic field, and $\mathcal K$ is the anisotropy constant along the $z$ direction.
The spin model described by Eq.~\eqref{eq:spin_hamiltonian} is the same as that used in a previous study on the intrinsic nonlinear thermal Hall effect~\cite{Li2024}, and we introduce it here to compare our results with those obtained in that study.

In the present calculations, we consider the case in which all exchange couplings are ferromagnetic, i.e., $J_1,J_2,J_3>0$.
In this case, the ground state of the model described by Eq.~\eqref{eq:spin_hamiltonian} is expected to be ferromagnetically ordered.
Accordingly, we assume that all spins are aligned along the $S^z$ direction in the ground state.

Under the ferromagnetic order, we briefly discuss the symmetry of the system.
For the undistorted case with $J_1=J_2=J_3=1$ in the absence of the staggered magnetic field, the system possesses both spatial inversion symmetry and the combined symmetries of time-reversal and mirror-reflection operations, $\mathcal M_x \mathcal T$ and $\mathcal M_y \mathcal T$.
The staggered field $B$ breaks the spatial inversion symmetry and $\mathcal M_x \mathcal T$, indicating that $\kappa_{y;xx}$ can be nonzero, whereas $\kappa_{x;yy}$ vanishes according to Table~\ref{tab:NLTHE_sym}~\cite{Li2024}.
Furthermore, when the $\mathcal C_3^z$ symmetry is broken by introducing anisotropy in the exchange constants, $\mathcal M_y \mathcal T$ is also broken, suggesting that $\kappa_{x;yy}$ can become nonzero.

To analyze the transport properties of this model with ferromagnetic order, we employ linear spin-wave theory, in which spin fluctuations around the ground state are described by bosonic excitations called magnons.
We represent the spin operators using bosonic operators via the Holstein-Primakoff (HP) transformation as
\begin{align}
  S_{i\rm A}^z=S-a_i^{\dagger}a_i,\quad
  S_{i\rm A}^x +iS_{i\rm A}^y=\sqrt{2S-a_i^{\dagger}a_i}a_i
\end{align}
for spins on the A sublattice and
\begin{align}
  S_{i\rm B}^z=S-b_i^{\dagger}b_i,\quad
  S_{i\rm B}^x +iS_{i\rm B}^y=\sqrt{2S-b_i^{\dagger}b_i}b_i
\end{align}
for spins on the B sublattice, where $a_i$ and $b_i$ ($a_i^{\dagger}$ and $b_i^{\dagger}$) denote the annihilation (creation) operators of bosons describing magnons on the A and B sublattices, respectively.
By neglecting magnon-magnon interactions, the Hamiltonian given in Eq.~\eqref{eq:spin_hamiltonian} reduces to a bilinear form in the magnon operators.
Higher-order terms in the HP expansion give rise to magnon-magnon interactions, which can renormalize magnon bands and quasiparticle wave functions, modify quantum-geometric quantities, and introduce magnon damping~\cite{Chernyshev2009,Mook2020,Koyama2024,Habel2024}.
A systematic treatment of these effects would require an extension to a Green's-function formalism~\cite{Shitade2014}, which is beyond the scope of the present work.

Here, we introduce the Fourier transforms of the magnon operators as
\begin{align}
  \begin{pmatrix}
    a_i \\
    b_i
  \end{pmatrix}
  =\frac{1}{\sqrt{N}}\sum_{\bm k}e^{i\bm k\cdot\bm r_i}
  \begin{pmatrix}
    a_{\bm k} \\
    b_{\bm k}
  \end{pmatrix},
\end{align}
where $N$ is the number of unit cells, and $a_{\bm k}$ and $b_{\bm k}$ denote the annihilation operators of magnons with wave vector $\bm k$ on the A and B sublattices, respectively.
The sum over $\bm k$ is taken over the first Brillouin zone.
Using these operators, the Hamiltonian in the wave-number representation is written as
\begin{align}\label{eq:hamil-model}
  \mathscr{H}=\sum_{\bm k}\Phi^\dagger(\bm k)H_0(\bm k)\Phi(\bm k),
\end{align}
where $\Phi^\dagger(\bm k)$ is a two-component vector defined as $\Phi^\dagger(\bm k)=(a_{\bm k}^\dagger,b_{\bm k}^\dagger)$.
Note that the relative momentum $\bm p$ introduced in the Wigner transformation is related to the wave vector by $\bm p=\hbar \bm k$.
Furthermore, since the number of magnons is conserved in Eq.~\eqref{eq:hamil-model}, $H_0(\bm k)$ reduces to a $2\times 2$ Hermitian matrix, whereas it would generally be represented by a $4\times 4$ matrix in the BdG formalism.
Thus, $H_0(\bm k)$ is related to $\mathsf H_0(\bm p)$ introduced in Sec.~\ref{sec:2_THE} as 
$\mathsf H_0(\hbar \bm k)=
\begin{psmallmatrix}
   H_0(\bm k) & 0 \\
   0 & H_0(\bm k)
\end{psmallmatrix}$.
The explicit form of $H_0(\bm k)$ is given by
\begin{align}
  H_0(\bm k)=h_0 I_2+h_x(\bm k)\sigma_x+h_y(\bm k)\sigma_y+h_z(\bm k)\sigma_z,
\end{align}
where $I_2$ is the $2\times 2$ identity matrix and $\sigma_{x,y,z}$ are the Pauli matrices defined as
$\sigma_x=\begin{psmallmatrix}0&1\\1&0\end{psmallmatrix}$, $\sigma_y=\begin{psmallmatrix}0&-i\\i&0\end{psmallmatrix}$, and $\sigma_z=\begin{psmallmatrix}1&0\\0&-1\end{psmallmatrix}$.
For simplicity, we assume that the primitive translation vectors have unit length.
The coefficients $h_0,h_x(\bm k),h_y(\bm k),h_z(\bm k)$ are given by 
$h_0=J_1 S+J_2 S+J_3 S+2\mathcal K S$, 
$h_x(\bm k)=-\sum_{\gamma=1}^{3}J_\gamma S\cos(\bm k\cdot \bm \delta_\gamma)$, 
$h_y(\bm k)=\sum_{\gamma=1}^{3}J_\gamma S\sin(\bm k\cdot \bm \delta_\gamma)$, and 
$h_z(\bm k)=B+2DS(\sin(\bm k\cdot \bm a_1)+\sin(-\bm k\cdot \bm a_2)+\sin(\bm k\cdot (\bm a_2-\bm a_1)))$.
By diagonalizing $H_0(\bm k)$, we obtain the magnon band energies $\varepsilon_n(\bm k)$ and the corresponding eigenvectors $\ket{\Phi_n(\bm k)}$, which are used to calculate the quantum geometric quantities introduced in Sec.~\ref{sec:2_THE}.
They are given by $\varepsilon_n(\bm k)=h_0+n\zeta(\bm k)$ with the branch index $n=\pm1$, and $\ket{\Phi_n(\bm k)}=\frac{1}{\sqrt{2}}
  \begin{pmatrix}
  \lambda^{+}_{n} \\
  n e^{i\beta(\bm k)}\lambda^{-}_{n}
  \end{pmatrix}$,
where $\zeta(\bm k)=\sqrt{h_x^2(\bm k)+h_y^2(\bm k)+h_z^2(\bm k)}$, $\lambda^{\pm}_{n}=\sqrt{1\pm n h_z(\bm k)/\zeta(\bm k)}$, and $\beta(\bm k)=\arctan(h_y(\bm k)/h_x(\bm k))$.
Note that the paraunitary matrix that diagonalizes $\mathsf H_0(\hbar \bm k)$ in the BdG formalism is in fact unitary because the original bosonic Hamiltonian conserves the number of magnons, and its matrix elements are determined from the explicit forms of $\ket{\Phi_{+}(\bm k)}$ and $\ket{\Phi_{-}(\bm k)}$.
Here and hereafter, we abbreviate the branch index $n=\pm1$ as $n=\pm$ for simplicity.

\subsection{Result}
\label{subsec:result}

In this section, we present the results for the nonlinear thermal Hall conductivity calculated using the formula derived in Sec.~\ref{sec:2_THE} for the spin model given by Eq.~\eqref{eq:spin_hamiltonian}.
Hereafter, the parameters are set to $S=1/2$, $D/J_1=0.2$, $B/J_1=0.1$, and $\mathcal{K}/J_1=0.2$, with $J_1$ taken as the unit.

\subsubsection{Undistorted case}

\begin{figure*}[t]
\centering
    \includegraphics[width=2\columnwidth,clip]{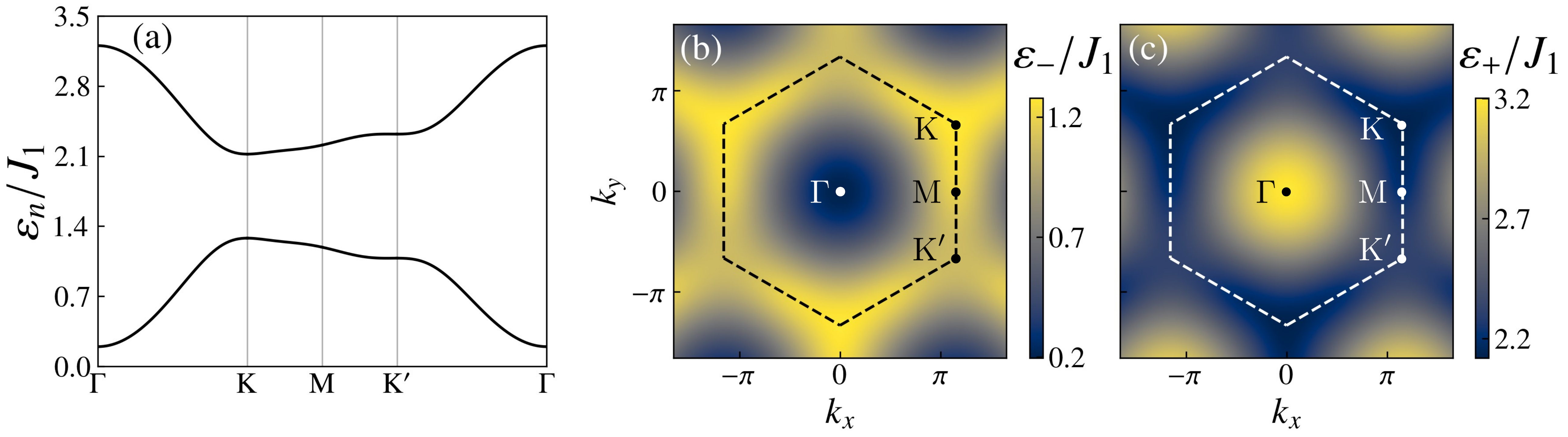}
    \caption{
Magnon band structure of the isotropic ferromagnetic honeycomb lattice model with $J_1=J_2=J_3$ under a staggered magnetic field.
(a) Momentum dependence of the magnon bands along high-symmetry lines in the first Brillouin zone.
(b), (c) Contour maps of the magnon band dispersions for the (b) low-energy and (c) high-energy bands.
In panels (b) and (c), the dashed hexagons indicate the first Brillouin zone.
The positions of the high-symmetry points shown in panel (a) within the first Brillouin zone are indicated in panels (b) and (c).
  }
    \label{fig:FM_ene}
\end{figure*}

First, we consider the case in which the nearest-neighbor exchange interactions are equal, i.e., $J_1=J_2=J_3$.
In this case, the system possesses threefold rotational symmetry $\mathcal C_3^z$ about the $z$ axis and the combined symmetry $\mathcal M_y \mathcal T$.

Figure~\ref{fig:FM_ene} shows the magnon band dispersion.
A nonzero excitation gap appears at the $\Gamma$ point, and a gap exists between the high-energy and low-energy bands, as shown in Fig.~\ref{fig:FM_ene}(a).
In particular, the gaps at the $\mathrm{K}$ and $\mathrm{K^{\prime}}$ points originate from the DM interaction and the staggered magnetic field.
Figures~\ref{fig:FM_ene}(b) and \ref{fig:FM_ene}(c) show the magnon dispersions of the low-energy and high-energy bands, respectively.
These results indicate that the band dispersion possesses threefold rotational symmetry and is an even function of $k_x$, which arises from the $\mathcal M_y \mathcal T$ symmetry.

\begin{figure*}[t]
\centering
    \includegraphics[width=2\columnwidth,clip]{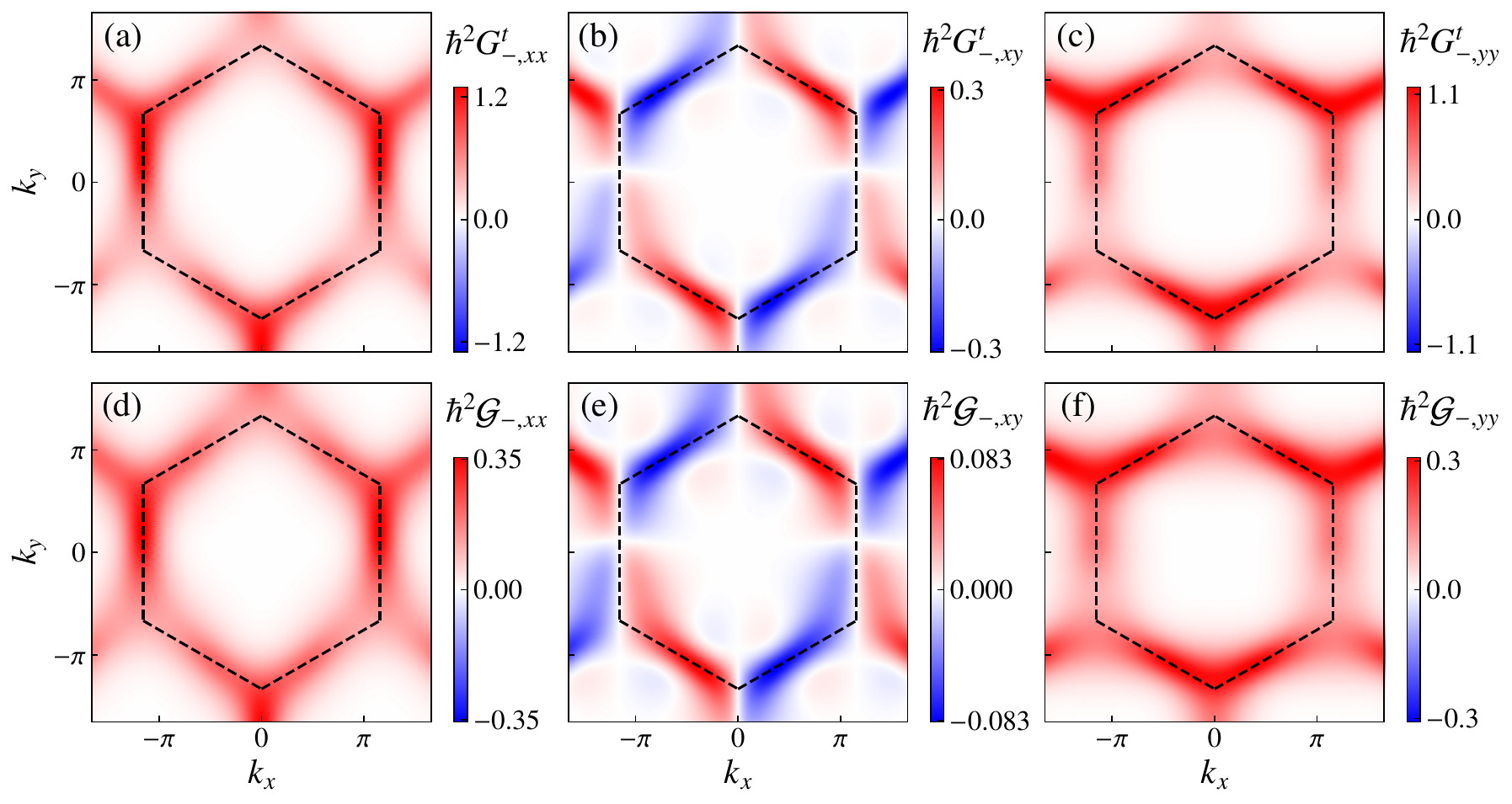}
    \caption{
Momentum dependence of the (a) $xx$, (b) $xy$, and (c) $yy$ components of TBCP and the (d) $xx$, (e) $xy$, and (f) $yy$ components of the quantum metric for the low-energy band in the isotropic ferromagnetic honeycomb lattice model with $J_1=J_2=J_3$ under a staggered magnetic field.
The dashed hexagons indicate the first Brillouin zone.
  }
    \label{fig:FM_geom}
\end{figure*}

Next, we calculate the quantum geometric quantities.
Since the present bosonic model is described by a two-band model, the TBCP and quantum metric can be represented in simple forms.
By replacing $\bm p$ with $\hbar\bm k$ in Eqs.~\eqref{eq:def_TBCP} and~\eqref{g_pp}, the quantum metric for the low-energy band is expressed as
\begin{align}
  \mathcal{G}_{-,ij}(\bm{k})=\mathcal{G}_{+,ij}(\bm{k})
  =\Re\left[\Lambda_{k_i,-+}^{(0)}(\bm{k})\Lambda_{k_j,+-}^{(0)}(\bm{k})\right]
\end{align}
with $\Lambda_{k_i,mn}^{(0)}(\bm{k})=-\frac{i}{\hbar}\braket{\Phi_m(\bm k)}{\partial_{k_i}\Phi_n(\bm k)}$, and the TBCP for the low-energy band is given by
\begin{align}\label{eq:FM_TBCP_model}
  G^t_{-,ij}(\bm{k}) =-G^t_{+,ij}(\bm{k})
  = \frac{2h_0\, \mathcal{G}_{-,ij}(\bm{k})}{\Delta\varepsilon(\bm k)},
\end{align}
where $\Delta\varepsilon(\bm k)=\varepsilon_+(\bm k)-\varepsilon_-(\bm k)$.
Figure~\ref{fig:FM_geom} shows the momentum dependence of the TBCP [Figs.~\ref{fig:FM_geom}(a)--\ref{fig:FM_geom}(c)] and the quantum metric [Figs.~\ref{fig:FM_geom}(d)--\ref{fig:FM_geom}(f)] for the low-energy band with branch index $n=-1$.
Due to the $\mathcal M_y \mathcal T$ symmetry, $G^t_{-,xx}(\bm{k})$ and $G^t_{-,yy}(\bm{k})$ are even functions of $k_x$, while $G^t_{-,xy}(\bm{k})$ is an odd function of $k_x$.
Similar symmetry properties hold for the components of the quantum metric.
On the other hand, the $\mathcal M_x \mathcal T$ symmetry is broken in the present model due to the staggered magnetic field, and thus there is no specific symmetry with respect to $k_y$ for each component of the TBCP and the quantum metric.
Moreover, from Eq.~\eqref{eq:FM_TBCP_model}, it is clear that the absolute values of the TBCP become larger than those of the quantum metric around the Brillouin zone boundary where the band gap $\Delta\varepsilon(\bm k)$ is small.
Thus, we expect that the TBCP contribution plays a more significant role in the nonlinear thermal Hall conductivity than the quantum metric contribution.
However, as discussed in Sec.~\ref{subsec:NLTHcoefficient} and Table~\ref{tab:NLTHE_sym}, the TBCP contribution $\kappa^{\mathrm{TBCP}}$ vanishes due to the $\mathcal C_3^z$ symmetry in the undistorted case, and thus the other two contributions, $\kappa^{\mathrm{QM}}$ and $\kappa^{\mathrm{disp}}$, are important for the nonlinear thermal Hall conductivity.

\begin{figure*}[t]
\centering
    \includegraphics[width=2\columnwidth,clip]{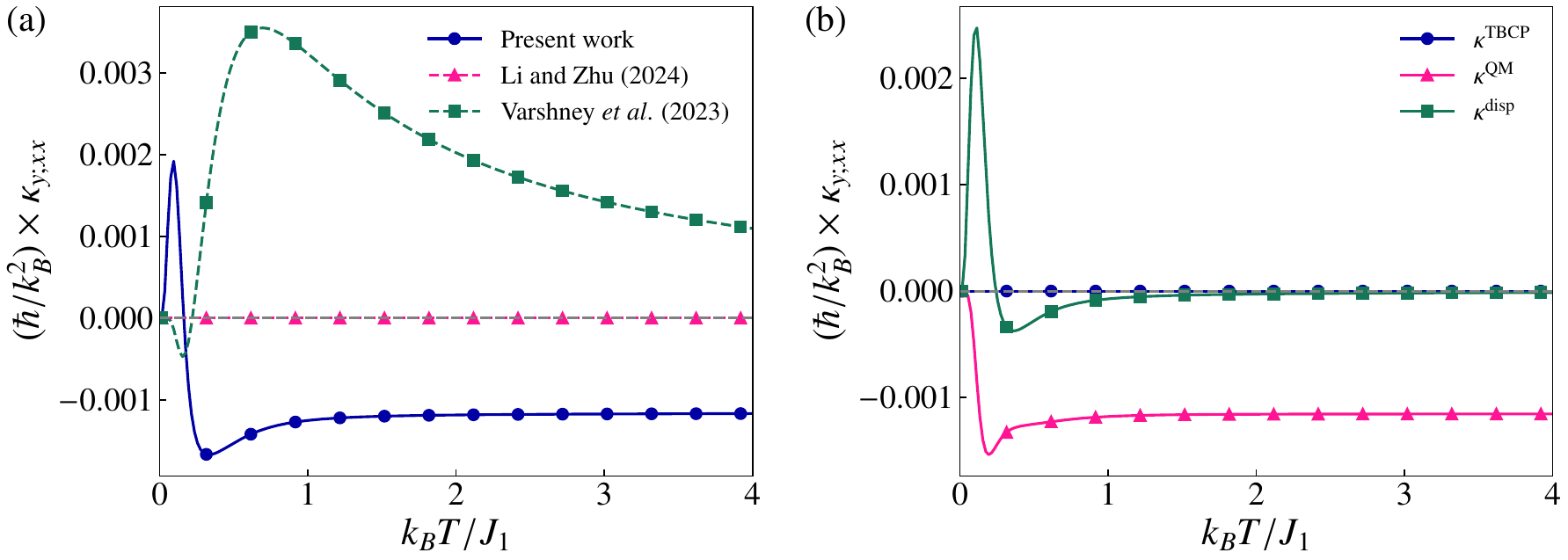}
    \caption{
(a) Temperature dependence of the nonlinear thermal Hall conductivity $\kappa_{y;xx}$ in the isotropic ferromagnetic honeycomb lattice model with $J_1=J_2=J_3$ under a staggered magnetic field.
The solid line represents the conductivity obtained from our theory, and the dashed lines represent those obtained by Li and Zhu (2024) using the semiclassical approach~\cite{Li2024} and by Varshney {\it et al}. (2023) using the quantum kinetic theory~\cite{Varshney2023}.
(b) Decomposition of $\kappa_{y;xx}$ into three contributions: $\kappa^{\mathrm{TBCP}}$, $\kappa^{\mathrm{QM}}$, and $\kappa^{\mathrm{disp}}$, which are introduced in Eqs.~\eqref{eq:kappa_ijk^TBCP-k}, \eqref{eq:kappa_ijk^QM-k}, and \eqref{eq:kappa_ijk^disp-k}, respectively.
    }
    \label{fig:FM_yxx}
\end{figure*}

Here, we calculate the nonlinear thermal Hall conductivity.
As discussed in Sec.~\ref{subsec:NLTHcoefficient} and Table~\ref{tab:NLTHE_sym}, owing to the presence of the $\mathcal M_y \mathcal T$ symmetry, $\kappa_{x;yy}$ vanishes, whereas $\kappa_{y;xx}$ can be nonzero; therefore, we focus on $\kappa_{y;xx}$.
Figure~\ref{fig:FM_yxx}(a) shows the temperature dependence of $\kappa_{y;xx}$ together with results obtained using formulas derived in previous studies~\cite{Varshney2023,Li2024}.
The nonlinear thermal Hall conductivity $\kappa_{y;xx}$ calculated within our theoretical framework exhibits nonmonotonic behavior at low temperatures and approaches a nonzero value in the high-temperature limit.
Such nonmonotonic behavior at low temperatures is also observed in results obtained from the quantum kinetic theory proposed in Ref.~\cite{Varshney2023}.
On the other hand, the nonlinear thermal Hall conductivity $\kappa^{\mathrm{SC}}$ obtained from the semiclassical theory in Ref.~\cite{Li2024} vanishes at all temperatures owing to the $\mathcal C_3^z$ symmetry, as discussed in Sec.~\ref{subsec:NLTHcoefficient} and Table~\ref{tab:NLTHE_sym}.

To elucidate the origin of the behavior of $\kappa_{y;xx}$ described above, we decompose it into three contributions: $\kappa^{\mathrm{TBCP}}$, $\kappa^{\mathrm{QM}}$, and $\kappa^{\mathrm{disp}}$.
From Eqs.~\eqref{eq:kappa_ijk^TBCP}, \eqref{eq:kappa_ijk^QM}, and \eqref{eq:kappa_ijk^disp}, these contributions can be expressed in the wave-vector representation as
\begin{widetext}
\begin{align}
  \label{eq:kappa_ijk^TBCP-k}  
  \kappa_{i;jl}^{\mathrm{TBCP}}
  &=\frac{\hbar k_B^2}{V}\sum_{n=\pm}\sum_{\bm k}\varepsilon_n^{-1}\left(\partial_{k_j}(\varepsilon_n G_{n,il}^t)-\partial_{k_i}(\varepsilon_n G_{n,jl}^t)\right)c_2[n_B(\varepsilon_n,T)],\\
  \label{eq:kappa_ijk^QM-k}
  \kappa_{i;jl}^{\mathrm{QM}}
  &=\frac{\hbar k_B^2}{V}\sum_{n=\pm}\sum_{\bm k}\left[\left(2\varepsilon_n^{-1}\partial_{k_i}\varepsilon_n \mathcal{G}_{n,jl}+\frac{5}{2}\partial_{k_i}\mathcal{G}_{n,jl}\right)c_2[n_B(\varepsilon_n,T)]
  -\frac{\varepsilon_n^3}{2 k_B^2 T^2}\partial_{k_i}\mathcal{G}_{n,jl}n_B^{\prime}(\varepsilon_n,T)\right],\\
  \label{eq:kappa_ijk^disp-k}
  \kappa_{i;jk}^{\mathrm{disp}}
  &=\frac{1}{24\hbar V}\sum_{n=\pm}\sum_{\bm k}\varepsilon_n^2\partial_{k_i k_j k_l}^3\varepsilon_n\frac{\varepsilon_n n_B^{\prime\prime}(\varepsilon_n,T)-n_B^{\prime}(\varepsilon_n,T)}{T^2}.
\end{align}
\end{widetext}
The temperature dependence of each contribution to $\kappa_{y;xx}$ is shown in Fig.~\ref{fig:FM_yxx}(b).
Similar to the semiclassical result~\cite{Li2024}, $\kappa^{\mathrm{TBCP}}$ vanishes due to the $\mathcal C_3^z$ symmetry, whereas both $\kappa^{\mathrm{QM}}$ and $\kappa^{\mathrm{disp}}$ are nonzero and contribute to the total nonlinear thermal Hall conductivity~\footnote{The ``Hall'' contributions defined in Eq.~\eqref{eq:kappa_Hall-def} for $\kappa^{\mathrm{QM}}$ and $\kappa^{\mathrm{disp}}$ also vanish owing to the $\mathcal C_3^z$ symmetry.}.
At low temperatures, $\kappa^{\mathrm{disp}}$ is the dominant contribution, whereas $\kappa^{\mathrm{QM}}$ becomes dominant at high temperatures.
The nonmonotonic behavior of the total $\kappa_{y;xx}$ originates mainly from the strong temperature dependence of $\kappa^{\mathrm{disp}}$ at low temperatures.
In addition, we find that $\kappa^{\mathrm{QM}}$ approaches a nonzero value, whereas $\kappa^{\mathrm{disp}}$ vanishes in the high-temperature limit, resulting in a nonzero value of the total $\kappa_{y;xx}$.

Here, we discuss the high-temperature behavior of $\kappa^{\mathrm{QM}}$ and $\kappa^{\mathrm{disp}}$.
In the high-temperature limit, the Bose distribution function can be approximated as $n_B(\varepsilon,T)\simeq k_B T/\varepsilon$, and $c_2[n_B]\to \pi^2/3$.
In Eq.~\eqref{eq:kappa_ijk^QM-k}, the terms containing $n_B'/T^2$ vanish, whereas the term proportional to $c_2[n_B]$ can remain nonzero in this limit.
Moreover, the term proportional to $\partial_{k_i}\mathcal{G}_{n,jl}$ in Eq.~\eqref{eq:kappa_ijk^QM-k} also vanishes because $\sum_{\bm k}\partial_{k_i}\mathcal{G}_{n,jl}=0$ due to the periodicity of the Brillouin zone.
The remaining term proportional to $\partial_{k_i}\varepsilon_n \mathcal{G}_{n,jl}$ can remain nonzero, leading to the dominant contribution to $\kappa^{\mathrm{QM}}$ in the high-temperature limit.
On the other hand, in Eq.~\eqref{eq:kappa_ijk^disp-k}, $n_B^{\prime}/T^2$ and $n_B^{\prime\prime}/T^2$ approach zero with increasing temperature, and hence, $\kappa_{y;xx}^{\rm disp}$ also vanishes in this limit.
Finally, we comment on the difference between the high-temperature behavior obtained in our theory and that derived from the quantum kinetic theory in Ref.~\cite{Varshney2023}.
As shown in Fig.~\ref{fig:FM_yxx}(a), the nonlinear thermal Hall conductivity derived from the quantum kinetic theory vanishes in the high-temperature limit.
This behavior can be understood from the fact that, within the quantum kinetic theory, the nonlinear thermal Hall conductivity is proportional to $n_B(\varepsilon,T)/T^2$, which approaches zero in the high-temperature limit.
In contrast, in our theory, $\kappa^{\mathrm{QM}}$ remains nonzero in the high-temperature limit because it depends on $c_2[n_B]$, leading to the difference in the high-temperature behavior compared with the quantum kinetic theory.
Note that, even in the linear thermal Hall effect, the conductivity is known to approach a nonzero value in the high-temperature limit~\cite{Matsumoto2011L,Matsumoto2011B,Matsumoto2014}.

\subsubsection{Distorted case}

\begin{figure*}[t]
\centering
    \includegraphics[width=2\columnwidth,clip]{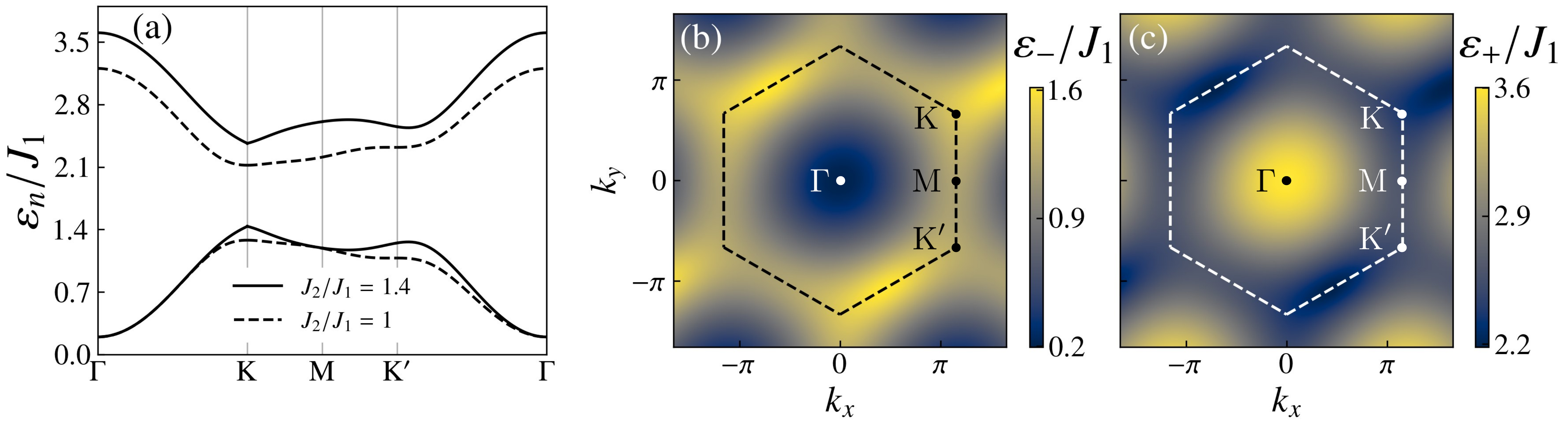}
    \caption{
Similar plots to those in Fig.~\ref{fig:FM_ene} for the distorted ferromagnetic honeycomb lattice model with $J_1=J_3$ and $J_2/J_1=1.4$ under a staggered magnetic field.
The dashed lines represent the magnon band dispersions for the undistorted case with $J_1=J_2=J_3$, as shown in Fig.~\ref{fig:FM_ene}(a).
  }
    \label{fig:st_FM_ene}
\end{figure*}

\begin{figure*}[t]
\centering
    \includegraphics[width=2\columnwidth,clip]{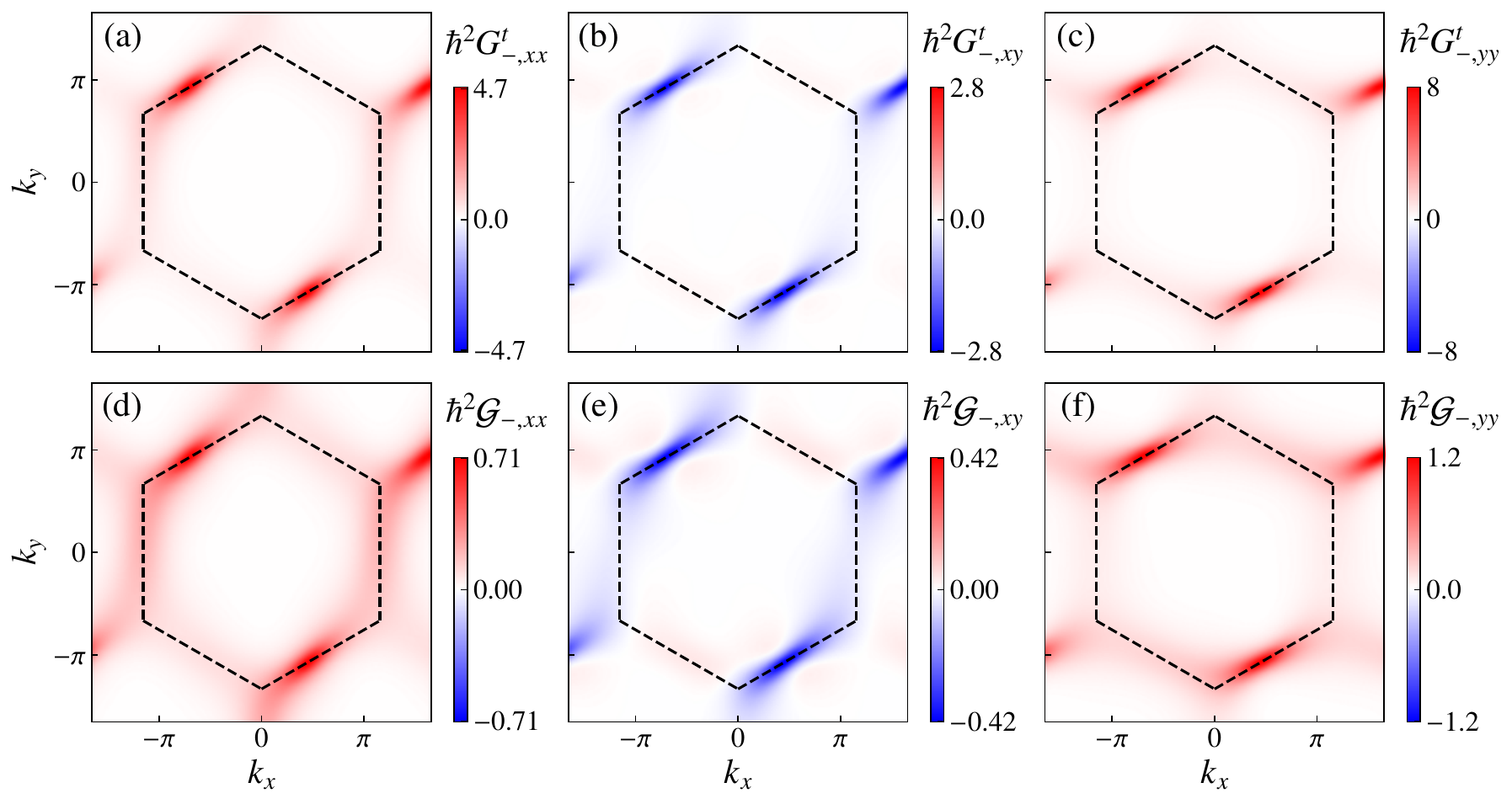}
    \caption{
Similar plots to those in Fig.~\ref{fig:FM_geom} for the distorted ferromagnetic honeycomb lattice model with $J_1=J_3$ and $J_2/J_1=1.4$ under a staggered magnetic field.
  }
    \label{fig:st_FM_geom}
\end{figure*}

In the previous section, we showed that in the undistorted ferromagnetic honeycomb-lattice model with $J_1=J_2=J_3$, the nonlinear thermal Hall conductivity $\kappa_{x;yy}$ vanishes because of the presence of $\mathcal M_y \mathcal T$ symmetry (Table~\ref{tab:NLTHE_sym}).
To break this symmetry, we introduce uniaxial strain along the $\bm \delta_2$ direction.
This uniaxial strain is incorporated by introducing anisotropy into the exchange couplings, with $J_1=J_3\neq J_2$~\cite{Owerre2018,Kondo2022,Li2024}.
In the present calculations, we set $J_1=J_3$ and $J_2/J_1=1.4$.

Figure~\ref{fig:st_FM_ene} shows the magnon band dispersion in the distorted case.
We find that the magnon band structure is modified by the uniaxial strain; in particular, the energy of the low-energy branch increases around the $\mathrm{K}$ and $\mathrm{K^{\prime}}$ points, as shown in Fig.~\ref{fig:st_FM_ene}(a).
Figures~\ref{fig:st_FM_ene}(b) and \ref{fig:st_FM_ene}(c) show the magnon dispersions of the low-energy and high-energy bands, respectively, in the $k_x$--$k_y$ plane.
The symmetry of the magnon dispersions $\varepsilon_{+}(\bm k)$ and $\varepsilon_{-}(\bm k)$ with respect to the $k_x=0$ line is broken by the uniaxial strain.

Figure~\ref{fig:st_FM_geom} shows the momentum dependence of the TBCP [Figs.~\ref{fig:st_FM_geom}(a)--\ref{fig:st_FM_geom}(c)] and the quantum metric [Figs.~\ref{fig:st_FM_geom}(d)--\ref{fig:st_FM_geom}(f)] for the low-energy band in the distorted case.
As in the magnon band dispersions, the symmetry with respect to the $k_x=0$ line in each component of the TBCP and the quantum metric is broken by the uniaxial strain.
We also find that, although the momentum dependence of the TBCP is similar to that of the quantum metric, the absolute values of the TBCP are greater than those of the quantum metric.
This suggests that the TBCP, rather than the quantum metric, plays a significant role in the nonlinear thermal Hall conductivity because $\kappa^{\mathrm{TBCP}}$ can be nonzero, unlike in the undistorted case.

\begin{figure*}[t]
\centering
    \includegraphics[width=2\columnwidth,clip]{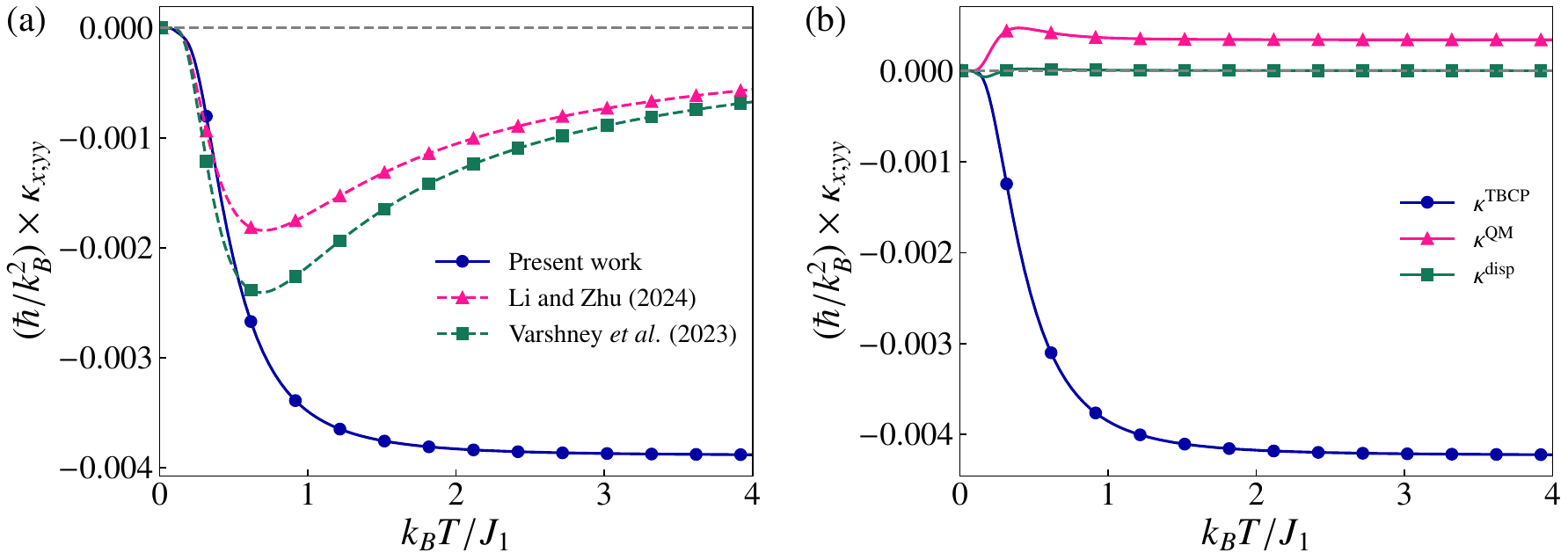}
    \caption{
Similar plots of $\kappa_{x;yy}$ to those in Fig.~\ref{fig:FM_yxx} for the distorted ferromagnetic honeycomb lattice model with $J_1=J_3$ and $J_2/J_1=1.4$ under a staggered magnetic field.
  }
    \label{fig:FM_xyy}
\end{figure*}

Here, we calculate the nonlinear thermal Hall conductivity.
In particular, we focus on the component $\kappa_{x;yy}$, in which a heat current along the $x$ direction is driven by a temperature gradient in the $y$ direction; this response is forbidden in the undistorted case.
Figure~\ref{fig:FM_xyy}(a) shows the temperature dependence of $\kappa_{x;yy}$ together with results calculated using formulas obtained in previous studies~\cite{Varshney2023,Li2024}.
As shown in Fig.~\ref{fig:FM_xyy}(a), the nonlinear thermal Hall conductivity $\kappa_{x;yy}$ calculated within our theory agrees well with the results of previous studies at low temperatures.
With increasing temperature, $\kappa_{x;yy}$ calculated within our theory decreases and approaches a nonzero negative value in the high-temperature limit, whereas the results of previous studies~\cite{Varshney2023,Li2024} instead increase and approach zero in this limit.
To clarify the origin of this behavior of $\kappa_{x;yy}$, we decompose it into three contributions, namely $\kappa^{\mathrm{TBCP}}$, $\kappa^{\mathrm{QM}}$, and $\kappa^{\mathrm{disp}}$, similar to the undistorted case.
As shown in Fig.~\ref{fig:FM_xyy}(b), $\kappa^{\mathrm{TBCP}}$ provides the dominant contribution to the total nonlinear thermal Hall conductivity $\kappa_{x;yy}$ over the entire temperature range; this can be understood from the larger absolute values of the TBCP compared with those of the quantum metric shown in Fig.~\ref{fig:st_FM_geom}.
As discussed in Sec.~\ref{subsec:NLTHcoefficient}, $\kappa^{\mathrm{TBCP}}$ is decomposed into two terms, $\kappa^{\mathrm{TBCP}}_{x;yy}=\kappa^{\mathrm{SC}}_{x;yy}+\kappa^{\mathrm{TBCP},2}_{x;yy}$, where $\kappa^{\mathrm{SC}}_{x;yy}$ and $\kappa^{\mathrm{TBCP},2}_{x;yy}$ are given by
\begin{widetext}
\begin{align}\label{eq:kappa_SC-k}
  \kappa^{\mathrm{SC}}_{x;yy}&=\frac{k_B^2\hbar}{V}\sum_{\bm k}\sum_{n=\pm1}(\partial_{k_y}G^t_{n,xy}-\partial_{k_x}G^t_{n,yy})c_2[n_B(\varepsilon_n,T)],\\
  \kappa^{\mathrm{TBCP},2}_{x;yy}&=\frac{k_B^2\hbar}{V}\sum_{\bm k}\sum_{n=\pm1}\varepsilon_n^{-1}(\partial_{k_y}\varepsilon_n G^t_{n,xy}-\partial_{k_x}\varepsilon_n G^t_{n,yy})c_2[n_B(\varepsilon_n,T)],
\end{align}
\end{widetext}
where $\kappa^{\mathrm{SC}}$ denotes the result obtained within the semiclassical theory~\cite{Li2024} and is derived from Eq.~\eqref{eq:kappa_SC} by replacing the momentum $\bm{p}$ with $\hbar \bm{k}$.
Thus, $\kappa^{\mathrm{TBCP},2}_{x;yy}$ can be regarded as a quantum correction to the semiclassical result $\kappa^{\mathrm{SC}}_{x;yy}$.

At low temperatures, magnons in the low-energy region predominantly contribute to the nonlinear thermal Hall conductivity due to the Bose distribution function.
As shown in Fig.~\ref{fig:st_FM_ene}, the magnon energy is minimized at the $\Gamma$ point; therefore, the primary contributions to the conductivity arise from quantum states in the vicinity of the $\Gamma$ point at low temperatures.
From Figs.~\ref{fig:st_FM_ene} and \ref{fig:st_FM_geom}, we find that both $G^t$ and $\partial_{k_i}\varepsilon_n$ are small near the $\Gamma$ point, whereas the momentum derivatives of $G^t$ are relatively large in this region.
These characteristics around the $\Gamma$ point lead to a dominant contribution of $\kappa^{\mathrm{SC}}$ to the total nonlinear thermal Hall conductivity at low temperatures and result in good agreement between the results obtained from our framework and those derived from the semiclassical theory~\cite{Li2024} in this temperature range.

On the other hand, in the high-temperature limit, $c_2[n_B(\varepsilon_n,T)]$ approaches $\frac{\pi^2}{3}$, which is independent of both temperature and the magnon energy $\varepsilon_n(\bm k)$.
In this limit, $\kappa^{\mathrm{SC}}_{x;yy}\to\frac{k_B^2\pi^2}{3\hbar V}\sum_{n,\bm k}(\partial_{k_y}G^t_{n,xy}-\partial_{k_x}G^t_{n,yy})$, which vanishes owing to the periodicity of the Brillouin zone.
Thus, as the temperature increases, the contribution from $\kappa^{\mathrm{TBCP},2}_{x;yy}$ becomes dominant in the total nonlinear thermal Hall conductivity, leading to a deviation from the semiclassical result at high temperatures.

As mentioned above, the low-temperature behavior of the nonlinear thermal Hall conductivity is in close agreement with that reported in previous studies~\cite{Li2024}, which has been attributed to the dominant contribution of $\kappa^{\mathrm{TBCP}}$.
Therefore, suppressing the TBCP contribution would facilitate the experimental verification of the present theory.
To this end, we consider systems with $\mathcal C_3^z$ symmetry, in which the TBCP contribution vanishes, as demonstrated in the undistorted case.
In such systems, the present study predicts a sign change in the nonlinear thermal Hall conductivity at low temperatures; therefore, experimental confirmation of this prediction is essential for establishing the validity of our theory.

\section{Summary and outlook}
\label{sec:summary}

In summary, we have formulated the intrinsic nonlinear thermal conductivity in bosonic systems, which is applicable to a wide range of systems including magnons and photons.
This formulation is based on the Wigner representation of the Hamiltonian and the density matrix, which allows us to systematically expand the heat current density with respect to spatial gradients in the star product.
Within this approach, we can avoid the use of Luttinger's gravitational potential, which requires the introduction of energy magnetization.
The resulting nonlinear thermal conductivity consists of three contributions: two of them are related to the quantum geometric properties of bosons, namely the thermal Berry connection polarizability and the quantum metric, while the other is determined solely by the band dispersion.
We have applied our theory to a ferromagnetic honeycomb lattice spin model with a staggered magnetic field, which serves as a minimal model for realizing the nonlinear thermal Hall effect in magnon systems.
As candidate platforms close to this model, two-dimensional van der Waals honeycomb ferromagnets, including chromium trihalides such as CrBr$_3$, CrI$_3$, and CrCl$_3$, have been discussed in previous studies~\cite{Varshney2023}.
We have shown that the contribution from the thermal Berry connection polarizability vanishes in the presence of threefold rotational symmetry and that the quantum metric contribution becomes dominant at high temperatures.
Such symmetry and temperature dependences may provide useful clues for
identifying the origin of the nonlinear thermal Hall effect in experiments.
The results obtained in this work significantly differ from those of previous theories based on the semiclassical approach and quantum kinetic theory, which rely on Luttinger's gravitational potential.
These differences are considered to arise from quantum corrections beyond semiclassical theory or from approximations used to address energy magnetization.
Therefore, we believe that our theory provides a more accurate description of the nonlinear thermal Hall effect in bosonic systems and is capable of clarifying the role of quantum geometry in thermal transport phenomena.

In the present study, we have focused on the intrinsic contribution to the nonlinear thermal conductivity, which is independent of the relaxation time.
In future work, it will be important to extend our theory to include extrinsic contributions that depend on the relaxation time and are relevant to experimental observations.
To this end, damping can be introduced into the quantum kinetic equation and treated within the relaxation-time approximation.
In disordered systems, translational symmetry is broken, so momentum-space quantities such as the quantum metric in the clean-limit formula should be replaced by quantities defined in the real-space representation.
Impurity effects on the nonlinear thermal Hall effect can also be included in the real-space formulation in a manner similar to that used for the linear thermal Hall effect~\cite{Mangeolle2026}.
Moreover, although we have formulated the nonlinear thermal conductivity for bosonic systems, it is also important to extend our theory to fermionic systems.
This extension is expected to be carried out easily because the fermionic Hamiltonian can be diagonalized by a simple unitary transformation, unlike bosonic systems.
A fermionic formulation for electronic systems is essential for describing the cross-correlation between the electric field (electric current) and the temperature gradient (heat current), which is an important topic in condensed matter physics.
In particular, our theory of heat-current generation may be applied to nonreciprocal transport phenomena~\cite{Takashima2018,Nomura2019,Nakai2019,Hirokane2020,Nomura2023,Sano2024,Terashima2025} and nonlinear cross-responses involving electric and heat currents~\cite{Zeng2019,Zeng2022,Yamaguchi2024,Nakazawa2024,Nomoto2025,Nakazawa2025}, where contributions that cannot be addressed by the Kubo formula or the Boltzmann equation are expected to be captured within our formulation.
The generation of spin current induced by a temperature gradient, i.e., the spin Nernst effect~\cite{Xiao2021,Koyama2025,Matsushita2025}, is also an important topic to which our theory can be applied. Although a semiclassical theory has been developed, a fully quantum-mechanical description is still lacking.
Thus, our theory can be applied to a wide range of phenomena in both bosonic and fermionic systems and is expected to provide new insights into the role of quantum geometry in transport phenomena.

\begin{acknowledgments}
  The authors thank A.~Ono, K.~Takasan, and S.~Okumura for fruitful discussions.
  This work was supported by Grant-in-Aid for Scientific Research from
  JSPS, KAKENHI Grant No.~JP20H00122, JP22H01175, JP23H01129, JP23H04865, JP24K00563.
  \end{acknowledgments}

\appendix

\section{Convolution in Wigner transformation}\label{conv_star}

In this appendix, we show that the Wigner transformation of the convolution of two functions, $A(\bm r_{1},\bm r_{2})$ and $B(\bm r_{1},\bm r_{2})$, can be expressed as the star product of their Wigner transformations.
Here, in the integral notation, we explicitly retain the integration variables to avoid confusion.

We define the Wigner transformation of a function $A(\bm r_{1},\bm r_{2})$ as
\begin{align}
  \mathsf A(\bm X,\bm p)=\int e^{-i\frac{\bm p\cdot\bm x}{\hbar}}A\left(\bm X+\frac{\bm x}{2},\bm X-\frac{\bm x}{2}\right)\,d\bm x,
\end{align}
where $\bm X=(\bm r_1+\bm r_2)/2$ and $\bm x=\bm r_1-\bm r_2$.
From this definition, the inverse transformation is given by
\begin{align}\label{eq:inv_Wigner}
  &\frac{1}{(2\pi\hbar)^d}\int \mathsf A(\bm X,\bm p)e^{i\frac{\bm p\cdot\bm x}{\hbar}}\,d\bm p\notag\\
  &=\frac{1}{(2\pi\hbar)^d}\iint e^{i\frac{\bm p\cdot(\bm x-\bm x^{\prime})}{\hbar}}A\left(\bm X+\frac{\bm x^{\prime}}{2},\bm X-\frac{\bm x^{\prime}}{2}\right)\,d\bm x^\prime d\bm p\notag\\
  &=A(\bm r_{1},\bm r_{2}).
\end{align}
In addition, we define $\tilde{\mathsf A}(\bm X,\bm p)$ by the following transformation:
\begin{align}
  \mathsf A(\bm X,\bm p)=\iint \tilde{\mathsf A}(\bm\alpha,\bm\beta)e^{\frac{i}{\hbar}(\bm X\cdot\bm \alpha+\bm p\cdot\bm\beta)}\,d\bm\alpha d\bm\beta.
\end{align}
From Eq.~\eqref{eq:inv_Wigner}, the function $A(\bm r_{1},\bm r_{2})$ can be expressed in terms of $\tilde{\mathsf A}$ as follows:
\begin{align}
  A(\bm r_{1},\bm r_{2})
  &=\frac{1}{(2\pi\hbar)^d}\iiint \tilde{\mathsf A}(\bm\alpha,\bm\beta)e^{\frac{i}{\hbar}(\bm X\cdot\bm \alpha+\bm p\cdot\bm\beta)}e^{i\frac{\bm p\cdot\bm x}{\hbar}}\,d\bm\alpha d\bm\beta d\bm p\notag\\
  &=\iint \delta(\bm\beta+\bm x)\tilde{\mathsf A}(\bm\alpha,\bm\beta)e^{i\frac{\bm X\cdot\bm\alpha}{\hbar}}\,d\bm\alpha d\bm\beta\notag\\
  &=\int \tilde{\mathsf A}(\bm \alpha,\bm r_2-\bm r_1)e^{\frac{i}{\hbar}\left(\frac{\bm r_1+\bm r_2}{2}\right)\cdot\bm\alpha}\,d\bm \alpha.
\end{align}
Thus, the Wigner transformation of the convolution $[A*B](\bm r_{1},\bm r_{2})\equiv\int A(\bm r_{1},\bm r)B(\bm r,\bm r_{2})\,d\bm r$ can be expressed as
\begin{widetext}
\begin{align}
  [A*B]^{W}(\bm X,\bm p)
  &=\iint A\left(\bm X+\frac{\bm x}{2},\bm r\right)B\left(\bm r,\bm X-\frac{\bm x}{2}\right)e^{-i\frac{\bm p\cdot\bm x}{\hbar}}\,d\bm r d\bm x\notag\\
  &=\iiiint \tilde{\mathsf A}\left(\bm \alpha,\bm r-\bm X-\frac{\bm x}{2}\right)e^{\frac{i}{\hbar}\bm\alpha\cdot\left(\frac{\bm X+\bm x/2+\bm r}{2}\right)}
  \tilde{\mathsf B}\left(\bm\beta,-\bm r+\bm X-\frac{\bm x}{2}\right)e^{\frac{i}{\hbar}\bm\beta\cdot\left(\frac{\bm X-\bm x/2+\bm r}{2}\right)}e^{-i\frac{\bm p\cdot\bm x}{\hbar}}\,d\bm r d\bm x d\bm\alpha d\bm\beta.
\end{align}
Here, we introduce $\bm X_1\equiv\bm r-\bm X-\frac{\bm x}{2}$ and $\bm X_2\equiv-\bm r+\bm X-\frac{\bm x}{2}$, and proceed with the calculation as follows:
\begin{align}
  [A*B]^{W}(\bm X,\bm p)&=\iiiint \tilde{\mathsf A}(\bm\alpha,\bm X_1)\tilde{\mathsf B}(\bm\beta,\bm X_2)e^{\frac{i}{\hbar}\bm\alpha\cdot\left(\frac{2\bm X-\bm X_2}{2}\right)}e^{\frac{i}{\hbar}\bm\beta\cdot\left(\frac{2\bm X+\bm X_2}{2}\right)}e^{\frac{i}{\hbar}\bm p\cdot(\bm X_1+\bm X_2)}\,d\bm X_1 d\bm X_2 d\bm\alpha d\bm\beta\notag\\
  &=\iiiint \tilde{\mathsf A}(\bm\alpha,\bm X_1)\tilde{\mathsf B}(\bm\beta,\bm X_2)e^{\frac{i}{\hbar}(\bm X\cdot(\bm\alpha+\bm\beta)+\bm p\cdot(\bm X_1+\bm X_2))}e^{\frac{i}{2\hbar}(\bm X_1\cdot\bm\beta-\bm\alpha\cdot\bm X_2)}\,d\bm X_1 d\bm X_2 d\bm\alpha d\bm\beta\notag\\
  &=\sum_{n=0}^{\infty}\frac{1}{n!}\iiiint e^{\frac{i}{\hbar}(\bm X\cdot\bm\alpha+\bm p\cdot\bm X_1)}\tilde{\mathsf A}(\bm\alpha,\bm X_1)\left(\frac{i}{2\hbar}(\bm X_1\cdot\bm\beta-\bm\alpha\cdot\bm X_2)\right)^n e^{\frac{i}{\hbar}(\bm X\cdot\bm\beta+\bm p\cdot\bm X_2)}\tilde{\mathsf B}(\bm\beta,\bm X_2)\,d\bm X_1 d\bm X_2 d\bm\alpha d\bm\beta\notag\\
  &=\sum_{n=0}^{\infty}\frac{1}{n!}\iiiint e^{\frac{i}{\hbar}(\bm X\cdot\bm\alpha+\bm p\cdot\bm X_1)}\tilde{\mathsf A}(\bm\alpha,\bm X_1)\left(\frac{\hbar}{2i}(\backvec{\partial}_{\bm p}\cdot\vec{\partial}_{\bm X}-\backvec{\partial}_{\bm X}\cdot\vec{\partial}_{\bm p})\right)^n e^{\frac{i}{\hbar}(\bm X\cdot\bm\beta+\bm p\cdot\bm X_2)}\tilde{\mathsf B}(\bm\beta,\bm X_2)\,d\bm X_1 d\bm X_2 d\bm\alpha d\bm\beta\notag\\
  &=\sum_{n=0}^{\infty}\frac{1}{n!}\mathsf A(\bm X,\bm p)\left(\frac{\hbar}{2i}(\backvec{\partial}_{\bm p}\cdot\vec{\partial}_{\bm X}-\backvec{\partial}_{\bm X}\cdot\vec{\partial}_{\bm p})\right)^n\mathsf B(\bm X,\bm p).
\end{align}
The differential operators with left (right) arrows act on the functions on the left (right) side, as introduced in Eq.~\eqref{def_star} for the definition of the star product.
Thus, the Wigner transformation of the convolution can be written using the star product as
\begin{align}\label{eq:Wigner_conv_star}
  [A*B]^W(\bm X,\bm p)=\mathsf A(\bm X,\bm p)\star\mathsf B(\bm X,\bm p).
\end{align}

The associative law of the star product and the cyclic property of the trace under the full phase-space integral can be demonstrated using Eq.~\eqref{eq:Wigner_conv_star} as follows.
The associativity of the convolution for the original functions $A(\bm r_1,\bm r_2)$, $B(\bm r_1,\bm r_2)$, and $C(\bm r_1,\bm r_2)$ is expressed as
$[[A*B]*C](\bm r_1,\bm r_2)=\int [A*B](\bm r_1,\bm r)C(\bm r,\bm r_2)\,d\bm r=\iint A(\bm r_1,\bm r^\prime)B(\bm r^\prime,\bm r)C(\bm r,\bm r_2)\,d\bm r^\prime d\bm r=[A*[B*C]](\bm r_1,\bm r_2)$.
From this associative law and Eq.~\eqref{eq:Wigner_conv_star}, the associativity of the star product is obtained as
$(\mathsf A(\bm X,\bm p)\star \mathsf B(\bm X,\bm p))\star \mathsf C(\bm X,\bm p)
  =[[A*B]*C]^W(\bm X,\bm p)=[A*[B*C]]^W(\bm X,\bm p)=\mathsf A(\bm X,\bm p)\star(\mathsf B(\bm X,\bm p)\star \mathsf C(\bm X,\bm p))$.

Next, we show the cyclic property of the trace under the full phase-space integral in the star product.
Given $A(\bm r_1,\bm r_2)$ and $B(\bm r_1,\bm r_2)$ as $n\times n$ matrix-valued functions, and their Wigner transformations $\mathsf A(\bm X,\bm p)$ and $\mathsf B(\bm X,\bm p)$, we obtain
\begin{align}
  \frac{1}{(2\pi\hbar)^d}\iint \mTr[\mathsf A(\bm X,\bm p)\star\mathsf B(\bm X,\bm p)]\,d\bm X d\bm p
  &=\frac{1}{(2\pi\hbar)^d}\iiiint \mTr\left[A\left(\bm X+\frac{\bm x}{2},\bm r\right)B\left(\bm r,\bm X-\frac{\bm x}{2}\right)\right]e^{-i\frac{\bm p\cdot\bm x}{\hbar}}\,d\bm r d\bm x d\bm X d\bm p\notag\\
  &=\iiint \mTr\left[A\left(\bm X+\frac{\bm x}{2},\bm r\right)B\left(\bm r,\bm X-\frac{\bm x}{2}\right)\right]\delta(\bm x)\,d\bm r d\bm x d\bm X\notag\\
  &=\iint \mTr[A(\bm X,\bm r)B(\bm r,\bm X)]\,d\bm r d\bm X\notag\\
  &=\iint \mTr[B(\bm X,\bm r)A(\bm r,\bm X)]\,d\bm r d\bm X\notag\\
  &=\frac{1}{(2\pi\hbar)^d}\iint \mTr[\mathsf B(\bm X,\bm p)\star\mathsf A(\bm X,\bm p)]\,d\bm X d\bm p,
\end{align}
which demonstrates that the cyclic property of the trace holds under the full phase-space integral.

\section{Derivations of gauge-invariant heat current density}\label{energy_current_calc}

In this section, we provide a detailed derivation of the gauge-invariant heat current density~\eqref{diag_J}.
Before presenting the derivation, however, we show in the following section that the Hamiltonian and the density matrix can be expressed in terms of gauge-invariant projection operators and the quantum geometric tensor.

\subsection{Representation of physical quantities using projection operators}\label{app:P_n}

In this section, we show that the Hamiltonian and the density matrix can be expressed as in Eqs.~\eqref{proj_h} and \eqref{FstarH} by using the projection operator $P_n={\mathsf T}\star p_n\star {\mathsf T}^{-1}$ defined in Eq.~\eqref{def_P_n}.
Since both $\tilde{h}$ and $p_n$ are diagonal matrices and $p_n$ is a constant independent of $\bm X$ and $\bm p$, the commutation relation $[\tilde h\stco p_n]=0$ holds.
Then, from $[\tilde h\stco p_n]=0$ and the expression of the Hamiltonian $\bar{\mathsf H}={\mathsf T}\star\tilde h\star {\mathsf T}^{-1}$, we find that $[\bar{\mathsf H}\stco P_n]=0$.
Moreover, from the relation $\sum_{n=1}^{2N}P_n=I_{2N}$, we obtain $\bar{\mathsf H}=\sum_{n=1}^{2N} P_n\star\bar{\mathsf H}\star P_n=\sum_{n=1}^{2N}{\mathsf T}\star\tilde h_np_n\star {\mathsf T}^{-1}$.
Here, we assume that there exists a matrix-valued function $\underline{h}_n$ satisfying
\begin{align}\label{eq:P_n_1}
  P_n\star\underline{h}_n\star P_n={\mathsf T}\star\tilde{h}_np_n\star {\mathsf T}^{-1}.
\end{align}
The following two relations hold for arbitrary matrix-valued functions $A(\bm X, \bm p)$ and $B(\bm X, \bm p)$ and a scalar-valued function $c(\bm X, \bm p)$:
\begin{align}
  \label{identity1}
  A\star c\star B
  &=c(A\star B)-\frac{i\hbar}{2}\epsilon_{\alpha\beta}\partial_\alpha c(\partial_\beta A\star B-A\star \partial_\beta B)-\frac{\hbar^2}{8}\epsilon_{\alpha\beta}\epsilon_{\sigma\lambda}\partial_{\alpha\sigma}^2c(\partial_{\beta\lambda}^2(A\star B)-4\partial_\beta A\star\partial_\lambda B)+\mathcal O(\hbar^3).\\
\label{identity2}
  \mTr(A\star B)
  &=\mTr(B\star A)-i\hbar\epsilon_{\alpha\beta}\mTr(\partial_\alpha B\star \partial_\beta A)-\frac{\hbar^2}{2}\epsilon_{\alpha\beta}\epsilon_{\sigma\lambda}\mTr(\partial_{\alpha\sigma}^2B\star \partial_{\beta\lambda}^2A)+\mathcal O(\hbar^3),
\end{align}
Here and in what follows, the Einstein summation convention is assumed for Greek indices such as $\alpha$ and $\beta$.
Using the above relations, we can calculate the left-hand and right-hand sides of Eq.~\eqref{eq:P_n_1} up to the second order in $\hbar$ as
\begin{align}
  \label{eq:appB1}
  \mTr(P_n\star\underline{h}_n\star P_n)&=\underline{h}_n\mTr(P_n)+\hbar^2\epsilon_{\alpha\beta}\epsilon_{\sigma\lambda}\partial_\sigma(\partial_\alpha\underline{h}_ng_{n,\beta\lambda})+\mathcal O(\hbar^3),\\
  \label{eq:appB2}
  \mTr({\mathsf T}\star\tilde{h}_np_n\star {\mathsf T}^{-1})
  &=\tilde{h}_n+\hbar\epsilon_{\alpha\beta}\partial_\alpha(\tilde{h}_nA_{n,\beta})-\frac{\hbar^2}{4}\epsilon_{\alpha\beta}\epsilon_{\sigma\lambda}\partial_{\alpha\sigma}^2(\tilde{h}_n\{\Lambda_\beta,\Lambda_\lambda\}_{nn})+\mathcal O(\hbar^3),
\end{align}
respectively.
From Eqs.~\eqref{identity2}, \eqref{eq:def_Curv}, and \eqref{eq:def_QM}, the trace of the projection operator $P_n$ is calculated as
\begin{align}\label{eq:trP_n}
  \mTr(P_n)
  &=\mTr({\mathsf T}\star p_n\star {\mathsf T}^{-1})=\mTr({\mathsf T}\star (p_n{\mathsf T}^{-1}))\notag\\
  &=\mTr(p_n)+\hbar\epsilon_{\alpha\beta}\partial_\alpha\mTr(p_n\Lambda_\beta)-\frac{\hbar^2}{2}\epsilon_{\alpha\beta}\epsilon_{\gamma\sigma}\partial_{\alpha\gamma}^2\mTr(p_n{\mathsf T}^{-1}\partial_{\beta\sigma}^2{\mathsf T})+\mathcal O(\hbar^3)\notag\\
  &=1+\hbar\epsilon_{\alpha\beta}\partial_\alpha A_{n,\beta}+\frac{\hbar^2}{2}\epsilon_{\alpha\beta}\epsilon_{\gamma\sigma}\partial_{\alpha\gamma}^2\left[\mathrm{diag}(\Lambda_\beta\Lambda_\sigma)\right]_n+\mathcal O(\hbar^3)\notag\\
  &=1+\hbar\epsilon_{\alpha\beta}\Omega_{n,\alpha\beta}+\frac{\hbar^2}{2}\epsilon_{\alpha\beta}\epsilon_{\gamma\sigma}\partial_{\alpha\gamma}^2g_{n,\beta\sigma}
  +\frac{\hbar^2}{2}\epsilon_{\alpha\beta}\epsilon_{\gamma\sigma}(\partial_\gamma A_{n,\alpha}\partial_\sigma A_{n,\beta}-\partial_\gamma(\Omega_{n,\alpha\beta}A_{n,\sigma})+\partial_{\alpha\gamma}^2(A_{n,\beta}A_{n,\sigma}))+\mathcal O(\hbar^3)\notag\\
  &=1+\hbar\epsilon_{\alpha\beta}\Omega_{n,\alpha\beta}+\frac{\hbar^2}{2}\epsilon_{\alpha\beta}\epsilon_{\gamma\sigma}\partial_{\alpha\gamma}^2g_{n,\beta\sigma}
  +\frac{\hbar^2}{4}\epsilon_{\alpha\beta}\epsilon_{\gamma\sigma}\left(\Omega_{n,\alpha\gamma}\Omega_{n,\beta\sigma}-\frac12\Omega_{n,\alpha\beta}\Omega_{n,\gamma\sigma}\right)+\mathcal O(\hbar^3).
\end{align}
Substituting Eq.~\eqref{eq:trP_n} into Eq.~\eqref{eq:appB1} and comparing it with Eq.~\eqref{eq:appB2}, we obtain $\underline{h}=h$ up to the second order in $\hbar$ [see also Eq.~\eqref{diag_h} for the relation between $h$ and $\tilde h$].
From this result, we derive Eq.~\eqref{proj_h}.

Next, let us derive Eq.~\eqref{FstarH}.
From the relation $\mathsf F\star\mathsf H={\mathsf T}\star\tilde f\star\tilde h\star {\mathsf T}^{-1}+{\mathsf T}\star\tilde{\mathcal F}\star\tilde h\star {\mathsf T}^{-1}$ and by performing similar calculations as in the case above, we obtain
\begin{align}\label{app:FstarH}
  \mathsf F\star\mathsf H=\sum_{n=1}^{2N}P_n\star{d}_n\star P_n
  +{\mathsf T}\star\tilde{\mathcal F}\star\tilde h\star {\mathsf T}^{-1},
\end{align}
and 
\begin{align}
  d
  &=\tilde{f}\star\tilde{h}+\hbar\epsilon_{\alpha\beta}\partial_{\alpha}(\tilde f\star\tilde h)(A_\beta+\hbar\epsilon_{\gamma\sigma}A_{\sigma}\partial_\gamma A_\beta)
  -\frac{\hbar^2}{4}\epsilon_{\alpha\beta}\epsilon_{\gamma\sigma}\partial_{\alpha\gamma}^2(\tilde f\star\tilde h)(\mathrm{diag}(\{\Lambda_\beta,\Lambda_\sigma\})-4A_\beta A_\sigma)
  +\mathcal O(\hbar^3).
\end{align}
By expanding the star product, $d$ can be expressed as follows.
\begin{align}
  d
  &=\tilde{f}\tilde{h}+\hbar\epsilon_{\alpha\beta}\partial_{\alpha}(\tilde f\tilde h)(A_\beta+\hbar\epsilon_{\gamma\sigma}A_{\sigma}\partial_\gamma A_\beta)
  -\frac{\hbar^2}{4}\epsilon_{\alpha\beta}\epsilon_{\gamma\sigma}\partial_{\alpha\gamma}^2(\tilde f\tilde h)(\mathrm{diag}(\{\Lambda_\beta,\Lambda_\sigma\})-4A_\beta A_\sigma)
  \notag\\
  &\qquad
  +i\frac{\hbar}{2}\epsilon_{\alpha\beta}(\partial_\alpha\tilde{f}\partial_\beta\tilde{h}+\hbar\epsilon_{\gamma\sigma}\partial_{\gamma}(\partial_\alpha\tilde f\partial_\beta\tilde h)A_\sigma)
  -\frac{\hbar^2}{8}\epsilon_{\alpha\beta}\epsilon_{\gamma\sigma}\partial_{\alpha\gamma}^2\tilde{f}\partial_{\beta\sigma}^2\tilde{h}
  +\mathcal O(\hbar^3)\notag\\
\label{eq:appB3}
  &=\tilde{f}h+\hbar\epsilon_{\alpha\beta}\partial_\alpha \tilde{f}(A_\beta+\hbar\epsilon_{\gamma\sigma}A_{\sigma}\partial_\gamma A_\beta)h
  -\frac{\hbar^2}{4}\epsilon_{\alpha\beta}\epsilon_{\gamma\sigma}\partial_{\alpha\gamma}^2\tilde{f}(\mathrm{diag}(\{\Lambda_\beta,\Lambda_\sigma\})-4A_\beta A_\sigma)h
  \notag\\
  &\qquad
  -\hbar^2\epsilon_{\alpha\beta}\epsilon_{\gamma\sigma}\left(\partial_\alpha\tilde{f}\partial_\gamma hA_\beta A_\sigma
  +\frac{1}{2}\partial_\alpha \tilde f \partial_\gamma h(\mathrm{diag}(\{\Lambda_\beta,\Lambda_\sigma\})-4A_\beta A_\sigma)\right)
  \notag\\
  &\qquad
  +i\frac{\hbar}{2}\epsilon_{\alpha\beta}(\partial_\alpha\tilde{f}\partial_\beta h+\hbar\epsilon_{\gamma\sigma}\partial_{\alpha\gamma}^2 \tilde{f}A_\sigma \partial_\beta h
  -\hbar\epsilon_{\gamma\sigma}\partial_\alpha\tilde{f}\partial_\gamma h\partial_\beta A_\sigma)
  -\frac{\hbar^2}{8}\epsilon_{\alpha\beta}\epsilon_{\gamma\sigma}\partial_{\alpha\gamma}^2\tilde{f}\partial_{\beta\sigma}^2h+\mathcal O(\hbar^3).
\end{align}
Here, we introduce $\underline f$ as follows to simplify the expression of $d$:
\begin{align}\label{eq:def_uline_f}
  \underline f
  \equiv\tilde{f}+\hbar\epsilon_{\alpha\beta}\partial_{\alpha}\tilde f(A_\beta+\hbar\epsilon_{\gamma\sigma}A_{\sigma}\partial_\gamma A_\beta)
  -\frac{\hbar^2}{4}\epsilon_{\alpha\beta}\epsilon_{\gamma\sigma}\partial_{\alpha\gamma}^2\tilde f(\mathrm{diag}(\{\Lambda_\beta,\Lambda_\sigma\})-4A_\beta A_\sigma)
  +\mathcal O(\hbar^3)
\end{align}
Using $\underline f$ defined above, Eq.~\eqref{eq:appB3} can be rewritten as
\begin{align}
  d
  &=\underline{f}h-\hbar^2\epsilon_{\alpha\beta}\epsilon_{\gamma\sigma}\left(\partial_\alpha \underline{f}\partial_\gamma h\left(\frac{1}{2}\mathrm{diag}(\{\Lambda_\beta,\Lambda_\sigma\})-A_\beta A_\sigma\right)\right)\notag\\
  &\qquad
  +i\frac{\hbar}{2}\epsilon_{\alpha\beta}(\partial_\alpha\underline{f}\partial_\beta h
  -\hbar\epsilon_{\gamma\sigma}(\partial_\gamma\underline{f}\partial_\beta h\partial_\alpha A_\sigma+\partial_\alpha\underline{f}\partial_\gamma h\partial_\beta A_\sigma))
  -\frac{\hbar^2}{8}\epsilon_{\alpha\beta}\epsilon_{\gamma\sigma}\partial_{\alpha\gamma}^2\underline{f}\partial_{\beta\sigma}^2h+\mathcal O(\hbar^3).
\end{align}
From Eqs.~\eqref{eq:def_Curv} and \eqref{eq:def_QM}, $d$ in Eq.~\eqref{app:FstarH} can be expressed as
\begin{align}\label{eq:qpp:d_expression}
  {d}
  &=\underline{f} h-\hbar^2\epsilon_{\alpha\beta}\epsilon_{\gamma\sigma}\left(\partial_\alpha\underline{f}\partial_\gamma h g_{\beta\sigma}+\frac{1}{8}\partial_{\alpha\gamma}^2\underline{f} \partial_{\beta\sigma}^2 h\right)
  +i\frac{\hbar}{2}\epsilon_{\alpha\beta}(\partial_\alpha\underline{f}\partial_\beta h-\hbar\epsilon_{\gamma\sigma}\partial_\alpha\underline{f}\partial_\gamma h \Omega_{\beta\sigma})+\mathcal O(\hbar^3),
\end{align}
which corresponds to Eq.~\eqref{eq:def_uline_d} in the main text.
Using Eqs.~\eqref{diag_f} and \eqref{eq:def_uline_f}, we can express $\underline f$ in terms of $f$ as follows:
\begin{align}
  \underline{f}
  &=f-\hbar\epsilon_{\alpha\beta}\partial_{\alpha}(f(A_\beta+\hbar\epsilon_{\gamma\sigma}A_{\sigma}\partial_\gamma A_\beta))
  -\frac{\hbar^2}{4}\epsilon_{\alpha\beta}\epsilon_{\gamma\sigma}\partial_{\alpha\gamma}^2(f(\mathrm{diag}(\{\Lambda_\beta,\Lambda_\sigma\})-4A_\beta A_\sigma))\notag\\
  &\qquad+\hbar\epsilon_{\alpha\beta}\partial_\alpha f(A_\beta+\hbar\epsilon_{\gamma\sigma}A_{\sigma}\partial_\gamma A_\beta)
  -\hbar^2\epsilon_{\alpha\beta}\epsilon_{\gamma\sigma}\partial_{\alpha\gamma}^2(fA_\sigma)A_\beta
  -\frac{\hbar^2}{4}\epsilon_{\alpha\beta}\epsilon_{\gamma\sigma}\partial_{\alpha\gamma}^2\tilde f(\mathrm{diag}(\{\Lambda_\beta,\Lambda_\sigma\})+4A_\beta A_\sigma)
  +\mathcal O(\hbar^3).
\end{align}
Using the Berry curvature [Eq.~\eqref{eq:def_Curv}] and the quantum metric [Eq.~\eqref{eq:def_QM}], $\underline f$ is finally expressed in terms of $f$ as
\begin{align}
  \underline{f}
  &=\left(1-\frac{\hbar}{2}\epsilon_{\alpha\beta}\Omega_{\alpha\beta}-\frac{\hbar^2}{2}\epsilon_{\alpha\beta}\epsilon_{\gamma\sigma}\left(\frac{1}{2}\Omega_{\alpha\gamma}\Omega_{\beta\sigma}-\frac{3}{4}\Omega_{\alpha\beta}\Omega_{\gamma\sigma}+\partial_{\alpha\gamma}^2 g_{\beta\sigma}\right)\right)f
  -\hbar^2\epsilon_{\alpha\beta}\epsilon_{\gamma\sigma}\partial_\gamma(\partial_\alpha f g_{\beta\sigma}),
\end{align}
which corresponds to Eq.~\eqref{eq:uline_f} in the main text.

\subsection{Heat current density}\label{app:J}

In this section, we derive a gauge-invariant expression for the heat current density given in Eq.~\eqref{diag_J}.
We start from the expression of the heat current density $\mathcal J_i(\bm X)$ given in Eq.~\eqref{energy_current}.
We introduce the following phase-space heat current density $\mathscr J_\alpha(\bm X,\bm p)$ such that the relation $\mathcal J_i(\bm X)=\int_{\bm p}\mathscr J_{X_i}(\bm X,\bm p)$ is satisfied:
\begin{align}\label{eq:energy_current_2}
  \mathscr{J}_\alpha(\bm X,\bm p)
  =\frac{\epsilon_{\alpha\beta}}{2}\Re\mTr(\partial_{\beta}\bar{\mathsf H}(\mathsf F\star\mathsf H))
  -\frac{\hbar^2}{48}\epsilon_{\alpha\beta}\epsilon_{\gamma\sigma}\epsilon_{\mu\nu}\partial^2_{\gamma\mu}\Re\mTr(\partial^3_{\beta\sigma\nu}\bar{\mathsf H}\,\mathsf F\mathsf H)
  +\mathcal O(\hbar^3).
\end{align}
Here, we derive a gauge-invariant expression for $\mathscr J_\alpha(\bm X,\bm p)$.
Using the following relation, which holds for arbitrary matrix-valued functions $A$ and $B$,
\begin{align}
  AB
  =A\star B+\frac{i\hbar}{2}\epsilon_{\alpha\beta}\partial_\alpha(\partial_\beta A\star B)-\frac{\hbar^2}{8}\epsilon_{\alpha\beta}\epsilon_{\sigma\lambda}\partial_{\alpha\sigma}^2(\partial_{\beta\lambda}^2A\star B)+\mathcal O(\hbar^3),
\end{align}
we find that Eq.~\eqref{eq:energy_current_2} can be rewritten as
\begin{align}
\label{eq:current_1}
  \mathscr{J}_\alpha
  &=\frac{\epsilon_{\alpha\beta}}{2}\Re\mTr(\partial_\beta\bar{\mathsf H}\star\mathsf F\star\mathsf H)
  -\frac{\hbar}{4}\epsilon_{\alpha\beta}\epsilon_{\gamma\sigma}\partial_\gamma\Im\mTr(\partial_{\beta\sigma}^2\bar{\mathsf H}\star\mathsf F\star\mathsf H)
  -\frac{\hbar^2}{12}\epsilon_{\alpha\beta}\epsilon_{\gamma\sigma}\epsilon_{\mu\nu}\partial^2_{\gamma\mu}\Re\mTr(\partial^3_{\beta\sigma\nu}\bar{\mathsf H}\,\mathsf F\mathsf H)
  +\mathcal O(\hbar^3).
\end{align}
As discussed in \ref{subsec:h_and_J}, the off-diagonal part of the density matrix, $\tilde{\mathcal{F}}$, does not contribute to the heat current density, and hence, we can simplify $\mathsf F\star\mathsf H$ as $\sum_{n=1}^{2N}P_n\star {d}_n\star P_n$.
From this relation and Eq.~\eqref{identity2}, we calculate the first term of Eq.~\eqref{eq:current_1} as follows.
\begin{align}
\frac{\epsilon_{\alpha\beta}}{2}\Re\mTr(\partial_\beta\bar{\mathsf H}\star\mathsf F\star\mathsf H)
  &=\frac{\epsilon_{\alpha\beta}}{2}\Re\sum_{n=1}^{2N}\mTr(\partial_\beta\bar{\mathsf H}\star P_n\star {d}_n\star P_n)\notag\\
  &=\frac{\epsilon_{\alpha\beta}}{2}\Re\sum_{n=1}^{2N} \left({d}_n\star \mTr(P_n\star \partial_\beta\bar{\mathsf H}\star P_n)
  -i\hbar\epsilon_{\gamma\sigma}\partial_\gamma\{{d}_n\star \mTr(P_n\star \partial_\sigma(\partial_\beta\bar{\mathsf H}\star P_n))\}\vphantom{\frac{\hbar^2}{2}}\right.\notag\\
  &\qquad\qquad\qquad
  \left.-\frac{\hbar^2}{2}\epsilon_{\gamma\sigma}\epsilon_{\mu\nu}\partial_{\gamma\mu}^2\{{d}_n\star \mTr(P_n\star\partial_{\sigma\nu}^2(\partial_\beta\bar{\mathsf H}\star P_n))\}\right)+\mathcal O(\hbar^3)\notag\\
  &=\frac{\epsilon_{\alpha\beta}}{2}\sum_{n=1}^{2N}\left(\Re({d}_n V_{n,\beta})+\hbar\epsilon_{\gamma\sigma}\partial_\gamma\Im\left(-{d}_n \mTr(\partial_\sigma P_n \star \partial_\beta\bar{\mathsf H}\star P_n)+\frac{1}{2}{d}_n \partial_\sigma V_{n,\beta}\right)\right.\notag\\
  &\qquad\qquad\qquad
  \left.-\hbar^2\epsilon_{\gamma\sigma}\epsilon_{\mu\nu}\partial_{\gamma\mu}^2\Re\left(\frac{1}{8}{d}_n \partial_{\sigma\nu}^2V_{n,\beta}-\frac{1}{2}{d}_n\mTr(\partial_\nu P_n\partial_\sigma(\partial_\beta\bar{\mathsf H}P_n))\right)\right)+\mathcal O(\hbar^3)
  \notag\\
  &=\frac{\epsilon_{\alpha\beta}}{2}\sum_{n=1}^{2N}\left(\Re d_n V_{n,\beta}+\hbar\epsilon_{\gamma\sigma}\partial_\gamma\left(\Re d_n V_{n,\beta\sigma}+\Im{d}_n V_{n,\beta\sigma}^{\prime}+\frac12\Im{d}_n\partial_\sigma V_{n,\beta}\right)\right.
  \notag\\
  &\qquad\qquad\qquad
  +\hbar^2\epsilon_{\gamma\sigma}\epsilon_{\mu\nu}\partial_{\gamma\mu}^2\left\{\Re d_n\left(-\frac{1}{8}\partial_{\beta\sigma\nu}^3 h_n + \frac{1}{4}\partial_\sigma\mTr(\partial_\nu P_n \partial_\beta\bar{\mathsf H})\right.\right.
  \notag\\
  &\qquad\qquad\qquad\qquad\qquad\qquad
  \left.\left.\left.-\frac{1}{4}\mTr(\partial_{\nu\sigma}^2P_n \partial_\beta\bar{\mathsf H})+\frac{1}{2}\Re\mTr(\partial_\nu P_n\partial_\beta\bar{\mathsf H}\partial_\sigma P_n)\right)\right\}\right)+\mathcal O(\hbar^3),
\end{align}
where we introduce $V_{n,\beta}\equiv\mTr(P_n\star \partial_\beta\bar{\mathsf H} \star P_n)$, $V_{n,\beta\sigma}\equiv-\Im\mTr(\partial_\sigma P_n\star\partial_\beta\bar{\mathsf H}\star P_n)$, and $V_{n,\beta\sigma}^{\prime}\equiv-\Re\mTr(\partial_\sigma P_n\star\partial_\beta\bar{\mathsf H}\star P_n)$.
Furthermore, using $\mTr(P_n\star \partial_{\beta\sigma}^2\bar{\mathsf H}\star P_n)=\partial_\sigma V_{n,\beta}+2V_{n,\beta\sigma}^{\prime}$, the second term of Eq.~\eqref{eq:current_1} is calculated as
\begin{align}
  -\frac{\hbar}{4}\epsilon_{\alpha\beta}\epsilon_{\gamma\sigma}\partial_\gamma\Im\mTr(\partial_{\beta\sigma}^2\bar{\mathsf H}\star\mathsf F\star\mathsf H)
  &=-\frac{\hbar}{4}\epsilon_{\alpha\beta}\epsilon_{\gamma\sigma}\partial_\gamma\Im\sum_{n=1}^{2N}({d}_n\star \mTr(P_n\star \partial_{\beta\sigma}^2\bar{\mathsf H}\star P_n)
  -i\hbar\epsilon_{\mu\nu}\partial_\mu\{{d}_n\mTr(P_n\partial_\nu(\partial_{\beta\sigma}^2\bar{\mathsf H}P_n))\})+\mathcal O(\hbar^3)\notag\\
  &=-\frac{\hbar}{4}\epsilon_{\alpha\beta}\epsilon_{\gamma\sigma}\partial_\gamma\sum_{n=1}^{2N}\left(\Im{d}_n(\partial_\sigma V_{n,\beta}+2V_{n,\beta\sigma}^{\prime})-\frac{1}{2}\hbar\epsilon_{\mu\nu}\partial_\mu(\Re d_n\mTr(\partial_{\beta\sigma\nu}^3\bar{\mathsf H}P_n))\right)+\mathcal O(\hbar^3)\notag\\
  &=-\frac{\hbar}{4}\epsilon_{\alpha\beta}\epsilon_{\gamma\sigma}\partial_\gamma\sum_{n=1}^{2N}\Im{d}_n(\partial_\sigma V_{n,\beta}+2V_{n,\beta\sigma}^{\prime})\notag\\
  &\qquad+\frac{\hbar^2}{8}\epsilon_{\alpha\beta}\epsilon_{\gamma\sigma}\epsilon_{\mu\nu}\partial^2_{\gamma\mu}\sum_{n=1}^{2N} \Re d_n(\partial_{\beta\sigma\nu}^3 h_n -2\partial_\sigma\mTr(\partial_\beta\bar{\mathsf H}\partial_\nu P_n)+\mTr(\partial_\beta\bar{\mathsf H}\partial_{\nu\sigma}^2 P_n))+\mathcal O(\hbar^3).
\end{align}
Finally, the third term of  Eq.~\eqref{eq:current_1} is calculated as
\begin{align}
  -\frac{\hbar^2}{12}\epsilon_{\alpha\beta}\epsilon_{\gamma\sigma}\epsilon_{\mu\nu}\partial^2_{\gamma\mu}\Re\mTr(\partial^3_{\beta\sigma\nu}\bar{\mathsf H}\star\mathsf F\star\mathsf H)
  &=-\frac{\hbar^2}{12}\epsilon_{\alpha\beta}\epsilon_{\gamma\sigma}\epsilon_{\mu\nu}\partial^2_{\gamma\mu}\sum_{n=1}^{2N} \Re d_n\mTr(P_n\partial_{\beta\sigma\nu}^3\bar{\mathsf H}P_n)+\mathcal O(\hbar^3)\notag\\
  &=-\frac{\hbar^2}{12}\epsilon_{\alpha\beta}\epsilon_{\gamma\sigma}\epsilon_{\mu\nu}\partial^2_{\gamma\mu}\sum_{n=1}^{2N} \Re d_n(\partial_{\beta\sigma\nu}^3 h_n -2\partial_\sigma\mTr(\partial_\beta\bar{\mathsf H}\partial_\nu P_n)+\mTr(\partial_\beta\bar{\mathsf H}\partial_{\nu\sigma}^2 P_n))+\mathcal O(\hbar^3).
\end{align}
Thus, by combining these three terms, the heat current density is expressed as
\begin{align}\label{eq:ene1}
  \mathcal{J}_\alpha = \frac{\epsilon_{\alpha\beta}}{2}\sum_{n=1}^{2N}\left(\Re d_n V_{n,\beta} + \hbar\epsilon_{\gamma\sigma}\partial_\gamma(\Re d_n V_{n,\beta\sigma})
  +\frac{\hbar^2}{12}\epsilon_{\gamma\sigma}\epsilon_{\mu\nu}\partial_{\gamma\mu}^2(\Re d_n V_{n,\beta\sigma\nu})\right)+\mathcal O(\hbar^3),
\end{align}
where we introduce $V_{n,\beta\sigma\nu}$ as
\begin{align}
  V_{n,\beta\sigma\nu}&\equiv-\frac{1}{2}\partial_{\beta\sigma\nu}^3 h_n+\frac{1}{2}\partial_\nu\mTr(\partial_\sigma P_n\partial_\beta\bar{\mathsf H})+\frac{1}{2}\partial_\sigma\mTr(\partial_\nu P_n\partial_\beta\bar{\mathsf H})
  -2\mTr(\partial_{\sigma\nu}^2 P_n \partial_\beta\bar{\mathsf H})+6\Re\mTr(\partial_\sigma P_n \partial_\beta\bar{\mathsf H}\partial_\nu P_n).
\end{align}
Using Eqs.~\eqref{proj_h}, \eqref{identity1}, \eqref{identity2}, and \eqref{eq:trP_n}, and the identities $\mTr(\partial_\beta P_n\star P_n)=i\hbar\epsilon_{\gamma\sigma}\partial_\gamma\mathcal{T}_{n,\sigma\beta}+\mathcal O(\hbar^2)$, $\mTr(\partial_\alpha P_n\partial_\beta P_n)=2g_{n,\alpha\beta}+\mathcal O(\hbar)$, and $\mTr(\partial_\beta P_n\partial_{\sigma\nu}^2P_n)=\partial_\sigma g_{n,\beta\nu}+\partial_\nu g_{n,\sigma\beta}-\partial_\beta g_{n,\sigma\nu}+\mathcal O(\hbar)$ that hold for $P_n$~\cite{Park2025}, we calculate $V_{n,\beta}$, $V_{n,\beta\sigma}$ as follows:
\begin{align}\label{eq:V1}
  V_{n,\beta}
  &=\mTr(P_n\star(\partial_\beta P_n \star h_n + h_n \star \partial_\beta P_n + \partial_\beta h_n)\star P_n)\notag\\
  &=-i\hbar\epsilon_{\gamma\sigma}\partial_{\gamma} h_n \mTr(\partial_\sigma P_n\star \partial_\beta P_n \star P_n - P_n \star \partial_\beta P_n \star \partial_\sigma P_n)
  +\frac{\hbar^2}{2}\epsilon_{\gamma\sigma}\epsilon_{\mu\nu}\partial_{\gamma\mu}^2 h_n \Re \mTr(P_n \partial_{\beta\sigma}^2 P_n \partial_\nu P_n + \partial_\sigma P_n \partial_{\beta\nu}^2 P_n P_n)
  \notag\\
  &\qquad+\partial_\beta h_n \mTr(P_n)
  -\frac{i\hbar}{2}\epsilon_{\gamma\sigma}\partial_{\beta\gamma}^2h_n \mTr(\partial_\sigma P_n \star P_n - P_n \star \partial_\sigma P_n)+\frac{\hbar^2}{2}\epsilon_{\gamma\sigma}\epsilon_{\mu\nu}\partial_{\beta\gamma\mu}^3 h_n \mTr(\partial_\sigma P_n\partial_\nu P_n)+\mathcal O(\hbar^3)\notag\\
  &=\partial_\beta h_n \mTr(P_n)-2\hbar\epsilon_{\gamma\sigma}\partial_\gamma h_n \Im \mTr(\partial_\beta P_n \star (I_{2N}-P_n) \star \partial_\sigma P_n)\notag\\
  &\qquad+\hbar^2\epsilon_{\gamma\sigma}\epsilon_{\mu\nu}\left(\frac{1}{4}\partial_\beta \Re \mTr(P_n\partial_\sigma P_n\partial_\nu P_n)+\partial_{\beta\gamma\mu}^3 h_n g_{n,\sigma\nu}+\partial_{\beta\gamma}^2h_n \partial_{\mu}g_{n,\sigma\nu}\right)+\mathcal O(\hbar^3)\notag\\
  &=\partial_{\beta}h_n\left(1+\frac{\hbar}{2}\epsilon_{\gamma\sigma}\Omega_{n,\gamma\sigma}+\frac{\hbar^2}{4}\epsilon_{\gamma\sigma}\epsilon_{\mu\nu}\left(\Omega_{n,\gamma\mu}\Omega_{n,\sigma\nu}-\frac{1}{2}\Omega_{n,\gamma\sigma}\Omega_{n,\mu\nu}\right)\right)
  +\frac{\hbar^2}{2}\epsilon_{\gamma\sigma}\epsilon_{\mu\nu}\partial_{\beta}h_n\partial_{\gamma\mu}^2 g_{n,\sigma\nu}-\hbar \epsilon_{\gamma\sigma}\partial_{\gamma}h_n\Omega_{n,\beta\sigma}\notag\\
  &\qquad+\hbar^2\epsilon_{\gamma\sigma}\epsilon_{\mu\nu}\left(\frac{1}{2}\partial_{\gamma\mu}^2h_n\partial_{\beta}g_{n,\sigma\nu}+\partial_{\beta\gamma\mu}^3 h_n g_{n,\sigma\nu}+\partial_{\beta\gamma}^2h_n \partial_{\mu}g_{n,\sigma\nu}\right)+\mathcal O(\hbar^3),
\end{align}
and
\begin{align}\label{eq:V2}
  V_{n,\beta\sigma}
  &=\Im\mTr(\partial_\sigma P_n\star\bar{\mathsf H}\star \partial_\beta P_n)-\Im\mTr(\partial_\sigma P_n\star\partial_\beta(\bar{\mathsf H}\star P_n))\notag\\
  &=m_{n,\sigma\beta}-\Im\mTr(\partial_\sigma P_n\star \partial_\beta h_n\star P_n)+\hbar\epsilon_{\mu\nu}\Re\mTr(\partial_\sigma P_n\partial_\beta(\partial_\sigma h_n\partial_\nu P_n P_n))+\mathcal O(\hbar^2)\notag\\
  &=-m_{n,\beta\sigma}-\partial_\beta h_n \Im\mTr(\partial_\sigma P_n \star P_n)
  +\hbar\epsilon_{\mu\nu}(-\partial_{\beta\mu}^2 h_n \Re\mTr(\partial_\sigma P_n\partial_\nu P_n)+\Re\mTr(\partial_\sigma P_n\partial_\beta(\partial_\mu h_n \partial_\nu P_nP_n)))+\mathcal O(\hbar^2)\notag\\
  &=-m_{n,\beta\sigma}-\hbar\epsilon_{\mu\nu}(\partial_\beta h_n\partial_\mu g_{n,\nu\sigma}+\partial_{\beta\mu}^2 h_n g_{n,\nu\sigma})
  +\frac{\hbar}{2}\epsilon_{\mu\nu}\partial_\mu h_n\Re\mTr(\partial_{\beta\nu}^2 P_n(P_n\partial_\sigma P_n+\partial_\sigma P_n P_n))+\mathcal O(\hbar^2)\notag\\
  &=-m_{n,\beta\sigma}-\hbar\epsilon_{\mu\nu}(\partial_\beta h_n\partial_\mu g_{n,\nu\sigma}+\partial_{\beta\mu}^2 h_n g_{n,\nu\sigma})
  +\frac{\hbar}{2}\epsilon_{\mu\nu}\partial_\mu h_n(\partial_\beta g_{n,\sigma\nu}-\partial_\sigma g_{n,\nu\beta}+\partial_\nu g_{n,\beta\sigma})+\mathcal O(\hbar^2).
\end{align}
Furthermore, to calculate $V_{n,\beta\sigma\nu}$, we use the following relations:
\begin{align}
  \partial_\nu \mTr(\partial_\sigma P_n \partial_\beta \bar{\mathsf H})
  &=2\partial_\nu \Re \mTr(\partial_\sigma P_n \partial_\beta \bar{\mathsf H} P_n)
  =-2\partial_\nu c_{n,\sigma\beta}+\mathcal O(\hbar),\\
  \Re \mTr(\partial_\sigma P_n \partial_\beta \bar{\mathsf H}\partial_\nu P_n)
  &=\partial_\beta(\Re \mTr(\partial_\sigma P_n \bar{\mathsf H}\partial_\nu P_n))-\Re\mTr(\partial_{\sigma\beta}^2 P_n \bar{\mathsf H}\partial_\nu P_n + \partial_\sigma P_n \bar{\mathsf H}\partial_{\nu\beta}^2 P_n)\notag\\
  &=2\partial_\beta h_n g_{n,\sigma\nu}+\partial_\beta c_{n,\sigma\nu}-c_{n,\sigma\beta;\nu}-c_{n,\nu\beta;\sigma}+\mathcal O(\hbar),\\
  \mTr(\partial_{\sigma\nu}^2 P_n \partial_\beta \bar{\mathsf H})
  &=2\Re\mTr(\partial_\sigma P_n \partial_\beta \bar{\mathsf H}\partial_\nu P_n)+2\Re\mTr(\partial_{\sigma\nu}^2 P_n \partial_\beta \bar{\mathsf H} P_n)\notag\\
  &=2\partial_\beta c_{n,\sigma\nu}-2c_{n,\sigma\beta;\nu}-2c_{n,\nu\beta;\sigma}-2c_{n,\sigma\nu;\beta}+\mathcal O(\hbar).
\end{align}
As a result, we obtain the explicit expression of $V_{n,\beta\sigma\nu}$ as
\begin{align}\label{eq:V3}
  V_{n,\beta\sigma\nu}
  &=-\frac{1}{2}\partial_{\beta\sigma\nu}^3 h_n+12\partial_{\beta}h_n g_{n,\sigma\nu}+2\partial_{\beta}c_{n,\sigma\nu}-\partial_{\nu}c_{n,\sigma\beta}-\partial_{\sigma}c_{n,\nu\beta}
  +4c_{n,\sigma\nu;\beta}-2c_{n,\sigma\beta;\nu}-2c_{n,\nu\beta;\sigma}+\mathcal O(\hbar),
\end{align}
where $m_{n,\beta\sigma}$, $c_{n,\beta\sigma}$, and $c_{n,\sigma\nu;\beta}$ are given below.
\begin{align}
  m_{n,\beta\sigma}&\equiv \Im\mTr(\partial_\beta P_n\star(\bar{\mathsf H}-h_n I_{2N})\star \partial_\sigma P_n),\\
  c_{n,\beta\sigma}&\equiv \Re\mTr(\partial_\beta P_n\star(\bar{\mathsf H}-h_n I_{2N})\star \partial_\sigma P_n),\\
  c_{n,\sigma\nu;\beta}&\equiv \Re\mTr(\partial_{\sigma\nu}^2 P_n\star(\bar{\mathsf H}-h_n I_{2N})\star \partial_\beta P_n).
\end{align}

From Eq.~\eqref{eq:qpp:d_expression}, $\Re d_n$ can be expressed as
\begin{align}\label{eq:Re_d}
  \Re d
  =\underline{f} h-\hbar^2\epsilon_{\alpha\beta}\epsilon_{\gamma\sigma}\left(\partial_\alpha\underline{f}\partial_\gamma h g_{\beta\sigma}+\frac{1}{8}\partial_{\alpha\gamma}^2\underline{f} \partial_{\beta\sigma}^2 h\right).
\end{align}
By substituting Eqs.~\eqref{eq:V1}, \eqref{eq:V2}, \eqref{eq:V3}, and \eqref{eq:Re_d} into Eq.~\eqref{eq:ene1}, we obtain the following expression:
\begin{align}\label{eq:app_J}
  \mathscr{J}_\alpha
  &=\frac{\epsilon_{\alpha\beta}}{2}\mTr\left[f\left(h\partial_\beta h-\hbar\epsilon_{\gamma\sigma}\partial_\gamma h\left(h\Omega_{\beta\sigma}\left(1-\frac{\hbar}{2}\epsilon_{\mu\nu}\Omega_{\mu\nu}\right)+\hbar\epsilon_{\mu\nu}\partial_{\beta\mu}^2 h g_{\sigma\nu}\right)
  +\frac{\hbar^2}{2}\epsilon_{\gamma\sigma}\epsilon_{\mu\nu}\partial_{\gamma\mu}^2 h\left(h\partial_{\beta}g_{\sigma\nu}-\frac{1}{4}\partial_{\beta\sigma\nu}^3 h\right)\right)\right]\notag\\
  &\qquad+\frac{\hbar}{2}\epsilon_{\alpha\beta}\epsilon_{\gamma\sigma}\partial_\gamma\mTr\left[f(b_{\beta\sigma}^s + b_{\beta\sigma}^a)\right]
  +\frac{\hbar^2}{24}\epsilon_{\alpha\beta}\epsilon_{\gamma\sigma}\epsilon_{\mu\nu}\partial_{\gamma\mu}^2\mTr\left[f(b_{\beta\sigma\nu}^s + b_{\beta\sigma;\nu}^a)\right],
\end{align}
where $b_{\beta\sigma}^s$, $b_{\beta\sigma}^a$, $b_{\beta\sigma\nu}^s$, and $b_{\beta\sigma;\nu}^a$ are defined as follows:
\begin{align}
  b_{\beta\sigma}^s&\equiv\frac{\hbar}{2}\epsilon_{\mu\nu}(h\partial_\mu h \partial_\nu g_{\beta\sigma}+(h\partial_{\beta\mu}^2h+\partial_\beta h\partial_\mu h)g_{\sigma\nu}+(h\partial_{\sigma\mu}^2h+\partial_\sigma h\partial_\mu h)g_{\beta\nu}),\\
  b_{\beta\sigma}^a&\equiv-m_{\beta\sigma}h\left(1-\frac{\hbar}{2}\epsilon_{\mu\nu}\Omega_{\mu\nu}\right)
  -\frac{\hbar}{2}\epsilon_{\mu\nu}\left(\partial_\sigma\left(h\partial_\mu h g_{\beta\nu}+\frac{1}{4}\partial_\mu h\partial_{\beta\nu}^2 h\right)-\partial_\beta\left(h\partial_\mu h g_{\sigma\nu}+\frac{1}{4}\partial_\mu h \partial_{\sigma\nu}^2 h\right)\right),\\
  b_{\beta\sigma\nu}^s&\equiv-\frac{1}{4}\partial_{\beta\sigma\nu}^3 h^2,\\
  b_{\beta\sigma;\nu}^a&\equiv h\partial_\beta c_{\nu\sigma}-h\partial_\sigma c_{\nu\beta}+2hc_{\nu\sigma;\beta}-2hc_{\nu\beta;\sigma}-\frac{1}{2}\partial_\beta h\partial_{\nu\sigma}^2 h+\frac{1}{2}\partial_\sigma h\partial_{\nu\beta}^2 h.
\end{align}
Finally, by integrating Eq.~\eqref{eq:app_J} over $\bm p$, we obtain the gauge-invariant expression for the heat current density given in the main text [Eq.~\eqref{diag_J}].

\section{Density matrix under global equilibrium}
\label{subsec:f_eq}

In Sec.~\ref{sec:2_THE}, when calculating the thermal response coefficients, we use the expression for the density matrix under local thermal equilibrium.
In this context, we need the explicit form of the diagonal part of the gauge-invariant density matrix $f_{\mathrm{eq}}$ for a system in global thermal equilibrium.
This appendix provides a derivation of the gauge-invariant density matrix for a system in global thermal equilibrium up to second order in $\hbar$, using the Matsubara formalism for bosonic systems.

First, we introduce the bosonic Matsubara Green's function as
\begin{align}
  G_{ab}(\bm r_1,\tau_1;\bm r_2,\tau_2)\equiv -\means{T_\tau \Psi_a(\bm r_1,\tau_1)\Psi_b^{+}(\bm r_2,\tau_2)}_{\mathrm{can}},
\end{align}
where $\means{\bullet}_{\mathrm{can}}$ denotes the average over the canonical ensemble, and $T_\tau$ is the imaginary-time ordering operator.
The field operators $\Psi(\bm r,\tau)$ and $\Psi^{+}(\bm r,\tau)$ are $2N$-component vectors composed of field operators, which are the imaginary-time counterparts of the operators $\Psi(\bm r)$ and $\Psi^{+}(\bm r)$ introduced in Sec.~\ref{sec:heat_current_qke}, and are defined as $\Psi(\bm r,\tau)\equiv[\beta_1(\bm r,\tau),\dots,\beta_N(\bm r,\tau),\beta^{+}_1(\bm r,\tau),\dots,\beta^{+}_N(\bm r,\tau)]$ and $\Psi^{+}(\bm r,\tau)\equiv[\beta^{+}_1(\bm r,\tau),\dots,\beta^{+}_N(\bm r,\tau),\beta_1(\bm r,\tau),\dots,\beta_N(\bm r,\tau)]$.
Here, $\beta_i(\bm r,\tau)$ and $\beta_i^{+}(\bm r,\tau)$ are defined in terms of the bosonic annihilation and creation operators $\beta_i(\bm r)$ and $\beta_i^{\dagger}(\bm r)$ as $\beta_i(\bm r,\tau)\equiv e^{\tau\mathscr H}\beta_i(\bm r)e^{-\tau\mathscr H}$ and $\beta^{+}_i(\bm r,\tau)\equiv e^{\tau\mathscr H}\beta^{\dagger}_i(\bm r)e^{-\tau\mathscr H}$.
Under global thermal equilibrium, the density matrix $F_{\mathrm{eq},ab}(\bm r_1,\bm r_2)\equiv\means{\Psi_b^{\dagger}(\bm r_1)\Psi_a(\bm r_2)}_{\mathrm{can}}$ can be expressed in terms of the Matsubara Green's function defined above as
\begin{align}\label{eq:F_eq}
  F_{\mathrm{eq},ab}(\bm r_1,\bm r_2)=G_{ab}(\bm r_1,\tau;\bm r_2,\tau+0^+)=k_B T\sum_{n}e^{i\omega_n 0^+}G_{ab}(\bm r_1,\bm r_2;i\omega_n).
\end{align}
Here, $G_{ab}(\bm r_1,\bm r_2;i\omega_n)$ denotes the Fourier transform of the Matsubara Green's function with respect to imaginary time, where $\omega_n=2n\pi k_B T$ is the Matsubara frequency and $n$ is an integer.
The Matsubara Green's function $G_{ab}(\bm r_1,\bm r_2;i\omega_n)$ satisfies the following equations:
\begin{align}
  \int_{\bm r}(i\omega_n\delta(\bm r_1-\bm r)I_{2N}-H(\bm r_1,\bm r))G(\bm r,\bm r_2;i\omega_n) &= \delta(\bm r_1,\bm r_2)I_{2N},\\
  \int_{\bm r}G(\bm r_1,\bm r;i\omega_n)(i\omega_n\delta(\bm r-\bm r_2)I_{2N}-H(\bm r,\bm r_2)) &= \delta(\bm r_1,\bm r_2)I_{2N}.
\end{align}
By performing the Wigner transformation on the above equations, we obtain
\begin{align}
  (i\omega_n I_{2N}-\mathsf H(\bm X,\bm p))\star\mathsf G(\bm X,\bm p;i\omega_n)&=I_{2N},\\
  \mathsf G(\bm X,\bm p;i\omega_n)\star(i\omega_n I_{2N}-\mathsf H(\bm X,\bm p))&=I_{2N},
\end{align}
where $\mathsf G(\bm X,\bm p;i\omega_n)$ is the Wigner transform of $G(\bm r_1,\bm r_2;i\omega_n)$.
These equations are transformed by the star-unitary matrix $\mathsf T$ introduced in \ref{diag_star} as
\begin{align}
  \label{eq:def_G_1}
  (i\omega_n I_{2N}-\tilde{h})\star\tilde{\mathsf G}&=I_{2N},\\
  \label{eq:def_G_2}
  \tilde{\mathsf G}\star(i\omega_n I_{2N}-\tilde{h})&=I_{2N},
\end{align}
where $\tilde{\mathsf G}\equiv {\mathsf T}^{-1}\star\mathsf G\star \sigma_3\star {\mathsf T}$.
Since $\tilde{h}$ is diagonal, $\tilde{\mathsf G}$ is also a diagonal matrix.
Here, we expand the star product in powers of $\hbar$ in Eqs.~\eqref{eq:def_G_1} and \eqref{eq:def_G_2}, and calculate the coefficients in the $\hbar$ expansion of $\tilde{\mathsf G}$, where $\tilde{\mathsf G}=\tilde{\mathsf G}^{(0)}+\tilde{\mathsf G}^{(1)}+\cdots$ and $\tilde{\mathsf G}^{(n)}$ is the coefficient of order $\hbar^n$.
At zeroth order in $\hbar$, we have
\begin{align}
  \tilde{\mathsf G}^{(0)}=\frac{1}{i\omega_n I_{2N}-h^{(0)}}.
\end{align}
At first order in $\hbar$, we obtain the following equations:
\begin{align}
  (i\omega_n I_{2N}-\tilde{h}^{(0)})\tilde{\mathsf G}^{(1)}-\tilde{h}^{(1)}\tilde{\mathsf G}^{(0)}-\frac{i\hbar}{2}\epsilon_{\alpha\beta}\partial_{\alpha}\tilde{h}^{(0)}\partial_\beta\tilde{\mathsf G}^{(0)}&=0,\\
  \tilde{\mathsf G}^{(1)}(i\omega_n I_{2N}-\tilde{h}^{(0)})-\tilde{\mathsf G}^{(0)}\tilde{h}^{(1)}-\frac{i\hbar}{2}\epsilon_{\alpha\beta}\partial_{\alpha}\tilde{\mathsf G}^{(0)}\partial_\beta\tilde{h}^{(0)}&=0.
\end{align}
Here, the third terms on the left-hand sides vanish because $\partial_\beta\tilde{\mathsf G}^{(0)}=\partial_\beta\tilde{h}^{(0)}(i\omega_n I_{2N}-h^{(0)})^{-2}$ and $\epsilon_{\alpha\beta}$ is antisymmetric.
Thus, the first-order correction to $\tilde{\mathsf G}$ is given by
\begin{align}
  \tilde{\mathsf G}^{(1)}=\frac{\tilde{h}^{(1)}}{(i\omega_n I_{2N}-h^{(0)})^2}.
\end{align}
Next, at second order in $\hbar$, we obtain the following equations:
\begin{align}
  &(i\omega_n I_{2N}-\tilde{h}^{(0)})\tilde{\mathsf G}^{(2)}-\tilde{h}^{(1)}\tilde{\mathsf G}^{(1)}-\tilde{h}^{(2)}\tilde{\mathsf G}^{(0)}-\frac{i\hbar}{2}\epsilon_{\alpha\beta}(\partial_{\alpha}\tilde{h}^{(0)}\partial_\beta\tilde{\mathsf G}^{(1)}+\partial_{\alpha}\tilde{h}^{(1)}\partial_\beta\tilde{\mathsf G}^{(0)})
  +\frac{\hbar^2}{8}\epsilon_{\alpha\beta}\epsilon_{\gamma\sigma}\partial_{\alpha\gamma}^2\tilde{h}^{(0)}\partial_{\beta\sigma}^2\tilde{\mathsf G}^{(0)}
  =0,\\
  &\tilde{\mathsf G}^{(2)}(i\omega_n I_{2N}-\tilde{h}^{(0)})-\tilde{\mathsf G}^{(1)}\tilde{h}^{(1)}-\tilde{\mathsf G}^{(0)}\tilde{h}^{(2)}-\frac{i\hbar}{2}\epsilon_{\alpha\beta}(\partial_{\alpha}\tilde{\mathsf G}^{(1)}\partial_\beta\tilde{h}^{(0)}+\partial_{\alpha}\tilde{\mathsf G}^{(0)}\partial_\beta\tilde{h}^{(1)})
  +\frac{\hbar^2}{8}\epsilon_{\alpha\beta}\epsilon_{\gamma\sigma}\partial_{\alpha\gamma}^2\tilde{\mathsf G}^{(0)}\partial_{\beta\sigma}^2\tilde{h}^{(0)}
  =0.
\end{align}
Similar to the first-order case, the fourth terms on the left-hand sides of the above equations vanish because the relation $\epsilon_{\alpha\beta}\partial_\alpha\tilde{h}^{(0)}\partial_\beta\tilde{\mathsf G}^{(1)}=\epsilon_{\alpha\beta}\partial_\alpha\tilde{h}^{(0)}\partial_\beta\tilde{h}^{(1)}(i\omega_n I_{2N}-\tilde{h}^{(0)})^{-2}$ holds.
Thus, the second-order correction to $\tilde{\mathsf G}$ is given by
\begin{align}
  \tilde{\mathsf G}^{(2)}=\frac{\tilde{h}^{(2)}}{(i\omega_n I_{2N}-h^{(0)})^2}+\frac{(\tilde{h}^{(1)})^2}{(i\omega_n I_{2N}-h^{(0)})^3}
  -\frac{1}{8}\epsilon_{\alpha\beta}\epsilon_{\gamma\sigma}\frac{\partial_{\alpha\gamma}^2h^{(0)}\partial_{\beta\sigma}^2h^{(0)}}{(i\omega_n I_{2N}-h^{(0)})^3}-\frac{1}{4}\epsilon_{\alpha\beta}\epsilon_{\gamma\sigma}\frac{\partial_{\alpha\gamma}^2h^{(0)}\partial_{\beta}h^{(0)}\partial_{\sigma}h^{(0)}}{(i\omega_n I_{2N}-h^{(0)})^4}.
\end{align}

Using the expansion of the Green's function obtained above, we can now calculate the gauge-invariant density matrix under global thermal equilibrium.
Similar to Eq.~\eqref{eq:tilde_F}, the diagonal part of the matrix obtained by transforming the density matrix $F_{\mathrm{eq}}$ using the star-paraunitary matrix ${\mathsf T}$ is written as $\tilde{f}_\mathrm{eq}={\mathsf T}^{-1}\star \mathsf F_{\mathrm{eq}}\star\sigma_3\star {\mathsf T}$.
From Eq.~\eqref{eq:F_eq}, we find that $\tilde{f}_\mathrm{eq}$ can be expressed in terms of the Green's function as $\tilde{f}_\mathrm{eq}=k_B T\sum_{n}e^{i\omega_n0^+}\tilde{\mathsf G}(i\omega_n)$.
Since we consider bosonic systems, the Matsubara summation can be carried out as
\begin{align}
  k_B T\sum_n\frac{e^{i\omega_n0^+}}{i\omega_n I_{2N}-h^{(0)}}=n_B(h^{(0)},T)+\frac12,
\end{align}
where the summation over $n$ runs over all integers, and $n_B(\varepsilon,T)=(e^{\varepsilon/(k_B T)}-1)^{-1}$ is the Bose distribution function.
By differentiating both sides of the above equation with respect to $h^{(0)}$, we obtain
\begin{align}
  k_B T\sum_n\frac{e^{i\omega_n0^+}}{(i\omega_n I_{2N}-h^{(0)})^k}=\frac{1}{(k-1)!}n_B^{(k-1)}(h^{(0)},T),
\end{align}
for $k\ge 2$, where $n_B^{(k-1)}(\varepsilon,T)$ denotes the $(k-1)$-th derivative of $n_B(\varepsilon,T)$ with respect to $\varepsilon$.
Using the above results, we obtain the expression for $\tilde{f}_{\mathrm{eq}}$ up to second order in $\hbar$ as
\begin{align}
  \tilde{f}_{\mathrm{eq}}=n_B(\tilde{h},T)+\frac{1}{2}-\frac{\hbar^2}{16}\epsilon_{\alpha\beta}\epsilon_{\gamma\sigma}n_B^{\prime\prime}(\tilde{h},T)\partial_{\alpha\gamma}^2\tilde{h}\partial_{\beta\sigma}^2\tilde{h}-\frac{\hbar^2}{24}\epsilon_{\alpha\beta}\epsilon_{\gamma\sigma}n_B^{\prime\prime\prime}(\tilde{h},T)\partial_{\alpha\gamma}^2\tilde{h}\partial_{\beta}\tilde{h}\partial_{\sigma}\tilde{h}+\mathcal O(\hbar^3).
\end{align}
Finally, from Eqs.~\eqref{diag_h} and \eqref{diag_f}, we find that the diagonal part of the gauge-invariant density matrix under global thermal equilibrium is given by
\begin{align}\label{eq:f_eq_app}
  f_{\mathrm{eq}}&=\left(1+\frac{\hbar}{2}\epsilon_{\alpha\beta}\left(\Omega_{\alpha\beta}+\hbar\epsilon_{\gamma\sigma}\left(\frac{1}{2}\Omega_{\alpha\gamma}\Omega_{\beta\sigma}-\frac{1}{4}\Omega_{\alpha\beta}\Omega_{\gamma\sigma}+\partial_{\alpha\gamma}^2g_{\beta\sigma}\right)\right)\right)\left(n_B(h,T)+\frac12\right)+\hbar^2\epsilon_{\alpha\beta}\epsilon_{\gamma\sigma}n_B^{\prime}(h,T)\partial_\gamma(\partial_\alpha hg_{\beta\sigma})\notag\\
  &\quad
  +\frac{\hbar^2}{2}\epsilon_{\alpha\beta}\epsilon_{\gamma\sigma}n_B^{\prime\prime}(h,T)\left(\partial_\alpha h\partial_\gamma hg_{\beta\sigma}-\frac{1}{8}\partial_{\alpha\gamma}^2 h\partial_{\beta\sigma}^2 h\right)-\frac{\hbar^2}{24}\epsilon_{\alpha\beta}\epsilon_{\gamma\sigma}n_B^{\prime\prime\prime}(h,T)\partial_{\alpha\gamma}^2 h\partial_\beta h\partial_\sigma h + \mathcal O(\hbar^3).
\end{align}
We also obtain the equilibrium heat current density $\mathcal J_{\mathrm{eq},i}(\bm X)$ using Eqs.~\eqref{diag_J} and \eqref{eq:f_eq_app}.
It can be written as
\begin{align}\label{eq:Jeq-M}
    \mathcal J_{\mathrm{eq},i}(\bm X)
    =\sum_{jk}\varepsilon_{ijk}\partial_{X_j}M_k(\bm X),
\end{align}
where $\varepsilon_{ijk}$ is the Levi-Civita symbol. The quantity $M_k$ is given by
\begin{align}
    M_{k}&=\frac{\hbar}{2}\sum_{ij}\varepsilon_{ijk}\int_{\bm p}\mTr\left[\left(n_B(h,T)+\frac{1}{2}\right)\left(h\partial_{p_i}h A_{p_j}+\hbar\epsilon_{\alpha\beta}h\partial_{\beta}h\left(\frac{1}{2}A_{p_j}\partial_{\alpha}A_{p_i}+A_{\alpha}\partial_{p_i}A_{p_j}\right)\right)\vphantom{\sum_{l}}\right.\notag\\
    &\qquad\qquad\qquad\qquad+\frac{\hbar}{4}\epsilon_{\alpha\beta}\left(n_B(h,T)+\frac{1}{2}\right)\partial_{\alpha p_j}^2h^2g_{p_i\beta}+\frac{\hbar}{12}\epsilon_{\alpha\beta}\left(2\left(n_B(h,T)+\frac{1}{2}\right)-h n_B^{\prime}(h,T)\right)\partial_{\alpha p_i}^2h\partial_{\beta p_j}^2h\notag\\
    &\qquad\qquad\qquad\qquad\left.+\frac{1}{2}\left(n_B(h,T)+\frac{1}{2}\right)b_{ij}^a+\frac{\hbar}{2}\sum_{l}\partial_{X_l}\left(\left(g_B(h,T)+\frac{h}{2}\right)\partial_{p_j}h g_{p_ip_l}+\frac{1}{6}\left(n_B(h,T)+\frac{1}{2}\right) b_{ij;l}^a\right)\right]+\mathcal O(\hbar^3),
\end{align}
where $g_B(\varepsilon,T)=k_B T\ln|1-e^{-\varepsilon/(k_BT)}|$.
Note that the relation $\partial_\varepsilon g_B(\varepsilon,T)=n_B(\varepsilon,T)$ holds between $g_B(\varepsilon,T)$ and $n_B(\varepsilon,T)$.
From Eq.~\eqref{eq:Jeq-M}, we find that the equilibrium heat current density $\bm{\mathcal J}_{\mathrm{eq}}(\bm X)$ can be expressed as the curl of a vector field.

\section{Derivation of second-order thermal conductivity tensor}\label{app:kappa}

In this appendix, we derive the second-order thermal conductivity tensor from the expression for the heat current density given in Sec.~\ref{subsec:NLTHcoefficient}.
We begin with the expression for the heat current density given in Eqs.~\eqref{J2QG1} and \eqref{J2disp}, and show that the spatial gradient in the expression of the heat current density is naturally replaced by the temperature gradient through integration by parts in a system confined within a finite region.

The second-order term of the heat current density with respect to the spatial gradient or $\hbar$ is represented as $J_i^{(2)}=J_i^{(2),\mathrm{QG}}+J_i^{(2),\mathrm{disp}}$ where $J_i^{(2),\mathrm{QG}}$ and $J_i^{(2),\mathrm{disp}}$ are given by Eqs.~\eqref{J2QG1} and \eqref{J2disp}, respectively.
First, we focus on $J_i^{(2),\mathrm{QG}}$.
The first two terms in Eq.~\eqref{J2QG1} are calculated as
\begin{align}
  \frac{\hbar^2}{2V}&\sum_{jk}\int_{\bm X,\bm p}\mTr(n_B \mathcal E (\partial_{p_i}\mathcal E G^t_{jk}-\partial_{p_j}\mathcal E G^t_{ik})\chi^2\partial_{X_j}\psi_k-n_B\mathcal{E}^2(\partial_{p_i}G^t_{jk}-\partial_{p_j}G^t_{ik})\chi^2\psi_j\psi_k)\notag\\
  &=\frac{\hbar^2}{2V}\sum_{jk}\int_{\bm X,\bm p}\mTr(n_B \mathcal E (\partial_{p_i}\mathcal E G^t_{jk}-\partial_{p_j}\mathcal E G^t_{ik})(\chi\partial_{X_j X_k}^2\chi-\partial_{X_j}\chi\partial_{X_k}\chi)
  -n_B\mathcal{E}^2(\partial_{p_i}G^t_{jk}-\partial_{p_j}G^t_{ik})\partial_{X_j}\chi\partial_{X_k}\chi)\notag\\
  &=\frac{\hbar^2}{2V}\sum_{jk}\int_{\bm X,\bm p}\mTr((\mathcal E G_{jk}^t\partial_{p_i}\tilde{g}_B-\mathcal E G_{ik}^t\partial_{p_j}\tilde{g}_B)\partial_{X_j X_k}^2\chi
  -(\mathcal E \partial_{p_i}(\mathcal E G_{jk}^t)-\mathcal E \partial_{p_j}(\mathcal E G_{ik}^t))n_B\partial_{X_j}\chi\partial_{X_k}\chi)\notag\\
  &=-\frac{\hbar^2}{2V}\sum_{jk}\int_{\bm p}\mTr\left((\mathcal{E}^{-1}\partial_{p_i}(\mathcal E G_{jk}^t)-\mathcal{E}^{-1}\partial_{p_j}(\mathcal E G_{ik}^t))\int_{\bm X} (\partial_{X_j X_k}^2(\chi\mathcal E)\tilde{g}_B+\partial_{X_j}(\chi\mathcal E)\partial_{X_k}(\chi\mathcal E)n_B)\right),
\end{align}
where we introduce $\tilde{g}_B=g_B(h^{(0)},T)|_{T(\bm X)}+\frac{1}{2}h^{(0)}$.
Here and in what follows, the Einstein summation convention is assumed also for indices such as $j$ and $k$.
The third and fourth terms in Eq.~\eqref{J2QG1} are calculated as
\begin{align}
  \frac{\hbar^2}{2V}\sum_{jk}\int_{\bm X,\bm p}\mTr(n_B^{\prime}\mathcal E^2\partial_{p_i}\mathcal E \mathcal{G}_{jk}\chi^2\partial_{X_j X_k}^2\chi)
  =\frac{\hbar^2}{2V}\sum_{jk}\int_{\bm X,\bm p}\mTr\left(2\partial_{p_i}\mathcal E\, \mathcal{G}_{jk}\tilde{g}_B\partial_{X_j X_k}^2\chi+2\mathcal E \partial_{p_i}\mathcal{G}_{jk}\tilde{g}_B\partial_{X_j X_k}^2\chi-\mathcal E^2\partial_{p_i}\mathcal{G}_{jk}n_B\chi\partial_{X_j X_k}^2\chi\right)
  \end{align}
  and
  \begin{align}
  &\frac{\hbar^2}{2V}\sum_{jk}\int_{\bm X,\bm p}\mTr\left(\frac{1}{2}n_B^{\prime\prime}\mathcal{E}^3\partial_{p_i}\mathcal E \mathcal{G}_{jk}\chi^2\partial_{X_j}\chi\partial_{X_k}\chi\right)\notag\\
  &=\frac{\hbar^2}{2V}\sum_{jk}\int_{\bm X,\bm p}\mTr\left(\left(3\mathcal E \partial_{p_i}\mathcal E \mathcal{G}_{jk}+\frac{3}{2}\mathcal{E}^2\partial_{p_i}\mathcal{G}_{jk}\right)n_B\partial_{X_j}\chi\partial_{X_k}\chi
  -\frac{1}{2}\mathcal{E}^3\partial_{p_i}\mathcal{G}_{jk}n_B^{\prime}\chi\partial_{X_j}\chi\partial_{X_k}\chi\right),
\end{align}
respectively.
The last four terms in Eq.~\eqref{J2QG1} can be combined as
\begin{align}
  &\frac{\hbar^2}{2V}\sum_{jk}\int_{\bm X,\bm p}\mTr\left(n_B^{\prime}\mathcal E^2\partial_{p_i}\mathcal E \mathcal{G}_{jk}\chi^2\partial_{X_jX_k}^2\chi+\frac{1}{2}n_B^{\prime\prime}\mathcal E^3\partial_{p_i}\mathcal E \mathcal{G}_{jk}\chi^4\psi_j\psi_k
  -n_B \mathcal E\partial_{p_i}\mathcal E \mathcal{G}_{jk}\chi^2\psi_j\psi_k+\frac{1}{2}n_B \mathcal{E}^2 \partial_{p_i}\mathcal{G}_{jk}\chi\partial_{X_j X_k}^2\chi\right)\notag\\
  &=\frac{\hbar^2}{2V}\sum_{jk}\int_{\bm p}\mTr\left(2\mathcal{E}^{-1}\partial_{p_i}(\mathcal E \mathcal{G}_{jk})\int_{\bm X}(\partial_{X_j X_k}^2(\chi\mathcal E)\tilde{g}_B+\partial_{X_j}(\chi\mathcal E)\partial_{X_k}(\chi\mathcal E)n_B)\right.\notag\\
  &\qquad\qquad\qquad\qquad\qquad\qquad
  \left.-\frac{1}{2}\partial_{p_i}\mathcal{G}_{jk}\int_{\bm X}(\partial_{X_j X_k}^2(\chi\mathcal E)\chi\mathcal E n_B+\partial_{X_j}(\chi\mathcal E)\partial_{X_k}(\chi\mathcal E)(n_B+\chi\mathcal E n_B^{\prime}))\right).
\end{align}
From the above calculations, the expression of $J_i^{(2),\mathrm{QG}}$ is given by
\begin{align}
  J_i^{(2),\mathrm{QG}}
  &=-\frac{\hbar^2}{2V}\sum_{jk}\int_{\bm p}\mTr\left[\left(\mathcal{E}^{-1}\partial_{p_i}(\mathcal E G_{jk}^t)-\mathcal{E}^{-1}\partial_{p_j}(\mathcal E G_{ik}^t)-2\mathcal{E}^{-1}\partial_{p_i}(\mathcal E \mathcal{G}_{jk})\right)\vphantom{\int}
  \int_{\bm X}(\partial_{X_j X_k}^2(\chi\mathcal E)\tilde{g}_B+\partial_{X_j}(\chi\mathcal E)\partial_{X_k}(\chi\mathcal E)n_B)\right.\notag\\
  &\qquad\qquad\left.+\frac{1}{2}\partial_{p_i}\mathcal{G}_{jk}\int_{\bm X}\left(\partial_{X_j X_k}^2(\chi\mathcal E)\chi\mathcal E n_B+\partial_{X_j}(\chi\mathcal E)\partial_{X_k}(\chi\mathcal E)(n_B+\chi\mathcal E n_B^{\prime})\right)\right].
\end{align}
In the above expression, $G^t_{p_ip_j}$ and $\mathcal{G}_{p_ip_j}$ are diagonal matrices with $2N$ components, similar to $\mathcal E$.
To reduce the sum over $1\leq n\leq 2N$ to the sum over $1\leq n\leq N$, we consider particle-hole symmetry intrinsic to bosonic BdG Hamiltonian matrix $H(\bm r_1,\bm r_2)$ introduced in Sec.~\ref{sec:heat_current_qke}.
This symmetry implies that the Hamiltonian matrix satisfies the relation $\sigma_1 H(\bm r_1,\bm r_2)\sigma_1 = H^*(\bm r_1,\bm r_2)$, which leads to the following relations in momentum space:
  $\sigma_1\mathsf H_0(\bm p)\sigma_1=\mathsf H_0(-\bm p)^*$,
  $\sigma_1{\mathsf T}_0(\bm p)\sigma_1={\mathsf T}_0(-\bm p)^*$, and
  $\sigma_1\mathcal E(\bm p)\sigma_1=-\mathcal E(-\bm p)$.
From these relations, we can derive the relations for the TBCP and the quantum metric for $1\leq n\leq N$ as
  $G^t_{n,p_ip_j}(\bm p)=G^t_{n+N,p_ip_j}(-\bm p)$ and 
  $\mathcal G_{n,p_ip_j}(\bm p)=\mathcal G_{n+N,p_ip_j}(-\bm p)$, respectively.
Using these relations, we obtain the following form of $J_i^{(2),\mathrm{QG}}$:
\begin{align}
  \label{eq:app_J2QG}
  J_i^{(2),\mathrm{QG}}
  &=-\hbar^2\sum_{n=1}^{N}\sum_{jk}\int_{\bm p}\left[\left(\varepsilon_n^{-1}\partial_{p_i}(\varepsilon_n G_{n,jk}^t)-\varepsilon_n^{-1}\partial_{p_j}(\varepsilon_n G_{n,ik}^t)-2\varepsilon_n^{-1}\partial_{p_i}(\varepsilon_n \mathcal{G}_{n,jk})\right) I_{n,jk}^{\mathrm{QG1}}
  +\frac{1}{2}\partial_{p_i}\mathcal{G}_{n,jk} I_{n,jk}^{\mathrm{QG2}}\right].
\end{align}
In a similar manner, we can derive the following form of $J_i^{(2),\mathrm{disp}}$ from Eq.~\eqref{J2disp}:
\begin{align}
  \label{eq:app_J2disp}
  J_i^{(2),\mathrm{disp}}
  &=\frac{\hbar^2}{24}\sum_{n=1}^{N}\sum_{jk}\int_{\bm p}\varepsilon_n^{-2}(-\varepsilon_n\partial_{p_i p_j p_k}^3\varepsilon_n - 2\partial_{p_i}\varepsilon_n\partial_{p_j p_k}^2\varepsilon_n + 3\partial_{p_i p_j}^2\varepsilon_n\partial_{p_k}\varepsilon_n)I_{n,jk}^{\mathrm{disp}},
\end{align}
where we introduce $n_B(\varepsilon,T(\bm X))=n_B(\varepsilon,T)|_{T(\bm X)}$ and $g_B(\varepsilon,T(\bm X))=g_B(\varepsilon,T)|_{T(\bm X)}$, and $I_{n,jk}^{\mathrm{QG1}}$, $I_{n,jk}^{\mathrm{QG2}}$, and $I_{n,jk}^{\mathrm{disp}}$ are factors including spatial integrals, which are defined as follows:
\begin{align}
  \label{eq:app_IQG1}
  I_{n,jk}^{\mathrm{QG1}}
  &=\frac{1}{V}\int_{\bm X}\left\{\partial_{X_j X_k}^2(\chi\varepsilon_n)\left(g_B(\chi\varepsilon_n,T(\bm X))+\frac{1}{2}\chi\varepsilon_n\right)
  +\partial_{X_j}(\chi\varepsilon_n)\partial_{X_k}(\chi\varepsilon_n)\left(n_B(\chi\varepsilon_n,T(\bm X))+\frac{1}{2}\right)\right\},\\
  \label{eq:app_IQG2}
  I_{n,jk}^{\mathrm{QG2}}
  &=\frac{1}{V} \int_{\bm X}\left\{\partial_{X_j X_k}^2(\chi\varepsilon_n)\chi\varepsilon_n \left(n_B(\chi\varepsilon_n,T(\bm X))+\frac{1}{2}\right)
  +\partial_{X_j}(\chi\varepsilon_n)\partial_{X_k}(\chi\varepsilon_n)
  \left(n_B(\chi\varepsilon_n,T(\bm X))+\frac{1}{2}
  +\chi\varepsilon_n n_B^{\prime}(\chi\varepsilon_n,T(\bm X))\right)\right\},\\
  \label{eq:app_Idisp}
  I_{n,jk}^{\mathrm{disp}}
  &=\frac{1}{V}\int_{\bm X}
  \left\{(\chi\varepsilon_n)^2\partial_{X_k}(\chi \varepsilon_n)\partial_T n_B^{\prime}(\chi\varepsilon_n,T(\bm X))\partial_{X_j}T
  -2\chi\varepsilon_n\partial_{X_k}(\chi \varepsilon_n)\partial_T n_B(\chi\varepsilon_n,T(\bm X))\partial_{X_j}T\right\}.
\end{align}

Since we are interested in the second-order nonlinear thermal conductivity, we focus on the terms proportional to $\partial_{X_j}T(\bm X)\partial_{X_k}T(\bm X)$ in Eqs.~\eqref{eq:app_J2QG} and \eqref{eq:app_J2disp}.
Here, we consider a two-dimensional system on the $xy$ plane with volume $V=L^2$ and assume that the temperature gradient is uniform and that the temperature depends on position only through the temperature gradient, as
\begin{align}\label{eq:app_T2}
  T(\bm X) = T+x\partial_xT+y\partial_yT,
\end{align}
where $\bm X = (x,y)$.
Note that $T$, $\partial_x T$, and $\partial_y T$ are constants with respect to $\bm X$.

First, we focus on $I_{n,jk}^{\mathrm{QG1}}$ given in Eq.~\eqref{eq:app_IQG1}.
This term can be rewritten as
\begin{align}\label{eq:app_IQG1_2}
  I_{n,jk}^{\mathrm{QG1}}
  =-\frac{1}{L^2}\int_{\bm X}\partial_{X_j}(\chi\varepsilon_n)\partial_{X_k}T(\bm X)\partial_Tg_B(\chi\varepsilon_n,T(\bm X)),
\end{align}
where $\partial_T g_B(\varepsilon, T(\bm X))=[g_B(\varepsilon, T(\bm X))-\varepsilon n_B(\varepsilon, T(\bm X))]/T(\bm X)$.

As introduced in Sec.~\ref{subsec:thermal_current}, $\chi(\bm X)$ was defined such that $\chi=1$ in the region $D=\left\{(x,y)\in\mathbb{R}^2\,\middle|\,-\frac{L}{2}\leq x\leq \frac{L}{2},-\frac{L}{2}\leq y\leq \frac{L}{2}\right\}$ and it decays to zero over a length scale of $\mathcal O(L^0)$, as presented in Fig.~\ref{fig:chi_xy}.
Note that we have also introduced the support of $\chi$ as $D'=\left\{(x,y)\in\mathbb{R}^2\,\middle|\,-\frac{L'}{2}\leq x\leq \frac{L'}{2},-\frac{L'}{2}\leq y\leq \frac{L'}{2}\right\}$, where $L'>L$ and $L'-L=\mathcal O(L^0)$.
From these characteristics of $\chi(\bm X)$, the integrand of Eq.~\eqref{eq:app_IQG1_2} is nonzero only within a finite region with a width of $\mathcal O(L)$ and it decays to zero as $|x|,|y|\to\infty$.
Here, we introduce a cutoff function $C(x,y)$ that takes the value $1$ for $(x,y)\in D'$ and decays to zero as $|x|,|y|\to\infty$, satisfying $\partial_xC(x,y)=\mathcal O(L^{-1})$ and $\partial_yC(x,y)=\mathcal O(L^{-1})$.
The result of the spatial integral does not change even if we multiply the integrand of Eq.~\eqref{eq:app_IQG1_2} by the cutoff function, and hence we obtain
\begin{align}\label{eq:appC6}
  I_{n,jk}^{\mathrm{QG1}}
  &=-\frac{1}{L^2}\int_{\bm X}\partial_{X_j}(\chi\varepsilon_n)\partial_{X_k}T(\bm X)\partial_Tg_B(\chi\varepsilon_n,T(\bm X))C(\bm X)\notag\\
  &=-\frac{1}{L^2}\int d\bm X\,\partial_{X_j}(\chi\varepsilon_n)\int d\eta\,\delta(\eta-\chi\varepsilon_n)\partial_{X_k}T(\bm X)\partial_Tg_B(\eta,T(\bm X))C(\bm X)\notag\\
  &=\frac{1}{L^2}\int d\bm X\int d\eta\,\partial_{X_j}\theta(\eta-\chi\varepsilon_n)\partial_{X_k}g_B(\eta,T(\bm X))C(\bm X),
\end{align}
where $\theta(x)$ is the Heaviside step function.
Since $\partial_{X_j}C(\bm X)=\mathcal O(L^{-1})$ and the surface term vanishes due to the cutoff function, we have
\begin{align}
  I_{n,jk}^{\mathrm{QG1}}=-\frac{1}{L^2}\int d\bm X\int d\eta\,\theta(\eta-\chi\varepsilon_n)\partial_{X_jX_k}^2g_B(\eta,T(\bm X))C(\bm X)+\mathcal O(L^{-1}).
\end{align}
Here, we assume that the temperature gradient is constant as in Eq.~\eqref{eq:app_T2}, and hence $\partial_{X_jX_k}^2T(\bm X)=0$, which leads to
\begin{align}
  \partial_{X_jX_k}^2g_B(\eta,T(\bm X))
  &=\frac{\partial_{X_j}T(\bm X)\partial_{X_k}T(\bm X)}{T(\bm X)^2}\eta^2\partial_\eta n_B(\eta,T(\bm X))
  +\frac{\partial_{X_jX_k}^2T(\bm X)}{T(\bm X)}(g_B(\eta,T(\bm X))-\eta n_B(\eta,T(\bm X)))\notag\\
  &=\frac{\partial_{X_j}T(\bm X)\partial_{X_k}T(\bm X)}{T(\bm X)^2}\eta^2\partial_\eta n_B(\eta,T(\bm X)).
\end{align}
By substituting this expression into Eq.~\eqref{eq:appC6}, we have
\begin{align}
  I_{n,jk}^{\mathrm{QG1}}
  =-\frac{1}{L^2}\int d\bm X\int d\eta\,\theta(\eta-\chi\varepsilon_n)\frac{\partial_{X_j}T\partial_{X_k}T}{T^2}\eta^2\partial_\eta n_B(\eta,T)+\mathcal O((\partial_{\bm X}T)^3)+\mathcal O(L^{-1})
\end{align}
To obtain $I_{n,jk}^{\mathrm{QG1}}$ up to $\mathcal O(L^0)$, it is sufficient to assume that $\partial_{X_j}T$ is a nonzero constant for $(x,y)\in D$ and $\partial_{X_j}T=0$ for $(x,y)\notin D$~\cite{Mangeolle2024}.
Using the fact that $\chi(x,y)=1$ for $(x,y)\in D$, we obtain
\begin{align}\label{eq:result_IQG1}
  I_{n,jk}^{\mathrm{QG1}}
  =k_B^2\partial_{X_j}T\partial_{X_k}T c_2[n_B(\varepsilon_n,T)]+\mathcal O((\partial_{\bm X}T)^3)+\mathcal O(L^{-1}).
\end{align}

Next, we focus on $I_{n,jk}^{\mathrm{QG2}}$ given in Eq.~\eqref{eq:app_IQG2}, which is calculated as
\begin{align}
  I_{n,jk}^{\mathrm{QG2}}
  =-\frac{1}{L^2}\int_{\bm X}\partial_{X_j}(\chi\varepsilon_n)\partial_{X_k}T(\bm X)\chi\varepsilon_n\partial_Tn_B(\chi\varepsilon_n,T(\bm X)).
\end{align}
Similar to the case of $I_{n,jk}^{\mathrm{QG1}}$, we can multiply the integrand of $I_{n,jk}^{\mathrm{QG2}}$ by the cutoff function $C(\bm X)$ without changing the result, and hence we have
\begin{align}\label{eq:app_IQG2_2}
  I_{n,jk}^{\mathrm{QG2}}
  &=-\frac{1}{L^2}\int_{\bm X}\partial_{X_j}(\chi\varepsilon_n)\partial_{X_k}T(\bm X)\chi\varepsilon_n\partial_Tn_B(\chi\varepsilon_n,T(\bm X))C(\bm X)\notag\\
  &=-\frac{1}{L^2}\int d\bm X\int d\eta\,\theta(\eta-\chi\varepsilon_n)\eta\partial_{X_jX_k}^2n_B(\eta,T(\bm X))C(\bm X)+\mathcal O(L^{-1})
\end{align}
Here, $\partial_{X_jX_k}^2 n_B(\eta,T(\bm X))$ can be written as
\begin{align}\label{eq:appC7}
  \partial_{X_jX_k}^2 n_B(\eta,T(\bm X))
  = \frac{\partial_{X_j}T\partial_{X_k}T}{T(\bm X)^2}\left[\eta\partial_\eta n_B(\eta,T(\bm X))+\eta\partial_\eta\left(\eta\partial_\eta n_B(\eta,T(\bm X))\right)\right].
\end{align}
Substituting this expression into Eq.~\eqref{eq:app_IQG2_2} and assuming a similar condition for $\partial_{X_j}T$, we obtain
\begin{align}\label{eq:result_IQG2}
  I_{n,jk}^{\mathrm{QG2}}
  =-k_B^2\partial_{X_j}T\partial_{X_k}T\left(c_2[n_B(\varepsilon_n,T)]-\frac{\varepsilon_n^3}{k_B^2 T^2}n_B^{\prime}(\varepsilon_n,T)\right)
  + \mathcal O((\partial_{\bm X}T)^3) + \mathcal O(L^{-1}).
\end{align}

Finally, we focus on $I_{n,jk}^{\mathrm{disp}}$ given in Eq.~\eqref{eq:app_Idisp}.
As in the previous cases, we multiply the integrand of $I_{n,jk}^{\mathrm{disp}}$ by the cutoff function $C(\bm X)$, and thus obtain
\begin{align}\label{eq:app_Idisp_2}
  I_{n,jk}^{\mathrm{disp}}
  &=\frac{1}{L^2}\int d\bm X\,\left\{(\chi\varepsilon_n)^2\partial_{X_k}(\chi \varepsilon_n)\partial_{X_j}T(\bm X)\partial_T n_B^{\prime}(\chi\varepsilon_n,T(\bm X))
  -2\chi\varepsilon_n\partial_{X_k}(\chi \varepsilon_n)\partial_{X_j}T(\bm X)\partial_T n_B(\chi\varepsilon_n,T(\bm X))\right\}C(\bm X)\notag\\
  &=\frac{1}{L^2}\int d\bm X\int d\eta\,\theta(\eta-\chi\varepsilon_n)
  \partial_{X_jX_k}^2\left\{\eta^2n_B^{\prime}(\eta,T(\bm X))-2\eta n_B(\eta,T(\bm X))\right\}C(\bm X)+\mathcal O(L^{-1}).
\end{align}
Moreover, we proceed to evaluate the second derivative with respect to $X_j$ and $X_k$ in the integrand of the above expression as
\begin{align}
  \partial_{X_jX_k}^2\left\{\eta^2n_B^{\prime}(\eta,T(\bm X))-2\eta n_B(\eta,T(\bm X))\right\}
  =\frac{\partial_{X_j}T\partial_{X_k}T}{T^2}
  \left(4\eta^2\partial_\eta n_B(\eta,T(\bm X))+\partial_\eta
  \left\{\eta^4\partial_\eta^2n_B(\eta,T(\bm X))-2\eta^3\partial_\eta n_B(\eta,T(\bm X))\right\}
  \right).
\end{align}
By substituting this expression into Eq.~\eqref{eq:app_Idisp_2} and assuming a similar condition for $\partial_{X_j}T$, we obtain
\begin{align}\label{eq:result_Idisp}
  I_{n,jk}^{\mathrm{disp}}
  =-k_B^2\partial_{X_j}T\partial_{X_k}T\left(4c_2[n_B(\eta,T)]+\frac{\varepsilon_n^3}{k_B^2T^2}\left(\varepsilon_n n_B^{\prime\prime}(\eta,T)-2 n_B^{\prime}(\eta,T)\right)\right)
  +\mathcal O((\partial_{\bm X}T)^3)+\mathcal O(L^{-1}).
\end{align}

By substituting the results of Eqs.~\eqref{eq:result_IQG1}, \eqref{eq:result_IQG2}, and \eqref{eq:result_Idisp} into Eqs.~\eqref{eq:app_J2QG} and \eqref{eq:app_J2disp}, we obtain an expression for the second-order nonlinear thermal conductivity tensor, which is given by the sum of Eqs.~\eqref{eq:kappa_ijk^TBCP}, \eqref{eq:kappa_ijk^QM}, and \eqref{eq:kappa_ijk^disp}.

\end{widetext}

\bibliography{./refs}

\begin{thebibliography}{122}%
\makeatletter
\providecommand \@ifxundefined [1]{%
 \@ifx{#1\undefined}
}%
\providecommand \@ifnum [1]{%
 \ifnum #1\expandafter \@firstoftwo
 \else \expandafter \@secondoftwo
 \fi
}%
\providecommand \@ifx [1]{%
 \ifx #1\expandafter \@firstoftwo
 \else \expandafter \@secondoftwo
 \fi
}%
\providecommand \natexlab [1]{#1}%
\providecommand \enquote  [1]{``#1''}%
\providecommand \bibnamefont  [1]{#1}%
\providecommand \bibfnamefont [1]{#1}%
\providecommand \citenamefont [1]{#1}%
\providecommand \href@noop [0]{\@secondoftwo}%
\providecommand \href [0]{\begingroup \@sanitize@url \@href}%
\providecommand \@href[1]{\@@startlink{#1}\@@href}%
\providecommand \@@href[1]{\endgroup#1\@@endlink}%
\providecommand \@sanitize@url [0]{\catcode `\\12\catcode `\$12\catcode `\&12\catcode `\#12\catcode `\^12\catcode `\_12\catcode `\%12\relax}%
\providecommand \@@startlink[1]{}%
\providecommand \@@endlink[0]{}%
\providecommand \url  [0]{\begingroup\@sanitize@url \@url }%
\providecommand \@url [1]{\endgroup\@href {#1}{\urlprefix }}%
\providecommand \urlprefix  [0]{URL }%
\providecommand \Eprint [0]{\href }%
\providecommand \doibase [0]{https://doi.org/}%
\providecommand \selectlanguage [0]{\@gobble}%
\providecommand \bibinfo  [0]{\@secondoftwo}%
\providecommand \bibfield  [0]{\@secondoftwo}%
\providecommand \translation [1]{[#1]}%
\providecommand \BibitemOpen [0]{}%
\providecommand \bibitemStop [0]{}%
\providecommand \bibitemNoStop [0]{.\EOS\space}%
\providecommand \EOS [0]{\spacefactor3000\relax}%
\providecommand \BibitemShut  [1]{\csname bibitem#1\endcsname}%
\let\auto@bib@innerbib\@empty
\bibitem [{\citenamefont {Thouless}\ \emph {et~al.}(1982)\citenamefont {Thouless}, \citenamefont {Kohmoto}, \citenamefont {Nightingale},\ and\ \citenamefont {den Nijs}}]{Thouless1982}%
  \BibitemOpen
  \bibfield  {author} {\bibinfo {author} {\bibfnamefont {D.~J.}\ \bibnamefont {Thouless}}, \bibinfo {author} {\bibfnamefont {M.}~\bibnamefont {Kohmoto}}, \bibinfo {author} {\bibfnamefont {M.~P.}\ \bibnamefont {Nightingale}},\ and\ \bibinfo {author} {\bibfnamefont {M.}~\bibnamefont {den Nijs}},\ }\bibfield  {title} {\bibinfo {title} {{Quantized Hall Conductance in a Two-Dimensional Periodic Potential}},\ }\href {https://doi.org/10.1103/PhysRevLett.49.405} {\bibfield  {journal} {\bibinfo  {journal} {Phys. Rev. Lett.}\ }\textbf {\bibinfo {volume} {49}},\ \bibinfo {pages} {405} (\bibinfo {year} {1982})}\BibitemShut {NoStop}%
\bibitem [{\citenamefont {Berry}(1984)}]{Berry1984}%
  \BibitemOpen
  \bibfield  {author} {\bibinfo {author} {\bibfnamefont {M.~V.}\ \bibnamefont {Berry}},\ }\bibfield  {title} {\bibinfo {title} {{Quantal Phase Factors Accompanying Adiabatic Changes}},\ }\href {http://www.jstor.org/stable/2397741} {\bibfield  {journal} {\bibinfo  {journal} {Proc. R. Soc. Lond. A}\ }\textbf {\bibinfo {volume} {392}},\ \bibinfo {pages} {45} (\bibinfo {year} {1984})}\BibitemShut {NoStop}%
\bibitem [{\citenamefont {Simon}(1983)}]{Simon1983}%
  \BibitemOpen
  \bibfield  {author} {\bibinfo {author} {\bibfnamefont {B.}~\bibnamefont {Simon}},\ }\bibfield  {title} {\bibinfo {title} {{Holonomy, the Quantum Adiabatic Theorem, and Berry's Phase}},\ }\href {https://doi.org/10.1103/PhysRevLett.51.2167} {\bibfield  {journal} {\bibinfo  {journal} {Phys. Rev. Lett.}\ }\textbf {\bibinfo {volume} {51}},\ \bibinfo {pages} {2167} (\bibinfo {year} {1983})}\BibitemShut {NoStop}%
\bibitem [{\citenamefont {Provost}\ and\ \citenamefont {Vallee}(1980)}]{Provost1980}%
  \BibitemOpen
  \bibfield  {author} {\bibinfo {author} {\bibfnamefont {J.~P.}\ \bibnamefont {Provost}}\ and\ \bibinfo {author} {\bibfnamefont {G.}~\bibnamefont {Vallee}},\ }\bibfield  {title} {\bibinfo {title} {{Riemannian structure on manifolds of quantum states}},\ }\href {https://doi.org/10.1007/BF02193559} {\bibfield  {journal} {\bibinfo  {journal} {Commun. Math. Phys.}\ }\textbf {\bibinfo {volume} {76}},\ \bibinfo {pages} {289} (\bibinfo {year} {1980})}\BibitemShut {NoStop}%
\bibitem [{\citenamefont {Resta}(2011)}]{Resta2011}%
  \BibitemOpen
  \bibfield  {author} {\bibinfo {author} {\bibfnamefont {R.}~\bibnamefont {Resta}},\ }\bibfield  {title} {\bibinfo {title} {{The insulating state of matter: a geometrical theory}},\ }\href {https://doi.org/10.1140/epjb/e2010-10874-4} {\bibfield  {journal} {\bibinfo  {journal} {The European Physical Journal B}\ }\textbf {\bibinfo {volume} {79}},\ \bibinfo {pages} {121} (\bibinfo {year} {2011})}\BibitemShut {NoStop}%
\bibitem [{\citenamefont {Yu}\ \emph {et~al.}(2025)\citenamefont {Yu}, \citenamefont {Bernevig}, \citenamefont {Queiroz}, \citenamefont {Rossi}, \citenamefont {T{\"o}rm{\"a}},\ and\ \citenamefont {Yang}}]{Yu2025}%
  \BibitemOpen
  \bibfield  {author} {\bibinfo {author} {\bibfnamefont {J.}~\bibnamefont {Yu}}, \bibinfo {author} {\bibfnamefont {B.~A.}\ \bibnamefont {Bernevig}}, \bibinfo {author} {\bibfnamefont {R.}~\bibnamefont {Queiroz}}, \bibinfo {author} {\bibfnamefont {E.}~\bibnamefont {Rossi}}, \bibinfo {author} {\bibfnamefont {P.}~\bibnamefont {T{\"o}rm{\"a}}},\ and\ \bibinfo {author} {\bibfnamefont {B.-J.}\ \bibnamefont {Yang}},\ }\bibfield  {title} {\bibinfo {title} {{Quantum geometry in quantum materials}},\ }\href {https://doi.org/10.1038/s41535-025-00801-3} {\bibfield  {journal} {\bibinfo  {journal} {npj Quantum Mater.}\ }\textbf {\bibinfo {volume} {10}},\ \bibinfo {pages} {101} (\bibinfo {year} {2025})}\BibitemShut {NoStop}%
\bibitem [{\citenamefont {Lhachemi}\ and\ \citenamefont {Cano}(2026)}]{Lhachemi2026}%
  \BibitemOpen
  \bibfield  {author} {\bibinfo {author} {\bibfnamefont {M.~N.~Y.}\ \bibnamefont {Lhachemi}}\ and\ \bibinfo {author} {\bibfnamefont {J.}~\bibnamefont {Cano}},\ }\href@noop {} {\bibinfo {title} {{A unifying framework for sum rules and bounds on optical, thermoelectric and thermal transport from quantum geometry}}} (\bibinfo {year} {2026}),\ \Eprint {https://arxiv.org/abs/2603.10121} {arXiv:2603.10121 [cond-mat.mes-hall]} \BibitemShut {NoStop}%
\bibitem [{\citenamefont {Peotta}\ and\ \citenamefont {T{\"o}rm{\"a}}(2015)}]{Peotta2015}%
  \BibitemOpen
  \bibfield  {author} {\bibinfo {author} {\bibfnamefont {S.}~\bibnamefont {Peotta}}\ and\ \bibinfo {author} {\bibfnamefont {P.}~\bibnamefont {T{\"o}rm{\"a}}},\ }\bibfield  {title} {\bibinfo {title} {{Superfluidity in topologically nontrivial flat bands}},\ }\href {https://doi.org/10.1038/ncomms9944} {\bibfield  {journal} {\bibinfo  {journal} {Nat. Commun.}\ }\textbf {\bibinfo {volume} {6}},\ \bibinfo {pages} {8944} (\bibinfo {year} {2015})}\BibitemShut {NoStop}%
\bibitem [{\citenamefont {Liang}\ \emph {et~al.}(2017)\citenamefont {Liang}, \citenamefont {Vanhala}, \citenamefont {Peotta}, \citenamefont {Siro}, \citenamefont {Harju},\ and\ \citenamefont {T\"orm\"a}}]{Liang2017}%
  \BibitemOpen
  \bibfield  {author} {\bibinfo {author} {\bibfnamefont {L.}~\bibnamefont {Liang}}, \bibinfo {author} {\bibfnamefont {T.~I.}\ \bibnamefont {Vanhala}}, \bibinfo {author} {\bibfnamefont {S.}~\bibnamefont {Peotta}}, \bibinfo {author} {\bibfnamefont {T.}~\bibnamefont {Siro}}, \bibinfo {author} {\bibfnamefont {A.}~\bibnamefont {Harju}},\ and\ \bibinfo {author} {\bibfnamefont {P.}~\bibnamefont {T\"orm\"a}},\ }\bibfield  {title} {\bibinfo {title} {{Band geometry, Berry curvature, and superfluid weight}},\ }\href {https://doi.org/10.1103/PhysRevB.95.024515} {\bibfield  {journal} {\bibinfo  {journal} {Phys. Rev. B}\ }\textbf {\bibinfo {volume} {95}},\ \bibinfo {pages} {024515} (\bibinfo {year} {2017})}\BibitemShut {NoStop}%
\bibitem [{\citenamefont {Wang}\ \emph {et~al.}(2021{\natexlab{a}})\citenamefont {Wang}, \citenamefont {Cano}, \citenamefont {Millis}, \citenamefont {Liu},\ and\ \citenamefont {Yang}}]{Wang2021APS}%
  \BibitemOpen
  \bibfield  {author} {\bibinfo {author} {\bibfnamefont {J.}~\bibnamefont {Wang}}, \bibinfo {author} {\bibfnamefont {J.}~\bibnamefont {Cano}}, \bibinfo {author} {\bibfnamefont {A.~J.}\ \bibnamefont {Millis}}, \bibinfo {author} {\bibfnamefont {Z.}~\bibnamefont {Liu}},\ and\ \bibinfo {author} {\bibfnamefont {B.}~\bibnamefont {Yang}},\ }\bibfield  {title} {\bibinfo {title} {{Exact Landau Level Description of Geometry and Interaction in a Flatband}},\ }\href {https://doi.org/10.1103/PhysRevLett.127.246403} {\bibfield  {journal} {\bibinfo  {journal} {Phys. Rev. Lett.}\ }\textbf {\bibinfo {volume} {127}},\ \bibinfo {pages} {246403} (\bibinfo {year} {2021}{\natexlab{a}})}\BibitemShut {NoStop}%
\bibitem [{\citenamefont {Gao}\ \emph {et~al.}(2014)\citenamefont {Gao}, \citenamefont {Yang},\ and\ \citenamefont {Niu}}]{Gao2014}%
  \BibitemOpen
  \bibfield  {author} {\bibinfo {author} {\bibfnamefont {Y.}~\bibnamefont {Gao}}, \bibinfo {author} {\bibfnamefont {S.~A.}\ \bibnamefont {Yang}},\ and\ \bibinfo {author} {\bibfnamefont {Q.}~\bibnamefont {Niu}},\ }\bibfield  {title} {\bibinfo {title} {{Field Induced Positional Shift of Bloch Electrons and Its Dynamical Implications}},\ }\href {https://doi.org/10.1103/PhysRevLett.112.166601} {\bibfield  {journal} {\bibinfo  {journal} {Phys. Rev. Lett.}\ }\textbf {\bibinfo {volume} {112}},\ \bibinfo {pages} {166601} (\bibinfo {year} {2014})}\BibitemShut {NoStop}%
\bibitem [{\citenamefont {Gao}\ \emph {et~al.}(2015)\citenamefont {Gao}, \citenamefont {Yang},\ and\ \citenamefont {Niu}}]{Gao2015}%
  \BibitemOpen
  \bibfield  {author} {\bibinfo {author} {\bibfnamefont {Y.}~\bibnamefont {Gao}}, \bibinfo {author} {\bibfnamefont {S.~A.}\ \bibnamefont {Yang}},\ and\ \bibinfo {author} {\bibfnamefont {Q.}~\bibnamefont {Niu}},\ }\bibfield  {title} {\bibinfo {title} {{Geometrical effects in orbital magnetic susceptibility}},\ }\href {https://doi.org/10.1103/PhysRevB.91.214405} {\bibfield  {journal} {\bibinfo  {journal} {Phys. Rev. B}\ }\textbf {\bibinfo {volume} {91}},\ \bibinfo {pages} {214405} (\bibinfo {year} {2015})}\BibitemShut {NoStop}%
\bibitem [{\citenamefont {Das}\ \emph {et~al.}(2023)\citenamefont {Das}, \citenamefont {Lahiri}, \citenamefont {Atencia}, \citenamefont {Culcer},\ and\ \citenamefont {Agarwal}}]{Das2023}%
  \BibitemOpen
  \bibfield  {author} {\bibinfo {author} {\bibfnamefont {K.}~\bibnamefont {Das}}, \bibinfo {author} {\bibfnamefont {S.}~\bibnamefont {Lahiri}}, \bibinfo {author} {\bibfnamefont {R.~B.}\ \bibnamefont {Atencia}}, \bibinfo {author} {\bibfnamefont {D.}~\bibnamefont {Culcer}},\ and\ \bibinfo {author} {\bibfnamefont {A.}~\bibnamefont {Agarwal}},\ }\bibfield  {title} {\bibinfo {title} {{Intrinsic nonlinear conductivities induced by the quantum metric}},\ }\href {https://doi.org/10.1103/PhysRevB.108.L201405} {\bibfield  {journal} {\bibinfo  {journal} {Phys. Rev. B}\ }\textbf {\bibinfo {volume} {108}},\ \bibinfo {pages} {L201405} (\bibinfo {year} {2023})}\BibitemShut {NoStop}%
\bibitem [{\citenamefont {Gao}\ \emph {et~al.}(2023)\citenamefont {Gao}, \citenamefont {Liu}, \citenamefont {Qiu}, \citenamefont {Ghosh}, \citenamefont {Trevisan}, \citenamefont {Onishi}, \citenamefont {Hu}, \citenamefont {Qian}, \citenamefont {Tien}, \citenamefont {Chen}, \citenamefont {Huang}, \citenamefont {B^^c3^^a9rub^^c3^^a9}, \citenamefont {Li}, \citenamefont {Tzschaschel}, \citenamefont {Dinh}, \citenamefont {Sun}, \citenamefont {Ho}, \citenamefont {Lien}, \citenamefont {Singh}, \citenamefont {Watanabe}, \citenamefont {Taniguchi}, \citenamefont {Bell}, \citenamefont {Lin}, \citenamefont {Chang}, \citenamefont {Du}, \citenamefont {Bansil}, \citenamefont {Fu}, \citenamefont {Ni}, \citenamefont {Orth}, \citenamefont {Ma},\ and\ \citenamefont {Xu}}]{Gao2023}%
  \BibitemOpen
  \bibfield  {author} {\bibinfo {author} {\bibfnamefont {A.}~\bibnamefont {Gao}}, \bibinfo {author} {\bibfnamefont {Y.-F.}\ \bibnamefont {Liu}}, \bibinfo {author} {\bibfnamefont {J.-X.}\ \bibnamefont {Qiu}}, \bibinfo {author} {\bibfnamefont {B.}~\bibnamefont {Ghosh}}, \bibinfo {author} {\bibfnamefont {T.~V.}\ \bibnamefont {Trevisan}}, \bibinfo {author} {\bibfnamefont {Y.}~\bibnamefont {Onishi}}, \bibinfo {author} {\bibfnamefont {C.}~\bibnamefont {Hu}}, \bibinfo {author} {\bibfnamefont {T.}~\bibnamefont {Qian}}, \bibinfo {author} {\bibfnamefont {H.-J.}\ \bibnamefont {Tien}}, \bibinfo {author} {\bibfnamefont {S.-W.}\ \bibnamefont {Chen}}, \bibinfo {author} {\bibfnamefont {M.}~\bibnamefont {Huang}}, \bibinfo {author} {\bibfnamefont {D.}~\bibnamefont {B^^c3^^a9rub^^c3^^a9}}, \bibinfo {author} {\bibfnamefont {H.}~\bibnamefont {Li}}, \bibinfo {author} {\bibfnamefont {C.}~\bibnamefont {Tzschaschel}}, \bibinfo {author} {\bibfnamefont {T.}~\bibnamefont {Dinh}}, \bibinfo {author} {\bibfnamefont {Z.}~\bibnamefont
  {Sun}}, \bibinfo {author} {\bibfnamefont {S.-C.}\ \bibnamefont {Ho}}, \bibinfo {author} {\bibfnamefont {S.-W.}\ \bibnamefont {Lien}}, \bibinfo {author} {\bibfnamefont {B.}~\bibnamefont {Singh}}, \bibinfo {author} {\bibfnamefont {K.}~\bibnamefont {Watanabe}}, \bibinfo {author} {\bibfnamefont {T.}~\bibnamefont {Taniguchi}}, \bibinfo {author} {\bibfnamefont {D.~C.}\ \bibnamefont {Bell}}, \bibinfo {author} {\bibfnamefont {H.}~\bibnamefont {Lin}}, \bibinfo {author} {\bibfnamefont {T.-R.}\ \bibnamefont {Chang}}, \bibinfo {author} {\bibfnamefont {C.~R.}\ \bibnamefont {Du}}, \bibinfo {author} {\bibfnamefont {A.}~\bibnamefont {Bansil}}, \bibinfo {author} {\bibfnamefont {L.}~\bibnamefont {Fu}}, \bibinfo {author} {\bibfnamefont {N.}~\bibnamefont {Ni}}, \bibinfo {author} {\bibfnamefont {P.~P.}\ \bibnamefont {Orth}}, \bibinfo {author} {\bibfnamefont {Q.}~\bibnamefont {Ma}},\ and\ \bibinfo {author} {\bibfnamefont {S.-Y.}\ \bibnamefont {Xu}},\ }\bibfield  {title} {\bibinfo {title} {{Quantum metric nonlinear Hall effect in
  a topological antiferromagnetic heterostructure}},\ }\href {https://doi.org/10.1126/science.adf1506} {\bibfield  {journal} {\bibinfo  {journal} {Science}\ }\textbf {\bibinfo {volume} {381}},\ \bibinfo {pages} {181} (\bibinfo {year} {2023})}\BibitemShut {NoStop}%
\bibitem [{\citenamefont {Wang}\ \emph {et~al.}(2023)\citenamefont {Wang}, \citenamefont {Kaplan}, \citenamefont {Zhang}, \citenamefont {Holder}, \citenamefont {Cao}, \citenamefont {Wang}, \citenamefont {Zhou}, \citenamefont {Zhou}, \citenamefont {Jiang}, \citenamefont {Zhang}, \citenamefont {Ru}, \citenamefont {Cai}, \citenamefont {Watanabe}, \citenamefont {Taniguchi}, \citenamefont {Yan},\ and\ \citenamefont {Gao}}]{Wang2023}%
  \BibitemOpen
  \bibfield  {author} {\bibinfo {author} {\bibfnamefont {N.}~\bibnamefont {Wang}}, \bibinfo {author} {\bibfnamefont {D.}~\bibnamefont {Kaplan}}, \bibinfo {author} {\bibfnamefont {Z.}~\bibnamefont {Zhang}}, \bibinfo {author} {\bibfnamefont {T.}~\bibnamefont {Holder}}, \bibinfo {author} {\bibfnamefont {N.}~\bibnamefont {Cao}}, \bibinfo {author} {\bibfnamefont {A.}~\bibnamefont {Wang}}, \bibinfo {author} {\bibfnamefont {X.}~\bibnamefont {Zhou}}, \bibinfo {author} {\bibfnamefont {F.}~\bibnamefont {Zhou}}, \bibinfo {author} {\bibfnamefont {Z.}~\bibnamefont {Jiang}}, \bibinfo {author} {\bibfnamefont {C.}~\bibnamefont {Zhang}}, \bibinfo {author} {\bibfnamefont {S.}~\bibnamefont {Ru}}, \bibinfo {author} {\bibfnamefont {H.}~\bibnamefont {Cai}}, \bibinfo {author} {\bibfnamefont {K.}~\bibnamefont {Watanabe}}, \bibinfo {author} {\bibfnamefont {T.}~\bibnamefont {Taniguchi}}, \bibinfo {author} {\bibfnamefont {B.}~\bibnamefont {Yan}},\ and\ \bibinfo {author} {\bibfnamefont {W.}~\bibnamefont {Gao}},\ }\bibfield  {title}
  {\bibinfo {title} {{Quantum-metric-induced nonlinear transport in a topological antiferromagnet}},\ }\href {https://doi.org/10.1038/s41586-023-06363-3} {\bibfield  {journal} {\bibinfo  {journal} {Nature}\ }\textbf {\bibinfo {volume} {621}},\ \bibinfo {pages} {487} (\bibinfo {year} {2023})}\BibitemShut {NoStop}%
\bibitem [{\citenamefont {Fang}\ \emph {et~al.}(2024)\citenamefont {Fang}, \citenamefont {Cano},\ and\ \citenamefont {Ghorashi}}]{Fang2024}%
  \BibitemOpen
  \bibfield  {author} {\bibinfo {author} {\bibfnamefont {Y.}~\bibnamefont {Fang}}, \bibinfo {author} {\bibfnamefont {J.}~\bibnamefont {Cano}},\ and\ \bibinfo {author} {\bibfnamefont {S.~A.~A.}\ \bibnamefont {Ghorashi}},\ }\bibfield  {title} {\bibinfo {title} {{Quantum Geometry Induced Nonlinear Transport in Altermagnets}},\ }\href {https://doi.org/10.1103/PhysRevLett.133.106701} {\bibfield  {journal} {\bibinfo  {journal} {Phys. Rev. Lett.}\ }\textbf {\bibinfo {volume} {133}},\ \bibinfo {pages} {106701} (\bibinfo {year} {2024})}\BibitemShut {NoStop}%
\bibitem [{\citenamefont {Sodemann}\ and\ \citenamefont {Fu}(2015)}]{Sodemann2015}%
  \BibitemOpen
  \bibfield  {author} {\bibinfo {author} {\bibfnamefont {I.}~\bibnamefont {Sodemann}}\ and\ \bibinfo {author} {\bibfnamefont {L.}~\bibnamefont {Fu}},\ }\bibfield  {title} {\bibinfo {title} {{Quantum Nonlinear Hall Effect Induced by Berry Curvature Dipole in Time-Reversal Invariant Materials}},\ }\href {https://doi.org/10.1103/PhysRevLett.115.216806} {\bibfield  {journal} {\bibinfo  {journal} {Phys. Rev. Lett.}\ }\textbf {\bibinfo {volume} {115}},\ \bibinfo {pages} {216806} (\bibinfo {year} {2015})}\BibitemShut {NoStop}%
\bibitem [{\citenamefont {Du}\ \emph {et~al.}(2021)\citenamefont {Du}, \citenamefont {Wang}, \citenamefont {Sun}, \citenamefont {Lu},\ and\ \citenamefont {Xie}}]{Du2021}%
  \BibitemOpen
  \bibfield  {author} {\bibinfo {author} {\bibfnamefont {Z.~Z.}\ \bibnamefont {Du}}, \bibinfo {author} {\bibfnamefont {C.~M.}\ \bibnamefont {Wang}}, \bibinfo {author} {\bibfnamefont {H.-P.}\ \bibnamefont {Sun}}, \bibinfo {author} {\bibfnamefont {H.-Z.}\ \bibnamefont {Lu}},\ and\ \bibinfo {author} {\bibfnamefont {X.~C.}\ \bibnamefont {Xie}},\ }\bibfield  {title} {\bibinfo {title} {Quantum theory of the nonlinear hall effect},\ }\href {https://doi.org/10.1038/s41467-021-25273-4} {\bibfield  {journal} {\bibinfo  {journal} {Nat. Commun.}\ }\textbf {\bibinfo {volume} {12}},\ \bibinfo {pages} {5038} (\bibinfo {year} {2021})}\BibitemShut {NoStop}%
\bibitem [{\citenamefont {Bandyopadhyay}\ \emph {et~al.}(2024)\citenamefont {Bandyopadhyay}, \citenamefont {Joseph},\ and\ \citenamefont {Narayan}}]{Bandyopadhyay2024}%
  \BibitemOpen
  \bibfield  {author} {\bibinfo {author} {\bibfnamefont {A.}~\bibnamefont {Bandyopadhyay}}, \bibinfo {author} {\bibfnamefont {N.~B.}\ \bibnamefont {Joseph}},\ and\ \bibinfo {author} {\bibfnamefont {A.}~\bibnamefont {Narayan}},\ }\bibfield  {title} {\bibinfo {title} {Non-linear hall effects: Mechanisms and materials},\ }\href {https://doi.org/10.1016/j.mtelec.2024.100101} {\bibfield  {journal} {\bibinfo  {journal} {Mater. Today Electron.}\ }\textbf {\bibinfo {volume} {8}},\ \bibinfo {pages} {100101} (\bibinfo {year} {2024})}\BibitemShut {NoStop}%
\bibitem [{\citenamefont {Ulrich}\ \emph {et~al.}(2025)\citenamefont {Ulrich}, \citenamefont {Mitscherling}, \citenamefont {Classen},\ and\ \citenamefont {Schnyder}}]{Ulrich2025}%
  \BibitemOpen
  \bibfield  {author} {\bibinfo {author} {\bibfnamefont {Y.}~\bibnamefont {Ulrich}}, \bibinfo {author} {\bibfnamefont {J.}~\bibnamefont {Mitscherling}}, \bibinfo {author} {\bibfnamefont {L.}~\bibnamefont {Classen}},\ and\ \bibinfo {author} {\bibfnamefont {A.~P.}\ \bibnamefont {Schnyder}},\ }\href@noop {} {\bibinfo {title} {{Quantum Geometric Origin of the Intrinsic Nonlinear Hall Effect}}} (\bibinfo {year} {2025}),\ \Eprint {https://arxiv.org/abs/2506.17386} {arXiv:2506.17386 [cond-mat.mes-hall]} \BibitemShut {NoStop}%
\bibitem [{\citenamefont {Sipe}\ and\ \citenamefont {Shkrebtii}(2000)}]{Sipe2000}%
  \BibitemOpen
  \bibfield  {author} {\bibinfo {author} {\bibfnamefont {J.~E.}\ \bibnamefont {Sipe}}\ and\ \bibinfo {author} {\bibfnamefont {A.~I.}\ \bibnamefont {Shkrebtii}},\ }\bibfield  {title} {\bibinfo {title} {{Second-order optical response in semiconductors}},\ }\href {https://doi.org/10.1103/PhysRevB.61.5337} {\bibfield  {journal} {\bibinfo  {journal} {Phys. Rev. B}\ }\textbf {\bibinfo {volume} {61}},\ \bibinfo {pages} {5337} (\bibinfo {year} {2000})}\BibitemShut {NoStop}%
\bibitem [{\citenamefont {Zhang}\ \emph {et~al.}(2018{\natexlab{a}})\citenamefont {Zhang}, \citenamefont {Sun},\ and\ \citenamefont {Yan}}]{Zhang2018B}%
  \BibitemOpen
  \bibfield  {author} {\bibinfo {author} {\bibfnamefont {Y.}~\bibnamefont {Zhang}}, \bibinfo {author} {\bibfnamefont {Y.}~\bibnamefont {Sun}},\ and\ \bibinfo {author} {\bibfnamefont {B.}~\bibnamefont {Yan}},\ }\bibfield  {title} {\bibinfo {title} {{Berry curvature dipole in Weyl semimetal materials: An ab initio study}},\ }\href {https://doi.org/10.1103/PhysRevB.97.041101} {\bibfield  {journal} {\bibinfo  {journal} {Phys. Rev. B}\ }\textbf {\bibinfo {volume} {97}},\ \bibinfo {pages} {041101} (\bibinfo {year} {2018}{\natexlab{a}})}\BibitemShut {NoStop}%
\bibitem [{\citenamefont {Zhang}\ \emph {et~al.}(2018{\natexlab{b}})\citenamefont {Zhang}, \citenamefont {van~den Brink}, \citenamefont {Felser},\ and\ \citenamefont {Yan}}]{Zhang20182D}%
  \BibitemOpen
  \bibfield  {author} {\bibinfo {author} {\bibfnamefont {Y.}~\bibnamefont {Zhang}}, \bibinfo {author} {\bibfnamefont {J.}~\bibnamefont {van~den Brink}}, \bibinfo {author} {\bibfnamefont {C.}~\bibnamefont {Felser}},\ and\ \bibinfo {author} {\bibfnamefont {B.}~\bibnamefont {Yan}},\ }\bibfield  {title} {\bibinfo {title} {{Electrically tuneable nonlinear anomalous Hall effect in two-dimensional transition-metal dichalcogenides WTe2 and MoTe2}},\ }\href {https://doi.org/10.1088/2053-1583/aad1ae} {\bibfield  {journal} {\bibinfo  {journal} {2D Materials}\ }\textbf {\bibinfo {volume} {5}},\ \bibinfo {pages} {044001} (\bibinfo {year} {2018}{\natexlab{b}})}\BibitemShut {NoStop}%
\bibitem [{\citenamefont {Ma}\ \emph {et~al.}(2019)\citenamefont {Ma}, \citenamefont {Xu}, \citenamefont {Shen}, \citenamefont {MacNeill}, \citenamefont {Fatemi}, \citenamefont {Chang}, \citenamefont {Mier~Valdivia}, \citenamefont {Wu}, \citenamefont {Du}, \citenamefont {Hsu}, \citenamefont {Fang}, \citenamefont {Gibson}, \citenamefont {Watanabe}, \citenamefont {Taniguchi}, \citenamefont {Cava}, \citenamefont {Kaxiras}, \citenamefont {Lu}, \citenamefont {Lin}, \citenamefont {Fu}, \citenamefont {Gedik},\ and\ \citenamefont {Jarillo-Herrero}}]{Ma2019}%
  \BibitemOpen
  \bibfield  {author} {\bibinfo {author} {\bibfnamefont {Q.}~\bibnamefont {Ma}}, \bibinfo {author} {\bibfnamefont {S.-Y.}\ \bibnamefont {Xu}}, \bibinfo {author} {\bibfnamefont {H.}~\bibnamefont {Shen}}, \bibinfo {author} {\bibfnamefont {D.}~\bibnamefont {MacNeill}}, \bibinfo {author} {\bibfnamefont {V.}~\bibnamefont {Fatemi}}, \bibinfo {author} {\bibfnamefont {T.-R.}\ \bibnamefont {Chang}}, \bibinfo {author} {\bibfnamefont {A.~M.}\ \bibnamefont {Mier~Valdivia}}, \bibinfo {author} {\bibfnamefont {S.}~\bibnamefont {Wu}}, \bibinfo {author} {\bibfnamefont {Z.}~\bibnamefont {Du}}, \bibinfo {author} {\bibfnamefont {C.-H.}\ \bibnamefont {Hsu}}, \bibinfo {author} {\bibfnamefont {S.}~\bibnamefont {Fang}}, \bibinfo {author} {\bibfnamefont {Q.~D.}\ \bibnamefont {Gibson}}, \bibinfo {author} {\bibfnamefont {K.}~\bibnamefont {Watanabe}}, \bibinfo {author} {\bibfnamefont {T.}~\bibnamefont {Taniguchi}}, \bibinfo {author} {\bibfnamefont {R.~J.}\ \bibnamefont {Cava}}, \bibinfo {author} {\bibfnamefont {E.}~\bibnamefont
  {Kaxiras}}, \bibinfo {author} {\bibfnamefont {H.-Z.}\ \bibnamefont {Lu}}, \bibinfo {author} {\bibfnamefont {H.}~\bibnamefont {Lin}}, \bibinfo {author} {\bibfnamefont {L.}~\bibnamefont {Fu}}, \bibinfo {author} {\bibfnamefont {N.}~\bibnamefont {Gedik}},\ and\ \bibinfo {author} {\bibfnamefont {P.}~\bibnamefont {Jarillo-Herrero}},\ }\bibfield  {title} {\bibinfo {title} {{Observation of the nonlinear Hall effect under time-reversal-symmetric conditions}},\ }\href {https://doi.org/10.1038/s41586-018-0807-6} {\bibfield  {journal} {\bibinfo  {journal} {Nature}\ }\textbf {\bibinfo {volume} {565}},\ \bibinfo {pages} {337} (\bibinfo {year} {2019})}\BibitemShut {NoStop}%
\bibitem [{\citenamefont {Kang}\ \emph {et~al.}(2019)\citenamefont {Kang}, \citenamefont {Li}, \citenamefont {Sohn}, \citenamefont {Shan},\ and\ \citenamefont {Mak}}]{Kang2019}%
  \BibitemOpen
  \bibfield  {author} {\bibinfo {author} {\bibfnamefont {K.}~\bibnamefont {Kang}}, \bibinfo {author} {\bibfnamefont {T.}~\bibnamefont {Li}}, \bibinfo {author} {\bibfnamefont {E.}~\bibnamefont {Sohn}}, \bibinfo {author} {\bibfnamefont {J.}~\bibnamefont {Shan}},\ and\ \bibinfo {author} {\bibfnamefont {K.~F.}\ \bibnamefont {Mak}},\ }\bibfield  {title} {\bibinfo {title} {{Nonlinear anomalous Hall effect in few-layer WTe2}},\ }\href {https://doi.org/10.1038/s41563-019-0294-7} {\bibfield  {journal} {\bibinfo  {journal} {Nat. Mater.}\ }\textbf {\bibinfo {volume} {18}},\ \bibinfo {pages} {324} (\bibinfo {year} {2019})}\BibitemShut {NoStop}%
\bibitem [{\citenamefont {Tiwari}\ \emph {et~al.}(2021)\citenamefont {Tiwari}, \citenamefont {Chen}, \citenamefont {Zhong}, \citenamefont {Drueke}, \citenamefont {Koo}, \citenamefont {Kaczmarek}, \citenamefont {Xiao}, \citenamefont {Gao}, \citenamefont {Luo}, \citenamefont {Niu}, \citenamefont {Sun}, \citenamefont {Yan}, \citenamefont {Zhao},\ and\ \citenamefont {Tsen}}]{Tiwari2021}%
  \BibitemOpen
  \bibfield  {author} {\bibinfo {author} {\bibfnamefont {A.}~\bibnamefont {Tiwari}}, \bibinfo {author} {\bibfnamefont {F.}~\bibnamefont {Chen}}, \bibinfo {author} {\bibfnamefont {S.}~\bibnamefont {Zhong}}, \bibinfo {author} {\bibfnamefont {E.}~\bibnamefont {Drueke}}, \bibinfo {author} {\bibfnamefont {J.}~\bibnamefont {Koo}}, \bibinfo {author} {\bibfnamefont {A.}~\bibnamefont {Kaczmarek}}, \bibinfo {author} {\bibfnamefont {C.}~\bibnamefont {Xiao}}, \bibinfo {author} {\bibfnamefont {J.}~\bibnamefont {Gao}}, \bibinfo {author} {\bibfnamefont {X.}~\bibnamefont {Luo}}, \bibinfo {author} {\bibfnamefont {Q.}~\bibnamefont {Niu}}, \bibinfo {author} {\bibfnamefont {Y.}~\bibnamefont {Sun}}, \bibinfo {author} {\bibfnamefont {B.}~\bibnamefont {Yan}}, \bibinfo {author} {\bibfnamefont {L.}~\bibnamefont {Zhao}},\ and\ \bibinfo {author} {\bibfnamefont {A.~W.}\ \bibnamefont {Tsen}},\ }\bibfield  {title} {\bibinfo {title} {{Giant c-axis nonlinear anomalous Hall effect in Td-MoTe2 and WTe2}},\ }\href
  {https://doi.org/10.1038/s41467-021-22343-5} {\bibfield  {journal} {\bibinfo  {journal} {Nat. Commun.}\ }\textbf {\bibinfo {volume} {12}},\ \bibinfo {pages} {2049} (\bibinfo {year} {2021})}\BibitemShut {NoStop}%
\bibitem [{\citenamefont {Lai}\ \emph {et~al.}(2021)\citenamefont {Lai}, \citenamefont {Liu}, \citenamefont {Zhang}, \citenamefont {Zhao}, \citenamefont {Feng}, \citenamefont {Wang}, \citenamefont {Tang}, \citenamefont {Liu}, \citenamefont {Novoselov}, \citenamefont {Yang},\ and\ \citenamefont {Gao}}]{Lai2021}%
  \BibitemOpen
  \bibfield  {author} {\bibinfo {author} {\bibfnamefont {S.}~\bibnamefont {Lai}}, \bibinfo {author} {\bibfnamefont {H.}~\bibnamefont {Liu}}, \bibinfo {author} {\bibfnamefont {Z.}~\bibnamefont {Zhang}}, \bibinfo {author} {\bibfnamefont {J.}~\bibnamefont {Zhao}}, \bibinfo {author} {\bibfnamefont {X.}~\bibnamefont {Feng}}, \bibinfo {author} {\bibfnamefont {N.}~\bibnamefont {Wang}}, \bibinfo {author} {\bibfnamefont {C.}~\bibnamefont {Tang}}, \bibinfo {author} {\bibfnamefont {Y.}~\bibnamefont {Liu}}, \bibinfo {author} {\bibfnamefont {K.~S.}\ \bibnamefont {Novoselov}}, \bibinfo {author} {\bibfnamefont {S.~A.}\ \bibnamefont {Yang}},\ and\ \bibinfo {author} {\bibfnamefont {W.-b.}\ \bibnamefont {Gao}},\ }\bibfield  {title} {\bibinfo {title} {{Third-order nonlinear Hall effect induced by the Berry-connection polarizability tensor}},\ }\href {https://doi.org/10.1038/s41565-021-00917-0} {\bibfield  {journal} {\bibinfo  {journal} {Nat. Nanotechnol.}\ }\textbf {\bibinfo {volume} {16}},\ \bibinfo {pages} {869} (\bibinfo
  {year} {2021})}\BibitemShut {NoStop}%
\bibitem [{\citenamefont {Kaplan}\ \emph {et~al.}(2024)\citenamefont {Kaplan}, \citenamefont {Holder},\ and\ \citenamefont {Yan}}]{Kaplan2024}%
  \BibitemOpen
  \bibfield  {author} {\bibinfo {author} {\bibfnamefont {D.}~\bibnamefont {Kaplan}}, \bibinfo {author} {\bibfnamefont {T.}~\bibnamefont {Holder}},\ and\ \bibinfo {author} {\bibfnamefont {B.}~\bibnamefont {Yan}},\ }\bibfield  {title} {\bibinfo {title} {{Unification of Nonlinear Anomalous Hall Effect and Nonreciprocal Magnetoresistance in Metals by the Quantum Geometry}},\ }\href {https://doi.org/10.1103/PhysRevLett.132.026301} {\bibfield  {journal} {\bibinfo  {journal} {Phys. Rev. Lett.}\ }\textbf {\bibinfo {volume} {132}},\ \bibinfo {pages} {026301} (\bibinfo {year} {2024})}\BibitemShut {NoStop}%
\bibitem [{\citenamefont {Xiao}\ \emph {et~al.}(2010)\citenamefont {Xiao}, \citenamefont {Chang},\ and\ \citenamefont {Niu}}]{Xiao2010}%
  \BibitemOpen
  \bibfield  {author} {\bibinfo {author} {\bibfnamefont {D.}~\bibnamefont {Xiao}}, \bibinfo {author} {\bibfnamefont {M.-C.}\ \bibnamefont {Chang}},\ and\ \bibinfo {author} {\bibfnamefont {Q.}~\bibnamefont {Niu}},\ }\bibfield  {title} {\bibinfo {title} {{Berry phase effects on electronic properties}},\ }\href {https://doi.org/10.1103/RevModPhys.82.1959} {\bibfield  {journal} {\bibinfo  {journal} {Rev. Mod. Phys.}\ }\textbf {\bibinfo {volume} {82}},\ \bibinfo {pages} {1959} (\bibinfo {year} {2010})}\BibitemShut {NoStop}%
\bibitem [{\citenamefont {Liu}\ \emph {et~al.}(2021)\citenamefont {Liu}, \citenamefont {Zhao}, \citenamefont {Huang}, \citenamefont {Wu}, \citenamefont {Sheng}, \citenamefont {Xiao},\ and\ \citenamefont {Yang}}]{Liu2021}%
  \BibitemOpen
  \bibfield  {author} {\bibinfo {author} {\bibfnamefont {H.}~\bibnamefont {Liu}}, \bibinfo {author} {\bibfnamefont {J.}~\bibnamefont {Zhao}}, \bibinfo {author} {\bibfnamefont {Y.-X.}\ \bibnamefont {Huang}}, \bibinfo {author} {\bibfnamefont {W.}~\bibnamefont {Wu}}, \bibinfo {author} {\bibfnamefont {X.-L.}\ \bibnamefont {Sheng}}, \bibinfo {author} {\bibfnamefont {C.}~\bibnamefont {Xiao}},\ and\ \bibinfo {author} {\bibfnamefont {S.~A.}\ \bibnamefont {Yang}},\ }\bibfield  {title} {\bibinfo {title} {{Intrinsic Second-Order Anomalous Hall Effect and Its Application in Compensated Antiferromagnets}},\ }\href {https://doi.org/10.1103/PhysRevLett.127.277202} {\bibfield  {journal} {\bibinfo  {journal} {Phys. Rev. Lett.}\ }\textbf {\bibinfo {volume} {127}},\ \bibinfo {pages} {277202} (\bibinfo {year} {2021})}\BibitemShut {NoStop}%
\bibitem [{\citenamefont {Wang}\ \emph {et~al.}(2021{\natexlab{b}})\citenamefont {Wang}, \citenamefont {Gao},\ and\ \citenamefont {Xiao}}]{Wang2021}%
  \BibitemOpen
  \bibfield  {author} {\bibinfo {author} {\bibfnamefont {C.}~\bibnamefont {Wang}}, \bibinfo {author} {\bibfnamefont {Y.}~\bibnamefont {Gao}},\ and\ \bibinfo {author} {\bibfnamefont {D.}~\bibnamefont {Xiao}},\ }\bibfield  {title} {\bibinfo {title} {{Intrinsic Nonlinear Hall Effect in Antiferromagnetic Tetragonal CuMnAs}},\ }\href {https://doi.org/10.1103/PhysRevLett.127.277201} {\bibfield  {journal} {\bibinfo  {journal} {Phys. Rev. Lett.}\ }\textbf {\bibinfo {volume} {127}},\ \bibinfo {pages} {277201} (\bibinfo {year} {2021}{\natexlab{b}})}\BibitemShut {NoStop}%
\bibitem [{\citenamefont {Kraut}\ and\ \citenamefont {von Baltz}(1979)}]{Kraut1979}%
  \BibitemOpen
  \bibfield  {author} {\bibinfo {author} {\bibfnamefont {W.}~\bibnamefont {Kraut}}\ and\ \bibinfo {author} {\bibfnamefont {R.}~\bibnamefont {von Baltz}},\ }\bibfield  {title} {\bibinfo {title} {{Anomalous bulk photovoltaic effect in ferroelectrics: A quadratic response theory}},\ }\href {https://doi.org/10.1103/PhysRevB.19.1548} {\bibfield  {journal} {\bibinfo  {journal} {Phys. Rev. B}\ }\textbf {\bibinfo {volume} {19}},\ \bibinfo {pages} {1548} (\bibinfo {year} {1979})}\BibitemShut {NoStop}%
\bibitem [{\citenamefont {Aversa}\ and\ \citenamefont {Sipe}(1995)}]{Aversa1995}%
  \BibitemOpen
  \bibfield  {author} {\bibinfo {author} {\bibfnamefont {C.}~\bibnamefont {Aversa}}\ and\ \bibinfo {author} {\bibfnamefont {J.~E.}\ \bibnamefont {Sipe}},\ }\bibfield  {title} {\bibinfo {title} {{Nonlinear optical susceptibilities of semiconductors: Results with a length-gauge analysis}},\ }\href {https://doi.org/10.1103/PhysRevB.52.14636} {\bibfield  {journal} {\bibinfo  {journal} {Phys. Rev. B}\ }\textbf {\bibinfo {volume} {52}},\ \bibinfo {pages} {14636} (\bibinfo {year} {1995})}\BibitemShut {NoStop}%
\bibitem [{\citenamefont {Luttinger}(1958)}]{Luttinger1958}%
  \BibitemOpen
  \bibfield  {author} {\bibinfo {author} {\bibfnamefont {J.~M.}\ \bibnamefont {Luttinger}},\ }\bibfield  {title} {\bibinfo {title} {{Theory of the Hall Effect in Ferromagnetic Substances}},\ }\href {https://doi.org/10.1103/PhysRev.112.739} {\bibfield  {journal} {\bibinfo  {journal} {Phys. Rev.}\ }\textbf {\bibinfo {volume} {112}},\ \bibinfo {pages} {739} (\bibinfo {year} {1958})}\BibitemShut {NoStop}%
\bibitem [{\citenamefont {Sinitsyn}\ \emph {et~al.}(2007)\citenamefont {Sinitsyn}, \citenamefont {MacDonald}, \citenamefont {Jungwirth}, \citenamefont {Dugaev},\ and\ \citenamefont {Sinova}}]{Sinitsyn2007}%
  \BibitemOpen
  \bibfield  {author} {\bibinfo {author} {\bibfnamefont {N.~A.}\ \bibnamefont {Sinitsyn}}, \bibinfo {author} {\bibfnamefont {A.~H.}\ \bibnamefont {MacDonald}}, \bibinfo {author} {\bibfnamefont {T.}~\bibnamefont {Jungwirth}}, \bibinfo {author} {\bibfnamefont {V.~K.}\ \bibnamefont {Dugaev}},\ and\ \bibinfo {author} {\bibfnamefont {J.}~\bibnamefont {Sinova}},\ }\bibfield  {title} {\bibinfo {title} {{Anomalous Hall effect in a two-dimensional Dirac band: The link between the Kubo-Streda formula and the semiclassical Boltzmann equation approach}},\ }\href {https://doi.org/10.1103/PhysRevB.75.045315} {\bibfield  {journal} {\bibinfo  {journal} {Phys. Rev. B}\ }\textbf {\bibinfo {volume} {75}},\ \bibinfo {pages} {045315} (\bibinfo {year} {2007})}\BibitemShut {NoStop}%
\bibitem [{\citenamefont {Sinitsyn}(2007)}]{Sinitsyn2008}%
  \BibitemOpen
  \bibfield  {author} {\bibinfo {author} {\bibfnamefont {N.~A.}\ \bibnamefont {Sinitsyn}},\ }\bibfield  {title} {\bibinfo {title} {{Semiclassical theories of the anomalous Hall effect}},\ }\href {https://doi.org/10.1088/0953-8984/20/02/023201} {\bibfield  {journal} {\bibinfo  {journal} {J. Phys.: Condens. Matter}\ }\textbf {\bibinfo {volume} {20}},\ \bibinfo {pages} {023201} (\bibinfo {year} {2007})}\BibitemShut {NoStop}%
\bibitem [{\citenamefont {Xiao}\ \emph {et~al.}(2019)\citenamefont {Xiao}, \citenamefont {Du},\ and\ \citenamefont {Niu}}]{Xiao2019}%
  \BibitemOpen
  \bibfield  {author} {\bibinfo {author} {\bibfnamefont {C.}~\bibnamefont {Xiao}}, \bibinfo {author} {\bibfnamefont {Z.~Z.}\ \bibnamefont {Du}},\ and\ \bibinfo {author} {\bibfnamefont {Q.}~\bibnamefont {Niu}},\ }\bibfield  {title} {\bibinfo {title} {{Theory of nonlinear Hall effects: Modified semiclassics from quantum kinetics}},\ }\href {https://doi.org/10.1103/PhysRevB.100.165422} {\bibfield  {journal} {\bibinfo  {journal} {Phys. Rev. B}\ }\textbf {\bibinfo {volume} {100}},\ \bibinfo {pages} {165422} (\bibinfo {year} {2019})}\BibitemShut {NoStop}%
\bibitem [{\citenamefont {Matsyshyn}\ and\ \citenamefont {Sodemann}(2019)}]{Matsyshyn2019}%
  \BibitemOpen
  \bibfield  {author} {\bibinfo {author} {\bibfnamefont {O.}~\bibnamefont {Matsyshyn}}\ and\ \bibinfo {author} {\bibfnamefont {I.}~\bibnamefont {Sodemann}},\ }\bibfield  {title} {\bibinfo {title} {{Nonlinear Hall Acceleration and the Quantum Rectification Sum Rule}},\ }\href {https://doi.org/10.1103/PhysRevLett.123.246602} {\bibfield  {journal} {\bibinfo  {journal} {Phys. Rev. Lett.}\ }\textbf {\bibinfo {volume} {123}},\ \bibinfo {pages} {246602} (\bibinfo {year} {2019})}\BibitemShut {NoStop}%
\bibitem [{\citenamefont {Watanabe}\ and\ \citenamefont {Yanase}(2020)}]{Watanabe2020}%
  \BibitemOpen
  \bibfield  {author} {\bibinfo {author} {\bibfnamefont {H.}~\bibnamefont {Watanabe}}\ and\ \bibinfo {author} {\bibfnamefont {Y.}~\bibnamefont {Yanase}},\ }\bibfield  {title} {\bibinfo {title} {{Nonlinear electric transport in odd-parity magnetic multipole systems: Application to Mn-based compounds}},\ }\href {https://doi.org/10.1103/PhysRevResearch.2.043081} {\bibfield  {journal} {\bibinfo  {journal} {Phys. Rev. Res.}\ }\textbf {\bibinfo {volume} {2}},\ \bibinfo {pages} {043081} (\bibinfo {year} {2020})}\BibitemShut {NoStop}%
\bibitem [{\citenamefont {Xiang}\ \emph {et~al.}(2024)\citenamefont {Xiang}, \citenamefont {Wang}, \citenamefont {Wei}, \citenamefont {Qiao},\ and\ \citenamefont {Wang}}]{Xiang2024}%
  \BibitemOpen
  \bibfield  {author} {\bibinfo {author} {\bibfnamefont {L.}~\bibnamefont {Xiang}}, \bibinfo {author} {\bibfnamefont {B.}~\bibnamefont {Wang}}, \bibinfo {author} {\bibfnamefont {Y.}~\bibnamefont {Wei}}, \bibinfo {author} {\bibfnamefont {Z.}~\bibnamefont {Qiao}},\ and\ \bibinfo {author} {\bibfnamefont {J.}~\bibnamefont {Wang}},\ }\bibfield  {title} {\bibinfo {title} {{Linear displacement current solely driven by the quantum metric}},\ }\href {https://doi.org/10.1103/PhysRevB.109.115121} {\bibfield  {journal} {\bibinfo  {journal} {Phys. Rev. B}\ }\textbf {\bibinfo {volume} {109}},\ \bibinfo {pages} {115121} (\bibinfo {year} {2024})}\BibitemShut {NoStop}%
\bibitem [{\citenamefont {Mandal}\ \emph {et~al.}(2024)\citenamefont {Mandal}, \citenamefont {Sarkar}, \citenamefont {Das},\ and\ \citenamefont {Agarwal}}]{Mandal2024}%
  \BibitemOpen
  \bibfield  {author} {\bibinfo {author} {\bibfnamefont {D.}~\bibnamefont {Mandal}}, \bibinfo {author} {\bibfnamefont {S.}~\bibnamefont {Sarkar}}, \bibinfo {author} {\bibfnamefont {K.}~\bibnamefont {Das}},\ and\ \bibinfo {author} {\bibfnamefont {A.}~\bibnamefont {Agarwal}},\ }\bibfield  {title} {\bibinfo {title} {{Quantum geometry induced third-order nonlinear transport responses}},\ }\href {https://doi.org/10.1103/PhysRevB.110.195131} {\bibfield  {journal} {\bibinfo  {journal} {Phys. Rev. B}\ }\textbf {\bibinfo {volume} {110}},\ \bibinfo {pages} {195131} (\bibinfo {year} {2024})}\BibitemShut {NoStop}%
\bibitem [{\citenamefont {Katsura}\ \emph {et~al.}(2010)\citenamefont {Katsura}, \citenamefont {Nagaosa},\ and\ \citenamefont {Lee}}]{Katsura2010}%
  \BibitemOpen
  \bibfield  {author} {\bibinfo {author} {\bibfnamefont {H.}~\bibnamefont {Katsura}}, \bibinfo {author} {\bibfnamefont {N.}~\bibnamefont {Nagaosa}},\ and\ \bibinfo {author} {\bibfnamefont {P.~A.}\ \bibnamefont {Lee}},\ }\bibfield  {title} {\bibinfo {title} {{Theory of the Thermal Hall Effect in Quantum Magnets}},\ }\href {https://doi.org/10.1103/PhysRevLett.104.066403} {\bibfield  {journal} {\bibinfo  {journal} {Phys. Rev. Lett.}\ }\textbf {\bibinfo {volume} {104}},\ \bibinfo {pages} {066403} (\bibinfo {year} {2010})}\BibitemShut {NoStop}%
\bibitem [{\citenamefont {Qin}\ \emph {et~al.}(2011)\citenamefont {Qin}, \citenamefont {Niu},\ and\ \citenamefont {Shi}}]{Qin2011}%
  \BibitemOpen
  \bibfield  {author} {\bibinfo {author} {\bibfnamefont {T.}~\bibnamefont {Qin}}, \bibinfo {author} {\bibfnamefont {Q.}~\bibnamefont {Niu}},\ and\ \bibinfo {author} {\bibfnamefont {J.}~\bibnamefont {Shi}},\ }\bibfield  {title} {\bibinfo {title} {{Energy Magnetization and the Thermal Hall Effect}},\ }\href {https://doi.org/10.1103/PhysRevLett.107.236601} {\bibfield  {journal} {\bibinfo  {journal} {Phys. Rev. Lett.}\ }\textbf {\bibinfo {volume} {107}},\ \bibinfo {pages} {236601} (\bibinfo {year} {2011})}\BibitemShut {NoStop}%
\bibitem [{\citenamefont {Matsumoto}\ and\ \citenamefont {Murakami}(2011{\natexlab{a}})}]{Matsumoto2011L}%
  \BibitemOpen
  \bibfield  {author} {\bibinfo {author} {\bibfnamefont {R.}~\bibnamefont {Matsumoto}}\ and\ \bibinfo {author} {\bibfnamefont {S.}~\bibnamefont {Murakami}},\ }\bibfield  {title} {\bibinfo {title} {{Theoretical Prediction of a Rotating Magnon Wave Packet in Ferromagnets}},\ }\href {https://doi.org/10.1103/PhysRevLett.106.197202} {\bibfield  {journal} {\bibinfo  {journal} {Phys. Rev. Lett.}\ }\textbf {\bibinfo {volume} {106}},\ \bibinfo {pages} {197202} (\bibinfo {year} {2011}{\natexlab{a}})}\BibitemShut {NoStop}%
\bibitem [{\citenamefont {Matsumoto}\ and\ \citenamefont {Murakami}(2011{\natexlab{b}})}]{Matsumoto2011B}%
  \BibitemOpen
  \bibfield  {author} {\bibinfo {author} {\bibfnamefont {R.}~\bibnamefont {Matsumoto}}\ and\ \bibinfo {author} {\bibfnamefont {S.}~\bibnamefont {Murakami}},\ }\bibfield  {title} {\bibinfo {title} {{Rotational motion of magnons and the thermal Hall effect}},\ }\href {https://doi.org/10.1103/PhysRevB.84.184406} {\bibfield  {journal} {\bibinfo  {journal} {Phys. Rev. B}\ }\textbf {\bibinfo {volume} {84}},\ \bibinfo {pages} {184406} (\bibinfo {year} {2011}{\natexlab{b}})}\BibitemShut {NoStop}%
\bibitem [{\citenamefont {Qin}\ \emph {et~al.}(2012)\citenamefont {Qin}, \citenamefont {Zhou},\ and\ \citenamefont {Shi}}]{Qin2012}%
  \BibitemOpen
  \bibfield  {author} {\bibinfo {author} {\bibfnamefont {T.}~\bibnamefont {Qin}}, \bibinfo {author} {\bibfnamefont {J.}~\bibnamefont {Zhou}},\ and\ \bibinfo {author} {\bibfnamefont {J.}~\bibnamefont {Shi}},\ }\bibfield  {title} {\bibinfo {title} {{Berry curvature and the phonon Hall effect}},\ }\href {https://doi.org/10.1103/PhysRevB.86.104305} {\bibfield  {journal} {\bibinfo  {journal} {Phys. Rev. B}\ }\textbf {\bibinfo {volume} {86}},\ \bibinfo {pages} {104305} (\bibinfo {year} {2012})}\BibitemShut {NoStop}%
\bibitem [{\citenamefont {Fujiwara}\ \emph {et~al.}(2022)\citenamefont {Fujiwara}, \citenamefont {Kitamura},\ and\ \citenamefont {Morimoto}}]{Fujiwara2022}%
  \BibitemOpen
  \bibfield  {author} {\bibinfo {author} {\bibfnamefont {K.}~\bibnamefont {Fujiwara}}, \bibinfo {author} {\bibfnamefont {S.}~\bibnamefont {Kitamura}},\ and\ \bibinfo {author} {\bibfnamefont {T.}~\bibnamefont {Morimoto}},\ }\bibfield  {title} {\bibinfo {title} {{Thermal Hall responses in frustrated honeycomb spin systems}},\ }\href {https://doi.org/10.1103/PhysRevB.106.035113} {\bibfield  {journal} {\bibinfo  {journal} {Phys. Rev. B}\ }\textbf {\bibinfo {volume} {106}},\ \bibinfo {pages} {035113} (\bibinfo {year} {2022})}\BibitemShut {NoStop}%
\bibitem [{\citenamefont {Onose}\ \emph {et~al.}(2010)\citenamefont {Onose}, \citenamefont {Ideue}, \citenamefont {Katsura}, \citenamefont {Shiomi}, \citenamefont {Nagaosa},\ and\ \citenamefont {Tokura}}]{Onose2010}%
  \BibitemOpen
  \bibfield  {author} {\bibinfo {author} {\bibfnamefont {Y.}~\bibnamefont {Onose}}, \bibinfo {author} {\bibfnamefont {T.}~\bibnamefont {Ideue}}, \bibinfo {author} {\bibfnamefont {H.}~\bibnamefont {Katsura}}, \bibinfo {author} {\bibfnamefont {Y.}~\bibnamefont {Shiomi}}, \bibinfo {author} {\bibfnamefont {N.}~\bibnamefont {Nagaosa}},\ and\ \bibinfo {author} {\bibfnamefont {Y.}~\bibnamefont {Tokura}},\ }\bibfield  {title} {\bibinfo {title} {{Observation of the Magnon Hall Effect}},\ }\href {https://doi.org/10.1126/science.1188260} {\bibfield  {journal} {\bibinfo  {journal} {Science}\ }\textbf {\bibinfo {volume} {329}},\ \bibinfo {pages} {297} (\bibinfo {year} {2010})}\BibitemShut {NoStop}%
\bibitem [{\citenamefont {Ideue}\ \emph {et~al.}(2012)\citenamefont {Ideue}, \citenamefont {Onose}, \citenamefont {Katsura}, \citenamefont {Shiomi}, \citenamefont {Ishiwata}, \citenamefont {Nagaosa},\ and\ \citenamefont {Tokura}}]{Ideue2012}%
  \BibitemOpen
  \bibfield  {author} {\bibinfo {author} {\bibfnamefont {T.}~\bibnamefont {Ideue}}, \bibinfo {author} {\bibfnamefont {Y.}~\bibnamefont {Onose}}, \bibinfo {author} {\bibfnamefont {H.}~\bibnamefont {Katsura}}, \bibinfo {author} {\bibfnamefont {Y.}~\bibnamefont {Shiomi}}, \bibinfo {author} {\bibfnamefont {S.}~\bibnamefont {Ishiwata}}, \bibinfo {author} {\bibfnamefont {N.}~\bibnamefont {Nagaosa}},\ and\ \bibinfo {author} {\bibfnamefont {Y.}~\bibnamefont {Tokura}},\ }\bibfield  {title} {\bibinfo {title} {{Effect of lattice geometry on magnon Hall effect in ferromagnetic insulators}},\ }\href {https://doi.org/10.1103/PhysRevB.85.134411} {\bibfield  {journal} {\bibinfo  {journal} {Phys. Rev. B}\ }\textbf {\bibinfo {volume} {85}},\ \bibinfo {pages} {134411} (\bibinfo {year} {2012})}\BibitemShut {NoStop}%
\bibitem [{\citenamefont {Hirschberger}\ \emph {et~al.}(2015{\natexlab{a}})\citenamefont {Hirschberger}, \citenamefont {Chisnell}, \citenamefont {Lee},\ and\ \citenamefont {Ong}}]{Hirschberger2015L}%
  \BibitemOpen
  \bibfield  {author} {\bibinfo {author} {\bibfnamefont {M.}~\bibnamefont {Hirschberger}}, \bibinfo {author} {\bibfnamefont {R.}~\bibnamefont {Chisnell}}, \bibinfo {author} {\bibfnamefont {Y.~S.}\ \bibnamefont {Lee}},\ and\ \bibinfo {author} {\bibfnamefont {N.~P.}\ \bibnamefont {Ong}},\ }\bibfield  {title} {\bibinfo {title} {{Thermal Hall Effect of Spin Excitations in a Kagome Magnet}},\ }\href {https://doi.org/10.1103/PhysRevLett.115.106603} {\bibfield  {journal} {\bibinfo  {journal} {Phys. Rev. Lett.}\ }\textbf {\bibinfo {volume} {115}},\ \bibinfo {pages} {106603} (\bibinfo {year} {2015}{\natexlab{a}})}\BibitemShut {NoStop}%
\bibitem [{\citenamefont {Hirschberger}\ \emph {et~al.}(2015{\natexlab{b}})\citenamefont {Hirschberger}, \citenamefont {Krizan}, \citenamefont {Cava},\ and\ \citenamefont {Ong}}]{Hirschberger2015Science}%
  \BibitemOpen
  \bibfield  {author} {\bibinfo {author} {\bibfnamefont {M.}~\bibnamefont {Hirschberger}}, \bibinfo {author} {\bibfnamefont {J.~W.}\ \bibnamefont {Krizan}}, \bibinfo {author} {\bibfnamefont {R.~J.}\ \bibnamefont {Cava}},\ and\ \bibinfo {author} {\bibfnamefont {N.~P.}\ \bibnamefont {Ong}},\ }\bibfield  {title} {\bibinfo {title} {{Large thermal Hall conductivity of neutral spin excitations in a frustrated quantum magnet}},\ }\href {https://doi.org/10.1126/science.1257340} {\bibfield  {journal} {\bibinfo  {journal} {Science}\ }\textbf {\bibinfo {volume} {348}},\ \bibinfo {pages} {106} (\bibinfo {year} {2015}{\natexlab{b}})},\ \Eprint {https://arxiv.org/abs/https://www.science.org/doi/pdf/10.1126/science.1257340} {https://www.science.org/doi/pdf/10.1126/science.1257340} \BibitemShut {NoStop}%
\bibitem [{\citenamefont {McClarty}\ \emph {et~al.}(2018)\citenamefont {McClarty}, \citenamefont {Dong}, \citenamefont {Gohlke}, \citenamefont {Rau}, \citenamefont {Pollmann}, \citenamefont {Moessner},\ and\ \citenamefont {Penc}}]{Mcclarty2018}%
  \BibitemOpen
  \bibfield  {author} {\bibinfo {author} {\bibfnamefont {P.~A.}\ \bibnamefont {McClarty}}, \bibinfo {author} {\bibfnamefont {X.-Y.}\ \bibnamefont {Dong}}, \bibinfo {author} {\bibfnamefont {M.}~\bibnamefont {Gohlke}}, \bibinfo {author} {\bibfnamefont {J.~G.}\ \bibnamefont {Rau}}, \bibinfo {author} {\bibfnamefont {F.}~\bibnamefont {Pollmann}}, \bibinfo {author} {\bibfnamefont {R.}~\bibnamefont {Moessner}},\ and\ \bibinfo {author} {\bibfnamefont {K.}~\bibnamefont {Penc}},\ }\bibfield  {title} {\bibinfo {title} {Topological magnons in {K}itaev magnets at high fields},\ }\href {https://doi.org/10.1103/PhysRevB.98.060404} {\bibfield  {journal} {\bibinfo  {journal} {Phys. Rev. B}\ }\textbf {\bibinfo {volume} {98}},\ \bibinfo {pages} {060404} (\bibinfo {year} {2018})}\BibitemShut {NoStop}%
\bibitem [{\citenamefont {Joshi}(2018)}]{Joshi2018}%
  \BibitemOpen
  \bibfield  {author} {\bibinfo {author} {\bibfnamefont {D.~G.}\ \bibnamefont {Joshi}},\ }\bibfield  {title} {\bibinfo {title} {Topological excitations in the ferromagnetic {K}itaev-{H}eisenberg model},\ }\href {https://doi.org/10.1103/PhysRevB.98.060405} {\bibfield  {journal} {\bibinfo  {journal} {Phys. Rev. B}\ }\textbf {\bibinfo {volume} {98}},\ \bibinfo {pages} {060405} (\bibinfo {year} {2018})}\BibitemShut {NoStop}%
\bibitem [{\citenamefont {Akazawa}\ \emph {et~al.}(2020)\citenamefont {Akazawa}, \citenamefont {Shimozawa}, \citenamefont {Kittaka}, \citenamefont {Sakakibara}, \citenamefont {Okuma}, \citenamefont {Hiroi}, \citenamefont {Lee}, \citenamefont {Kawashima}, \citenamefont {Han},\ and\ \citenamefont {Yamashita}}]{Akazawa2020}%
  \BibitemOpen
  \bibfield  {author} {\bibinfo {author} {\bibfnamefont {M.}~\bibnamefont {Akazawa}}, \bibinfo {author} {\bibfnamefont {M.}~\bibnamefont {Shimozawa}}, \bibinfo {author} {\bibfnamefont {S.}~\bibnamefont {Kittaka}}, \bibinfo {author} {\bibfnamefont {T.}~\bibnamefont {Sakakibara}}, \bibinfo {author} {\bibfnamefont {R.}~\bibnamefont {Okuma}}, \bibinfo {author} {\bibfnamefont {Z.}~\bibnamefont {Hiroi}}, \bibinfo {author} {\bibfnamefont {H.-Y.}\ \bibnamefont {Lee}}, \bibinfo {author} {\bibfnamefont {N.}~\bibnamefont {Kawashima}}, \bibinfo {author} {\bibfnamefont {J.~H.}\ \bibnamefont {Han}},\ and\ \bibinfo {author} {\bibfnamefont {M.}~\bibnamefont {Yamashita}},\ }\bibfield  {title} {\bibinfo {title} {{Thermal Hall Effects of Spins and Phonons in Kagome Antiferromagnet Cd-Kapellasite}},\ }\href {https://doi.org/10.1103/PhysRevX.10.041059} {\bibfield  {journal} {\bibinfo  {journal} {Phys. Rev. X}\ }\textbf {\bibinfo {volume} {10}},\ \bibinfo {pages} {041059} (\bibinfo {year} {2020})}\BibitemShut {NoStop}%
\bibitem [{\citenamefont {Koyama}\ and\ \citenamefont {Nasu}(2021)}]{Koyama2021}%
  \BibitemOpen
  \bibfield  {author} {\bibinfo {author} {\bibfnamefont {S.}~\bibnamefont {Koyama}}\ and\ \bibinfo {author} {\bibfnamefont {J.}~\bibnamefont {Nasu}},\ }\bibfield  {title} {\bibinfo {title} {Field-angle dependence of thermal {H}all conductivity in a magnetically ordered {K}itaev-{H}eisenberg system},\ }\href {https://doi.org/10.1103/PhysRevB.104.075121} {\bibfield  {journal} {\bibinfo  {journal} {Phys. Rev. B}\ }\textbf {\bibinfo {volume} {104}},\ \bibinfo {pages} {075121} (\bibinfo {year} {2021})}\BibitemShut {NoStop}%
\bibitem [{\citenamefont {Zhang}\ \emph {et~al.}(2021)\citenamefont {Zhang}, \citenamefont {Xu}, \citenamefont {Carnahan}, \citenamefont {Sretenovic}, \citenamefont {Suri}, \citenamefont {Xiao},\ and\ \citenamefont {Ke}}]{Zhang2021L}%
  \BibitemOpen
  \bibfield  {author} {\bibinfo {author} {\bibfnamefont {H.}~\bibnamefont {Zhang}}, \bibinfo {author} {\bibfnamefont {C.}~\bibnamefont {Xu}}, \bibinfo {author} {\bibfnamefont {C.}~\bibnamefont {Carnahan}}, \bibinfo {author} {\bibfnamefont {M.}~\bibnamefont {Sretenovic}}, \bibinfo {author} {\bibfnamefont {N.}~\bibnamefont {Suri}}, \bibinfo {author} {\bibfnamefont {D.}~\bibnamefont {Xiao}},\ and\ \bibinfo {author} {\bibfnamefont {X.}~\bibnamefont {Ke}},\ }\bibfield  {title} {\bibinfo {title} {Anomalous {T}hermal {H}all {E}ffect in an {I}nsulating van der {W}aals {M}agnet},\ }\href {https://doi.org/10.1103/PhysRevLett.127.247202} {\bibfield  {journal} {\bibinfo  {journal} {Phys. Rev. Lett.}\ }\textbf {\bibinfo {volume} {127}},\ \bibinfo {pages} {247202} (\bibinfo {year} {2021})}\BibitemShut {NoStop}%
\bibitem [{\citenamefont {Czajka}\ \emph {et~al.}(2023)\citenamefont {Czajka}, \citenamefont {Gao}, \citenamefont {Hirschberger}, \citenamefont {Lampen-Kelley}, \citenamefont {Banerjee}, \citenamefont {Quirk}, \citenamefont {Mandrus}, \citenamefont {Nagler},\ and\ \citenamefont {Ong}}]{Czajka2023}%
  \BibitemOpen
  \bibfield  {author} {\bibinfo {author} {\bibfnamefont {P.}~\bibnamefont {Czajka}}, \bibinfo {author} {\bibfnamefont {T.}~\bibnamefont {Gao}}, \bibinfo {author} {\bibfnamefont {M.}~\bibnamefont {Hirschberger}}, \bibinfo {author} {\bibfnamefont {P.}~\bibnamefont {Lampen-Kelley}}, \bibinfo {author} {\bibfnamefont {A.}~\bibnamefont {Banerjee}}, \bibinfo {author} {\bibfnamefont {N.}~\bibnamefont {Quirk}}, \bibinfo {author} {\bibfnamefont {D.~G.}\ \bibnamefont {Mandrus}}, \bibinfo {author} {\bibfnamefont {S.~E.}\ \bibnamefont {Nagler}},\ and\ \bibinfo {author} {\bibfnamefont {N.~P.}\ \bibnamefont {Ong}},\ }\bibfield  {title} {\bibinfo {title} {Planar thermal {H}all effect of topological bosons in the {K}itaev magnet $\alpha$-{RuCl}$_3$},\ }\href {https://doi.org/10.1038/s41563-022-01397-w} {\bibfield  {journal} {\bibinfo  {journal} {Nat. Mater.}\ }\textbf {\bibinfo {volume} {22}},\ \bibinfo {pages} {36} (\bibinfo {year} {2023})}\BibitemShut {NoStop}%
\bibitem [{\citenamefont {Kitaev}(2006)}]{Kitaev2006}%
  \BibitemOpen
  \bibfield  {author} {\bibinfo {author} {\bibfnamefont {A.}~\bibnamefont {Kitaev}},\ }\bibfield  {title} {\bibinfo {title} {{Anyons in an exactly solved model and beyond}},\ }\href {https://doi.org/https://doi.org/10.1016/j.aop.2005.10.005} {\bibfield  {journal} {\bibinfo  {journal} {Ann. Phys.}\ }\textbf {\bibinfo {volume} {321}},\ \bibinfo {pages} {2} (\bibinfo {year} {2006})}\BibitemShut {NoStop}%
\bibitem [{\citenamefont {Nasu}\ \emph {et~al.}(2017)\citenamefont {Nasu}, \citenamefont {Yoshitake},\ and\ \citenamefont {Motome}}]{Nasu2017}%
  \BibitemOpen
  \bibfield  {author} {\bibinfo {author} {\bibfnamefont {J.}~\bibnamefont {Nasu}}, \bibinfo {author} {\bibfnamefont {J.}~\bibnamefont {Yoshitake}},\ and\ \bibinfo {author} {\bibfnamefont {Y.}~\bibnamefont {Motome}},\ }\bibfield  {title} {\bibinfo {title} {{Thermal Transport in the Kitaev Model}},\ }\href {https://doi.org/10.1103/PhysRevLett.119.127204} {\bibfield  {journal} {\bibinfo  {journal} {Phys. Rev. Lett.}\ }\textbf {\bibinfo {volume} {119}},\ \bibinfo {pages} {127204} (\bibinfo {year} {2017})}\BibitemShut {NoStop}%
\bibitem [{\citenamefont {Kasahara}\ \emph {et~al.}(2018{\natexlab{a}})\citenamefont {Kasahara}, \citenamefont {Sugii}, \citenamefont {Ohnishi}, \citenamefont {Shimozawa}, \citenamefont {Yamashita}, \citenamefont {Kurita}, \citenamefont {Tanaka}, \citenamefont {Nasu}, \citenamefont {Motome}, \citenamefont {Shibauchi},\ and\ \citenamefont {Matsuda}}]{Kasahara2018L}%
  \BibitemOpen
  \bibfield  {author} {\bibinfo {author} {\bibfnamefont {Y.}~\bibnamefont {Kasahara}}, \bibinfo {author} {\bibfnamefont {K.}~\bibnamefont {Sugii}}, \bibinfo {author} {\bibfnamefont {T.}~\bibnamefont {Ohnishi}}, \bibinfo {author} {\bibfnamefont {M.}~\bibnamefont {Shimozawa}}, \bibinfo {author} {\bibfnamefont {M.}~\bibnamefont {Yamashita}}, \bibinfo {author} {\bibfnamefont {N.}~\bibnamefont {Kurita}}, \bibinfo {author} {\bibfnamefont {H.}~\bibnamefont {Tanaka}}, \bibinfo {author} {\bibfnamefont {J.}~\bibnamefont {Nasu}}, \bibinfo {author} {\bibfnamefont {Y.}~\bibnamefont {Motome}}, \bibinfo {author} {\bibfnamefont {T.}~\bibnamefont {Shibauchi}},\ and\ \bibinfo {author} {\bibfnamefont {Y.}~\bibnamefont {Matsuda}},\ }\bibfield  {title} {\bibinfo {title} {{Unusual Thermal Hall Effect in a Kitaev Spin Liquid Candidate $\alpha$-${\mathrm{RuCl}}_{3}$}},\ }\href {https://doi.org/10.1103/PhysRevLett.120.217205} {\bibfield  {journal} {\bibinfo  {journal} {Phys. Rev. Lett.}\ }\textbf {\bibinfo {volume} {120}},\ \bibinfo
  {pages} {217205} (\bibinfo {year} {2018}{\natexlab{a}})}\BibitemShut {NoStop}%
\bibitem [{\citenamefont {Kasahara}\ \emph {et~al.}(2018{\natexlab{b}})\citenamefont {Kasahara}, \citenamefont {Ohnishi}, \citenamefont {Mizukami}, \citenamefont {Tanaka}, \citenamefont {Ma}, \citenamefont {Sugii}, \citenamefont {Kurita}, \citenamefont {Tanaka}, \citenamefont {Nasu}, \citenamefont {Motome}, \citenamefont {Shibauchi},\ and\ \citenamefont {Matsuda}}]{Kasahara2018N}%
  \BibitemOpen
  \bibfield  {author} {\bibinfo {author} {\bibfnamefont {Y.}~\bibnamefont {Kasahara}}, \bibinfo {author} {\bibfnamefont {T.}~\bibnamefont {Ohnishi}}, \bibinfo {author} {\bibfnamefont {Y.}~\bibnamefont {Mizukami}}, \bibinfo {author} {\bibfnamefont {O.}~\bibnamefont {Tanaka}}, \bibinfo {author} {\bibfnamefont {S.}~\bibnamefont {Ma}}, \bibinfo {author} {\bibfnamefont {K.}~\bibnamefont {Sugii}}, \bibinfo {author} {\bibfnamefont {N.}~\bibnamefont {Kurita}}, \bibinfo {author} {\bibfnamefont {H.}~\bibnamefont {Tanaka}}, \bibinfo {author} {\bibfnamefont {J.}~\bibnamefont {Nasu}}, \bibinfo {author} {\bibfnamefont {Y.}~\bibnamefont {Motome}}, \bibinfo {author} {\bibfnamefont {T.}~\bibnamefont {Shibauchi}},\ and\ \bibinfo {author} {\bibfnamefont {Y.}~\bibnamefont {Matsuda}},\ }\bibfield  {title} {\bibinfo {title} {{Majorana quantization and half-integer thermal quantum Hall effect in a Kitaev spin liquid}},\ }\href {https://doi.org/10.1038/s41586-018-0274-0} {\bibfield  {journal} {\bibinfo  {journal} {Nature}\ }\textbf
  {\bibinfo {volume} {559}},\ \bibinfo {pages} {227} (\bibinfo {year} {2018}{\natexlab{b}})}\BibitemShut {NoStop}%
\bibitem [{\citenamefont {Ye}\ \emph {et~al.}(2018)\citenamefont {Ye}, \citenamefont {Hal\'asz}, \citenamefont {Savary},\ and\ \citenamefont {Balents}}]{Ye2018}%
  \BibitemOpen
  \bibfield  {author} {\bibinfo {author} {\bibfnamefont {M.}~\bibnamefont {Ye}}, \bibinfo {author} {\bibfnamefont {G.~B.}\ \bibnamefont {Hal\'asz}}, \bibinfo {author} {\bibfnamefont {L.}~\bibnamefont {Savary}},\ and\ \bibinfo {author} {\bibfnamefont {L.}~\bibnamefont {Balents}},\ }\bibfield  {title} {\bibinfo {title} {Quantization of the thermal hall conductivity at small hall angles},\ }\href {https://doi.org/10.1103/PhysRevLett.121.147201} {\bibfield  {journal} {\bibinfo  {journal} {Phys. Rev. Lett.}\ }\textbf {\bibinfo {volume} {121}},\ \bibinfo {pages} {147201} (\bibinfo {year} {2018})}\BibitemShut {NoStop}%
\bibitem [{\citenamefont {Yokoi}\ \emph {et~al.}(2021)\citenamefont {Yokoi}, \citenamefont {Ma}, \citenamefont {Kasahara}, \citenamefont {Kasahara}, \citenamefont {Shibauchi}, \citenamefont {Kurita}, \citenamefont {Tanaka}, \citenamefont {Nasu}, \citenamefont {Motome}, \citenamefont {Hickey}, \citenamefont {Trebst},\ and\ \citenamefont {Matsuda}}]{Yokoi2021}%
  \BibitemOpen
  \bibfield  {author} {\bibinfo {author} {\bibfnamefont {T.}~\bibnamefont {Yokoi}}, \bibinfo {author} {\bibfnamefont {S.}~\bibnamefont {Ma}}, \bibinfo {author} {\bibfnamefont {Y.}~\bibnamefont {Kasahara}}, \bibinfo {author} {\bibfnamefont {S.}~\bibnamefont {Kasahara}}, \bibinfo {author} {\bibfnamefont {T.}~\bibnamefont {Shibauchi}}, \bibinfo {author} {\bibfnamefont {N.}~\bibnamefont {Kurita}}, \bibinfo {author} {\bibfnamefont {H.}~\bibnamefont {Tanaka}}, \bibinfo {author} {\bibfnamefont {J.}~\bibnamefont {Nasu}}, \bibinfo {author} {\bibfnamefont {Y.}~\bibnamefont {Motome}}, \bibinfo {author} {\bibfnamefont {C.}~\bibnamefont {Hickey}}, \bibinfo {author} {\bibfnamefont {S.}~\bibnamefont {Trebst}},\ and\ \bibinfo {author} {\bibfnamefont {Y.}~\bibnamefont {Matsuda}},\ }\bibfield  {title} {\bibinfo {title} {{Half-integer quantized anomalous thermal Hall effect in the Kitaev material candidate $\alpha$-RuCl$_3$}},\ }\href {https://doi.org/10.1126/science.aay5551} {\bibfield  {journal} {\bibinfo  {journal}
  {Science}\ }\textbf {\bibinfo {volume} {373}},\ \bibinfo {pages} {568} (\bibinfo {year} {2021})}\BibitemShut {NoStop}%
\bibitem [{\citenamefont {Hwang}\ \emph {et~al.}(2022)\citenamefont {Hwang}, \citenamefont {Go}, \citenamefont {Seong}, \citenamefont {Shibauchi},\ and\ \citenamefont {Moon}}]{Hwang2022}%
  \BibitemOpen
  \bibfield  {author} {\bibinfo {author} {\bibfnamefont {K.}~\bibnamefont {Hwang}}, \bibinfo {author} {\bibfnamefont {A.}~\bibnamefont {Go}}, \bibinfo {author} {\bibfnamefont {J.~H.}\ \bibnamefont {Seong}}, \bibinfo {author} {\bibfnamefont {T.}~\bibnamefont {Shibauchi}},\ and\ \bibinfo {author} {\bibfnamefont {E.-G.}\ \bibnamefont {Moon}},\ }\bibfield  {title} {\bibinfo {title} {Identification of a kitaev quantum spin liquid by magnetic field angle dependence},\ }\href {https://doi.org/10.1038/s41467-021-27943-9} {\bibfield  {journal} {\bibinfo  {journal} {Nat. Commun.}\ }\textbf {\bibinfo {volume} {13}},\ \bibinfo {pages} {1} (\bibinfo {year} {2022})}\BibitemShut {NoStop}%
\bibitem [{\citenamefont {Imamura}\ \emph {et~al.}(2024)\citenamefont {Imamura}, \citenamefont {Suetsugu}, \citenamefont {Mizukami}, \citenamefont {Yoshida}, \citenamefont {Hashimoto}, \citenamefont {Ohtsuka}, \citenamefont {Kasahara}, \citenamefont {Kurita}, \citenamefont {Tanaka}, \citenamefont {Noh}, \citenamefont {Nasu}, \citenamefont {Moon}, \citenamefont {Matsuda},\ and\ \citenamefont {Shibauchi}}]{Imamura2024}%
  \BibitemOpen
  \bibfield  {author} {\bibinfo {author} {\bibfnamefont {K.}~\bibnamefont {Imamura}}, \bibinfo {author} {\bibfnamefont {S.}~\bibnamefont {Suetsugu}}, \bibinfo {author} {\bibfnamefont {Y.}~\bibnamefont {Mizukami}}, \bibinfo {author} {\bibfnamefont {Y.}~\bibnamefont {Yoshida}}, \bibinfo {author} {\bibfnamefont {K.}~\bibnamefont {Hashimoto}}, \bibinfo {author} {\bibfnamefont {K.}~\bibnamefont {Ohtsuka}}, \bibinfo {author} {\bibfnamefont {Y.}~\bibnamefont {Kasahara}}, \bibinfo {author} {\bibfnamefont {N.}~\bibnamefont {Kurita}}, \bibinfo {author} {\bibfnamefont {H.}~\bibnamefont {Tanaka}}, \bibinfo {author} {\bibfnamefont {P.}~\bibnamefont {Noh}}, \bibinfo {author} {\bibfnamefont {J.}~\bibnamefont {Nasu}}, \bibinfo {author} {\bibfnamefont {E.-G.}\ \bibnamefont {Moon}}, \bibinfo {author} {\bibfnamefont {Y.}~\bibnamefont {Matsuda}},\ and\ \bibinfo {author} {\bibfnamefont {T.}~\bibnamefont {Shibauchi}},\ }\bibfield  {title} {\bibinfo {title} {{Majorana-fermion origin of the planar thermal Hall effect in the Kitaev
  magnet $\alpha$-RuCl$_3$}},\ }\href {https://doi.org/10.1126/sciadv.adk3539} {\bibfield  {journal} {\bibinfo  {journal} {Sci. Adv.}\ }\textbf {\bibinfo {volume} {10}},\ \bibinfo {pages} {eadk3539} (\bibinfo {year} {2024})}\BibitemShut {NoStop}%
\bibitem [{\citenamefont {Sheng}\ \emph {et~al.}(2006)\citenamefont {Sheng}, \citenamefont {Sheng},\ and\ \citenamefont {Ting}}]{Sheng2006}%
  \BibitemOpen
  \bibfield  {author} {\bibinfo {author} {\bibfnamefont {L.}~\bibnamefont {Sheng}}, \bibinfo {author} {\bibfnamefont {D.~N.}\ \bibnamefont {Sheng}},\ and\ \bibinfo {author} {\bibfnamefont {C.~S.}\ \bibnamefont {Ting}},\ }\bibfield  {title} {\bibinfo {title} {{Theory of the Phonon Hall Effect in Paramagnetic Dielectrics}},\ }\href {https://doi.org/10.1103/PhysRevLett.96.155901} {\bibfield  {journal} {\bibinfo  {journal} {Phys. Rev. Lett.}\ }\textbf {\bibinfo {volume} {96}},\ \bibinfo {pages} {155901} (\bibinfo {year} {2006})}\BibitemShut {NoStop}%
\bibitem [{\citenamefont {Zhang}\ \emph {et~al.}(2010)\citenamefont {Zhang}, \citenamefont {Ren}, \citenamefont {Wang},\ and\ \citenamefont {Li}}]{Zhang2010}%
  \BibitemOpen
  \bibfield  {author} {\bibinfo {author} {\bibfnamefont {L.}~\bibnamefont {Zhang}}, \bibinfo {author} {\bibfnamefont {J.}~\bibnamefont {Ren}}, \bibinfo {author} {\bibfnamefont {J.-S.}\ \bibnamefont {Wang}},\ and\ \bibinfo {author} {\bibfnamefont {B.}~\bibnamefont {Li}},\ }\bibfield  {title} {\bibinfo {title} {{Topological Nature of the Phonon Hall Effect}},\ }\href {https://doi.org/10.1103/PhysRevLett.105.225901} {\bibfield  {journal} {\bibinfo  {journal} {Phys. Rev. Lett.}\ }\textbf {\bibinfo {volume} {105}},\ \bibinfo {pages} {225901} (\bibinfo {year} {2010})}\BibitemShut {NoStop}%
\bibitem [{\citenamefont {Zhang}\ \emph {et~al.}(2011)\citenamefont {Zhang}, \citenamefont {Ren}, \citenamefont {Wang},\ and\ \citenamefont {Li}}]{Zhang2011}%
  \BibitemOpen
  \bibfield  {author} {\bibinfo {author} {\bibfnamefont {L.}~\bibnamefont {Zhang}}, \bibinfo {author} {\bibfnamefont {J.}~\bibnamefont {Ren}}, \bibinfo {author} {\bibfnamefont {J.-S.}\ \bibnamefont {Wang}},\ and\ \bibinfo {author} {\bibfnamefont {B.}~\bibnamefont {Li}},\ }\bibfield  {title} {\bibinfo {title} {The phonon hall effect: theory and application},\ }\href {https://doi.org/10.1088/0953-8984/23/30/305402} {\bibfield  {journal} {\bibinfo  {journal} {J. Phys.: Condens. Matter}\ }\textbf {\bibinfo {volume} {23}},\ \bibinfo {pages} {305402} (\bibinfo {year} {2011})}\BibitemShut {NoStop}%
\bibitem [{\citenamefont {Vinkler-Aviv}\ and\ \citenamefont {Rosch}(2018)}]{Vinkler-Aviv2018}%
  \BibitemOpen
  \bibfield  {author} {\bibinfo {author} {\bibfnamefont {Y.}~\bibnamefont {Vinkler-Aviv}}\ and\ \bibinfo {author} {\bibfnamefont {A.}~\bibnamefont {Rosch}},\ }\bibfield  {title} {\bibinfo {title} {Approximately quantized thermal hall effect of chiral liquids coupled to phonons},\ }\href {https://doi.org/10.1103/PhysRevX.8.031032} {\bibfield  {journal} {\bibinfo  {journal} {Phys. Rev. X}\ }\textbf {\bibinfo {volume} {8}},\ \bibinfo {pages} {031032} (\bibinfo {year} {2018})}\BibitemShut {NoStop}%
\bibitem [{\citenamefont {Chen}\ \emph {et~al.}(2022)\citenamefont {Chen}, \citenamefont {Boulanger}, \citenamefont {Wang}, \citenamefont {Tafti},\ and\ \citenamefont {Taillefer}}]{Chen2022}%
  \BibitemOpen
  \bibfield  {author} {\bibinfo {author} {\bibfnamefont {L.}~\bibnamefont {Chen}}, \bibinfo {author} {\bibfnamefont {M.-E.}\ \bibnamefont {Boulanger}}, \bibinfo {author} {\bibfnamefont {Z.-C.}\ \bibnamefont {Wang}}, \bibinfo {author} {\bibfnamefont {F.}~\bibnamefont {Tafti}},\ and\ \bibinfo {author} {\bibfnamefont {L.}~\bibnamefont {Taillefer}},\ }\bibfield  {title} {\bibinfo {title} {{Large phonon thermal Hall conductivity in the antiferromagnetic insulator Cu$_3$TeO$_6$}},\ }\href {https://doi.org/10.1073/pnas.2208016119} {\bibfield  {journal} {\bibinfo  {journal} {Proc. Natl. Acad. Sci.}\ }\textbf {\bibinfo {volume} {119}},\ \bibinfo {pages} {e2208016119} (\bibinfo {year} {2022})}\BibitemShut {NoStop}%
\bibitem [{\citenamefont {Chen}\ \emph {et~al.}(2024)\citenamefont {Chen}, \citenamefont {Lefran{\c{c}}ois}, \citenamefont {Vallipuram}, \citenamefont {Barth{\'e}lemy}, \citenamefont {Ataei}, \citenamefont {Yao}, \citenamefont {Li},\ and\ \citenamefont {Taillefer}}]{Chen2024}%
  \BibitemOpen
  \bibfield  {author} {\bibinfo {author} {\bibfnamefont {L.}~\bibnamefont {Chen}}, \bibinfo {author} {\bibfnamefont {{\'E}.}~\bibnamefont {Lefran{\c{c}}ois}}, \bibinfo {author} {\bibfnamefont {A.}~\bibnamefont {Vallipuram}}, \bibinfo {author} {\bibfnamefont {Q.}~\bibnamefont {Barth{\'e}lemy}}, \bibinfo {author} {\bibfnamefont {A.}~\bibnamefont {Ataei}}, \bibinfo {author} {\bibfnamefont {W.}~\bibnamefont {Yao}}, \bibinfo {author} {\bibfnamefont {Y.}~\bibnamefont {Li}},\ and\ \bibinfo {author} {\bibfnamefont {L.}~\bibnamefont {Taillefer}},\ }\bibfield  {title} {\bibinfo {title} {{Planar thermal Hall effect from phonons in a Kitaev candidate material}},\ }\href {https://doi.org/10.1038/s41467-024-47858-5} {\bibfield  {journal} {\bibinfo  {journal} {Nat. Commun.}\ }\textbf {\bibinfo {volume} {15}},\ \bibinfo {pages} {3513} (\bibinfo {year} {2024})}\BibitemShut {NoStop}%
\bibitem [{\citenamefont {Sharma}\ \emph {et~al.}(2024)\citenamefont {Sharma}, \citenamefont {Valldor},\ and\ \citenamefont {Lorenz}}]{Sharma2024}%
  \BibitemOpen
  \bibfield  {author} {\bibinfo {author} {\bibfnamefont {R.}~\bibnamefont {Sharma}}, \bibinfo {author} {\bibfnamefont {M.}~\bibnamefont {Valldor}},\ and\ \bibinfo {author} {\bibfnamefont {T.}~\bibnamefont {Lorenz}},\ }\bibfield  {title} {\bibinfo {title} {{Phonon thermal Hall effect in nonmagnetic ${\mathrm{Y}}_{2}{\mathrm{Ti}}_{2}{\mathrm{O}}_{7}$}},\ }\href {https://doi.org/10.1103/PhysRevB.110.L100301} {\bibfield  {journal} {\bibinfo  {journal} {Phys. Rev. B}\ }\textbf {\bibinfo {volume} {110}},\ \bibinfo {pages} {L100301} (\bibinfo {year} {2024})}\BibitemShut {NoStop}%
\bibitem [{\citenamefont {Oh}\ and\ \citenamefont {Nagaosa}(2025)}]{Oh2025}%
  \BibitemOpen
  \bibfield  {author} {\bibinfo {author} {\bibfnamefont {T.}~\bibnamefont {Oh}}\ and\ \bibinfo {author} {\bibfnamefont {N.}~\bibnamefont {Nagaosa}},\ }\bibfield  {title} {\bibinfo {title} {{Phonon Thermal Hall Effect in Mott Insulators via Skew Scattering by the Scalar Spin Chirality}},\ }\href {https://doi.org/10.1103/PhysRevX.15.011036} {\bibfield  {journal} {\bibinfo  {journal} {Phys. Rev. X}\ }\textbf {\bibinfo {volume} {15}},\ \bibinfo {pages} {011036} (\bibinfo {year} {2025})}\BibitemShut {NoStop}%
\bibitem [{\citenamefont {Mukherjee}\ \emph {et~al.}(2023)\citenamefont {Mukherjee}, \citenamefont {Verma},\ and\ \citenamefont {Kundu}}]{Mukherjee2023}%
  \BibitemOpen
  \bibfield  {author} {\bibinfo {author} {\bibfnamefont {R.}~\bibnamefont {Mukherjee}}, \bibinfo {author} {\bibfnamefont {S.}~\bibnamefont {Verma}},\ and\ \bibinfo {author} {\bibfnamefont {A.}~\bibnamefont {Kundu}},\ }\bibfield  {title} {\bibinfo {title} {{Nonlinear magnon transport in bilayer van der Waals antiferromagnets}},\ }\href {https://doi.org/10.1103/PhysRevB.107.245426} {\bibfield  {journal} {\bibinfo  {journal} {Phys. Rev. B}\ }\textbf {\bibinfo {volume} {107}},\ \bibinfo {pages} {245426} (\bibinfo {year} {2023})}\BibitemShut {NoStop}%
\bibitem [{\citenamefont {Varshney}\ \emph {et~al.}(2023)\citenamefont {Varshney}, \citenamefont {Mukherjee}, \citenamefont {Kundu},\ and\ \citenamefont {Agarwal}}]{Varshney2023}%
  \BibitemOpen
  \bibfield  {author} {\bibinfo {author} {\bibfnamefont {H.}~\bibnamefont {Varshney}}, \bibinfo {author} {\bibfnamefont {R.}~\bibnamefont {Mukherjee}}, \bibinfo {author} {\bibfnamefont {A.}~\bibnamefont {Kundu}},\ and\ \bibinfo {author} {\bibfnamefont {A.}~\bibnamefont {Agarwal}},\ }\bibfield  {title} {\bibinfo {title} {{Intrinsic nonlinear thermal Hall transport of magnons: A quantum kinetic theory approach}},\ }\href {https://doi.org/10.1103/PhysRevB.108.165412} {\bibfield  {journal} {\bibinfo  {journal} {Phys. Rev. B}\ }\textbf {\bibinfo {volume} {108}},\ \bibinfo {pages} {165412} (\bibinfo {year} {2023})}\BibitemShut {NoStop}%
\bibitem [{\citenamefont {Ni}\ \emph {et~al.}(2025)\citenamefont {Ni}, \citenamefont {Jin}, \citenamefont {Du},\ and\ \citenamefont {Chang}}]{Ni2025}%
  \BibitemOpen
  \bibfield  {author} {\bibinfo {author} {\bibfnamefont {J.}~\bibnamefont {Ni}}, \bibinfo {author} {\bibfnamefont {Y.}~\bibnamefont {Jin}}, \bibinfo {author} {\bibfnamefont {Q.}~\bibnamefont {Du}},\ and\ \bibinfo {author} {\bibfnamefont {G.}~\bibnamefont {Chang}},\ }\bibfield  {title} {\bibinfo {title} {{Magnon nonlinear Hall effect in two-dimensional antiferromagnetic insulators}},\ }\href {https://doi.org/10.1103/72sf-6km4} {\bibfield  {journal} {\bibinfo  {journal} {Phys. Rev. B}\ }\textbf {\bibinfo {volume} {112}},\ \bibinfo {pages} {054424} (\bibinfo {year} {2025})}\BibitemShut {NoStop}%
\bibitem [{\citenamefont {Li}\ and\ \citenamefont {Zhu}(2024)}]{Li2024}%
  \BibitemOpen
  \bibfield  {author} {\bibinfo {author} {\bibfnamefont {J.-C.}\ \bibnamefont {Li}}\ and\ \bibinfo {author} {\bibfnamefont {Z.-G.}\ \bibnamefont {Zhu}},\ }\bibfield  {title} {\bibinfo {title} {{Intrinsic second-order magnon thermal Hall effect}},\ }\href {https://doi.org/10.1088/1361-648X/ad5bb0} {\bibfield  {journal} {\bibinfo  {journal} {J. Phys.: Condens. Matter}\ }\textbf {\bibinfo {volume} {36}},\ \bibinfo {pages} {395802} (\bibinfo {year} {2024})}\BibitemShut {NoStop}%
\bibitem [{\citenamefont {Zhang}\ \emph {et~al.}(2025)\citenamefont {Zhang}, \citenamefont {Zhang}, \citenamefont {Zhu},\ and\ \citenamefont {Su}}]{Zhang2025}%
  \BibitemOpen
  \bibfield  {author} {\bibinfo {author} {\bibfnamefont {Y.-F.}\ \bibnamefont {Zhang}}, \bibinfo {author} {\bibfnamefont {Z.-F.}\ \bibnamefont {Zhang}}, \bibinfo {author} {\bibfnamefont {Z.-G.}\ \bibnamefont {Zhu}},\ and\ \bibinfo {author} {\bibfnamefont {G.}~\bibnamefont {Su}},\ }\bibfield  {title} {\bibinfo {title} {{Second-order intrinsic Wiedemann-Franz law}},\ }\href {https://doi.org/10.1103/PhysRevB.111.165424} {\bibfield  {journal} {\bibinfo  {journal} {Phys. Rev. B}\ }\textbf {\bibinfo {volume} {111}},\ \bibinfo {pages} {165424} (\bibinfo {year} {2025})}\BibitemShut {NoStop}%
\bibitem [{\citenamefont {Barman}(2025)}]{Barman2025}%
  \BibitemOpen
  \bibfield  {author} {\bibinfo {author} {\bibfnamefont {C.~K.}\ \bibnamefont {Barman}},\ }\href@noop {} {\bibinfo {title} {{Intrinsic Nonlinear Planar Thermal Hall Effect}}} (\bibinfo {year} {2025}),\ \Eprint {https://arxiv.org/abs/2511.01748} {arXiv:2511.01748 [cond-mat.mtrl-sci]} \BibitemShut {NoStop}%
\bibitem [{\citenamefont {Luttinger}(1964)}]{Luttinger1964}%
  \BibitemOpen
  \bibfield  {author} {\bibinfo {author} {\bibfnamefont {J.~M.}\ \bibnamefont {Luttinger}},\ }\bibfield  {title} {\bibinfo {title} {{Theory of Thermal Transport Coefficients}},\ }\href {https://doi.org/10.1103/PhysRev.135.A1505} {\bibfield  {journal} {\bibinfo  {journal} {Phys. Rev.}\ }\textbf {\bibinfo {volume} {135}},\ \bibinfo {pages} {A1505} (\bibinfo {year} {1964})}\BibitemShut {NoStop}%
\bibitem [{\citenamefont {Moreno}\ and\ \citenamefont {Coleman}(1996)}]{Moreno1996}%
  \BibitemOpen
  \bibfield  {author} {\bibinfo {author} {\bibfnamefont {J.}~\bibnamefont {Moreno}}\ and\ \bibinfo {author} {\bibfnamefont {P.}~\bibnamefont {Coleman}},\ }\href@noop {} {\bibinfo {title} {{Thermal currents in highly correlated systems}}} (\bibinfo {year} {1996}),\ \Eprint {https://arxiv.org/abs/cond-mat/9603079} {arXiv:cond-mat/9603079 [cond-mat]} \BibitemShut {NoStop}%
\bibitem [{\citenamefont {Tatara}(2015{\natexlab{a}})}]{Tatara2015L}%
  \BibitemOpen
  \bibfield  {author} {\bibinfo {author} {\bibfnamefont {G.}~\bibnamefont {Tatara}},\ }\bibfield  {title} {\bibinfo {title} {{Thermal Vector Potential Theory of Transport Induced by a Temperature Gradient}},\ }\href {https://doi.org/10.1103/PhysRevLett.114.196601} {\bibfield  {journal} {\bibinfo  {journal} {Phys. Rev. Lett.}\ }\textbf {\bibinfo {volume} {114}},\ \bibinfo {pages} {196601} (\bibinfo {year} {2015}{\natexlab{a}})}\BibitemShut {NoStop}%
\bibitem [{\citenamefont {Tatara}(2015{\natexlab{b}})}]{Tatara2015B}%
  \BibitemOpen
  \bibfield  {author} {\bibinfo {author} {\bibfnamefont {G.}~\bibnamefont {Tatara}},\ }\bibfield  {title} {\bibinfo {title} {{Thermal vector potential theory of magnon-driven magnetization dynamics}},\ }\href {https://doi.org/10.1103/PhysRevB.92.064405} {\bibfield  {journal} {\bibinfo  {journal} {Phys. Rev. B}\ }\textbf {\bibinfo {volume} {92}},\ \bibinfo {pages} {064405} (\bibinfo {year} {2015}{\natexlab{b}})}\BibitemShut {NoStop}%
\bibitem [{\citenamefont {Mangeolle}\ \emph {et~al.}(2024)\citenamefont {Mangeolle}, \citenamefont {Savary},\ and\ \citenamefont {Balents}}]{Mangeolle2024}%
  \BibitemOpen
  \bibfield  {author} {\bibinfo {author} {\bibfnamefont {L.}~\bibnamefont {Mangeolle}}, \bibinfo {author} {\bibfnamefont {L.}~\bibnamefont {Savary}},\ and\ \bibinfo {author} {\bibfnamefont {L.}~\bibnamefont {Balents}},\ }\bibfield  {title} {\bibinfo {title} {{Quantum kinetic equation and thermal conductivity tensor for bosons}},\ }\href {https://doi.org/10.1103/PhysRevB.109.235137} {\bibfield  {journal} {\bibinfo  {journal} {Phys. Rev. B}\ }\textbf {\bibinfo {volume} {109}},\ \bibinfo {pages} {235137} (\bibinfo {year} {2024})}\BibitemShut {NoStop}%
\bibitem [{\citenamefont {Park}\ \emph {et~al.}(2026)\citenamefont {Park}, \citenamefont {Huang}, \citenamefont {Savary},\ and\ \citenamefont {Balents}}]{Park2025}%
  \BibitemOpen
  \bibfield  {author} {\bibinfo {author} {\bibfnamefont {T.}~\bibnamefont {Park}}, \bibinfo {author} {\bibfnamefont {X.}~\bibnamefont {Huang}}, \bibinfo {author} {\bibfnamefont {L.}~\bibnamefont {Savary}},\ and\ \bibinfo {author} {\bibfnamefont {L.}~\bibnamefont {Balents}},\ }\bibfield  {title} {\bibinfo {title} {{Quantum geometry from the Moyal product: Quantum kinetic equation and nonlinear response}},\ }\href {https://doi.org/10.1103/x9pk-lb4x} {\bibfield  {journal} {\bibinfo  {journal} {Phys. Rev. B}\ }\textbf {\bibinfo {volume} {113}},\ \bibinfo {pages} {045146} (\bibinfo {year} {2026})}\BibitemShut {NoStop}%
\bibitem [{\citenamefont {Mangeolle}\ and\ \citenamefont {Knolle}(2026)}]{Mangeolle2026}%
  \BibitemOpen
  \bibfield  {author} {\bibinfo {author} {\bibfnamefont {L.}~\bibnamefont {Mangeolle}}\ and\ \bibinfo {author} {\bibfnamefont {J.}~\bibnamefont {Knolle}},\ }\bibfield  {title} {\bibinfo {title} {Extrinsic contribution to bosonic thermal hall transport},\ }\href {https://doi.org/10.1103/grzx-v6sj} {\bibfield  {journal} {\bibinfo  {journal} {Phys. Rev. X}\ }\textbf {\bibinfo {volume} {16}},\ \bibinfo {pages} {011048} (\bibinfo {year} {2026})}\BibitemShut {NoStop}%
\bibitem [{\citenamefont {Hardy}(1963)}]{Hardy1963}%
  \BibitemOpen
  \bibfield  {author} {\bibinfo {author} {\bibfnamefont {R.~J.}\ \bibnamefont {Hardy}},\ }\bibfield  {title} {\bibinfo {title} {{Energy-Flux Operator for a Lattice}},\ }\href {https://doi.org/10.1103/PhysRev.132.168} {\bibfield  {journal} {\bibinfo  {journal} {Phys. Rev.}\ }\textbf {\bibinfo {volume} {132}},\ \bibinfo {pages} {168} (\bibinfo {year} {1963})}\BibitemShut {NoStop}%
\bibitem [{\citenamefont {Kapustin}\ and\ \citenamefont {Spodyneiko}(2020)}]{Kapustin2020}%
  \BibitemOpen
  \bibfield  {author} {\bibinfo {author} {\bibfnamefont {A.}~\bibnamefont {Kapustin}}\ and\ \bibinfo {author} {\bibfnamefont {L.}~\bibnamefont {Spodyneiko}},\ }\bibfield  {title} {\bibinfo {title} {{Thermal Hall conductance and a relative topological invariant of gapped two-dimensional systems}},\ }\href {https://doi.org/10.1103/PhysRevB.101.045137} {\bibfield  {journal} {\bibinfo  {journal} {Phys. Rev. B}\ }\textbf {\bibinfo {volume} {101}},\ \bibinfo {pages} {045137} (\bibinfo {year} {2020})}\BibitemShut {NoStop}%
\bibitem [{Note1()}]{Note1}%
  \BibitemOpen
  \bibinfo {note} {The expression given in Eq.~\protect \eqref {def_energy_density} does not completely eliminate the ambiguity in the definition of the energy density, as an uncertainty remains that is related to derivatives with respect to the position $\protect \bm {X}$. The energy density is defined as a function $\protect \mathcal H(\protect \bm {X})=\protect \mathcal H^*(\protect \bm {X})$ that satisfies $\langle \protect \mathscr H\rangle =\DOTSI \intop \ilimits@ _{\protect \bm {X}}\protect \mathcal H(\protect \bm {X})$. Under this definition, the energy density remains ambiguous in the sense that it can be modified to $\protect \mathcal H(\protect \bm {X})+\partial _{\protect \bm {X}}\cdot \protect \bm {F}(\protect \bm {X})$ by adding any real function $\protect \bm {F}(\protect \bm {X})$ that vanishes as $|X_i|\to \infty $.}\BibitemShut {Stop}%
\bibitem [{\citenamefont {Colpa}(1978)}]{Colpa1978}%
  \BibitemOpen
  \bibfield  {author} {\bibinfo {author} {\bibfnamefont {J.}~\bibnamefont {Colpa}},\ }\bibfield  {title} {\bibinfo {title} {{Diagonalization of the quadratic boson hamiltonian}},\ }\href {https://doi.org/https://doi.org/10.1016/0378-4371(78)90160-7} {\bibfield  {journal} {\bibinfo  {journal} {Physica A}\ }\textbf {\bibinfo {volume} {93}},\ \bibinfo {pages} {327} (\bibinfo {year} {1978})}\BibitemShut {NoStop}%
\bibitem [{\citenamefont {wen Xiao}(2009)}]{Xiao2009}%
  \BibitemOpen
  \bibfield  {author} {\bibinfo {author} {\bibfnamefont {M.}~\bibnamefont {wen Xiao}},\ }\href@noop {} {\bibinfo {title} {{Theory of transformation for the diagonalization of quadratic Hamiltonians}}} (\bibinfo {year} {2009}),\ \Eprint {https://arxiv.org/abs/0908.0787} {arXiv:0908.0787 [math-ph]} \BibitemShut {NoStop}%
\bibitem [{\citenamefont {Shindou}\ \emph {et~al.}(2013)\citenamefont {Shindou}, \citenamefont {Matsumoto}, \citenamefont {Murakami},\ and\ \citenamefont {Ohe}}]{Shindou2013}%
  \BibitemOpen
  \bibfield  {author} {\bibinfo {author} {\bibfnamefont {R.}~\bibnamefont {Shindou}}, \bibinfo {author} {\bibfnamefont {R.}~\bibnamefont {Matsumoto}}, \bibinfo {author} {\bibfnamefont {S.}~\bibnamefont {Murakami}},\ and\ \bibinfo {author} {\bibfnamefont {J.-i.}\ \bibnamefont {Ohe}},\ }\bibfield  {title} {\bibinfo {title} {{Topological chiral magnonic edge mode in a magnonic crystal}},\ }\href {https://doi.org/10.1103/PhysRevB.87.174427} {\bibfield  {journal} {\bibinfo  {journal} {Phys. Rev. B}\ }\textbf {\bibinfo {volume} {87}},\ \bibinfo {pages} {174427} (\bibinfo {year} {2013})}\BibitemShut {NoStop}%
\bibitem [{\citenamefont {Koyama}\ and\ \citenamefont {Nasu}(2025)}]{Koyama2025}%
  \BibitemOpen
  \bibfield  {author} {\bibinfo {author} {\bibfnamefont {S.}~\bibnamefont {Koyama}}\ and\ \bibinfo {author} {\bibfnamefont {J.}~\bibnamefont {Nasu}},\ }\bibfield  {title} {\bibinfo {title} {{Formulation of the spin Nernst effect for spin-nonconserving insulating magnets}},\ }\href {https://doi.org/10.1103/8xgc-d9ch} {\bibfield  {journal} {\bibinfo  {journal} {Phys. Rev. B}\ }\textbf {\bibinfo {volume} {112}},\ \bibinfo {pages} {014447} (\bibinfo {year} {2025})}\BibitemShut {NoStop}%
\bibitem [{\citenamefont {Tesfaye}\ and\ \citenamefont {Eckardt}(2025)}]{Tesfaye2025}%
  \BibitemOpen
  \bibfield  {author} {\bibinfo {author} {\bibfnamefont {I.}~\bibnamefont {Tesfaye}}\ and\ \bibinfo {author} {\bibfnamefont {A.}~\bibnamefont {Eckardt}},\ }\bibfield  {title} {\bibinfo {title} {{Quantum geometry of bosonic Bogoliubov quasiparticles}},\ }\href {https://doi.org/10.1103/1sz1-nwzf} {\bibfield  {journal} {\bibinfo  {journal} {Phys. Rev. Res.}\ }\textbf {\bibinfo {volume} {7}},\ \bibinfo {pages} {L042052} (\bibinfo {year} {2025})}\BibitemShut {NoStop}%
\bibitem [{\citenamefont {Cooper}\ \emph {et~al.}(1997)\citenamefont {Cooper}, \citenamefont {Halperin},\ and\ \citenamefont {Ruzin}}]{Cooper1997}%
  \BibitemOpen
  \bibfield  {author} {\bibinfo {author} {\bibfnamefont {N.~R.}\ \bibnamefont {Cooper}}, \bibinfo {author} {\bibfnamefont {B.~I.}\ \bibnamefont {Halperin}},\ and\ \bibinfo {author} {\bibfnamefont {I.~M.}\ \bibnamefont {Ruzin}},\ }\bibfield  {title} {\bibinfo {title} {{Thermoelectric response of an interacting two-dimensional electron gas in a quantizing magnetic field}},\ }\href {https://doi.org/10.1103/PhysRevB.55.2344} {\bibfield  {journal} {\bibinfo  {journal} {Phys. Rev. B}\ }\textbf {\bibinfo {volume} {55}},\ \bibinfo {pages} {2344} (\bibinfo {year} {1997})}\BibitemShut {NoStop}%
\bibitem [{Note2()}]{Note2}%
  \BibitemOpen
  \bibinfo {note} {$2N\times 2N$ matrices $A$ and $B$ are called pseudo-Hermitian and anti-pseudo-Hermitian, respectively, if they satisfy the relations $A^{\dagger }=\sigma _3 A\sigma _3$ and $B^{\dagger }=-\sigma _3 B\sigma _3$. An arbitrary $2N\times 2N$ matrix $C$ can be decomposed into pseudo-Hermitian and anti-pseudo-Hermitian components as $C=\protect \frac {1}{2}(C+\sigma _3 C^{\dagger }\sigma _3)+\protect \frac {1}{2}(C-\sigma _3 C^{\dagger }\sigma _3)$.}\BibitemShut {Stop}%
\bibitem [{\citenamefont {Matsumoto}\ \emph {et~al.}(2014)\citenamefont {Matsumoto}, \citenamefont {Shindou},\ and\ \citenamefont {Murakami}}]{Matsumoto2014}%
  \BibitemOpen
  \bibfield  {author} {\bibinfo {author} {\bibfnamefont {R.}~\bibnamefont {Matsumoto}}, \bibinfo {author} {\bibfnamefont {R.}~\bibnamefont {Shindou}},\ and\ \bibinfo {author} {\bibfnamefont {S.}~\bibnamefont {Murakami}},\ }\bibfield  {title} {\bibinfo {title} {{Thermal Hall effect of magnons in magnets with dipolar interaction}},\ }\href {https://doi.org/10.1103/PhysRevB.89.054420} {\bibfield  {journal} {\bibinfo  {journal} {Phys. Rev. B}\ }\textbf {\bibinfo {volume} {89}},\ \bibinfo {pages} {054420} (\bibinfo {year} {2014})}\BibitemShut {NoStop}%
\bibitem [{\citenamefont {Tsirkin}\ and\ \citenamefont {Souza}(2022)}]{Stepan2022}%
  \BibitemOpen
  \bibfield  {author} {\bibinfo {author} {\bibfnamefont {S.~S.}\ \bibnamefont {Tsirkin}}\ and\ \bibinfo {author} {\bibfnamefont {I.}~\bibnamefont {Souza}},\ }\bibfield  {title} {\bibinfo {title} {{On the separation of Hall and Ohmic nonlinear responses}},\ }\href {https://doi.org/10.21468/SciPostPhysCore.5.3.039} {\bibfield  {journal} {\bibinfo  {journal} {SciPost Phys. Core}\ }\textbf {\bibinfo {volume} {5}},\ \bibinfo {pages} {039} (\bibinfo {year} {2022})}\BibitemShut {NoStop}%
\bibitem [{\citenamefont {Zhang}\ \emph {et~al.}(2023)\citenamefont {Zhang}, \citenamefont {Zhu},\ and\ \citenamefont {Su}}]{Zhang2023}%
  \BibitemOpen
  \bibfield  {author} {\bibinfo {author} {\bibfnamefont {Z.-F.}\ \bibnamefont {Zhang}}, \bibinfo {author} {\bibfnamefont {Z.-G.}\ \bibnamefont {Zhu}},\ and\ \bibinfo {author} {\bibfnamefont {G.}~\bibnamefont {Su}},\ }\bibfield  {title} {\bibinfo {title} {{Symmetry dictionary on charge and spin nonlinear responses for all magnetic point groups with nontrivial topological nature}},\ }\href {https://doi.org/10.1093/nsr/nwad104} {\bibfield  {journal} {\bibinfo  {journal} {Natl. Sci. Rev.}\ }\textbf {\bibinfo {volume} {10}},\ \bibinfo {pages} {nwad104} (\bibinfo {year} {2023})}\BibitemShut {NoStop}%
\bibitem [{\citenamefont {Chernyshev}\ and\ \citenamefont {Zhitomirsky}(2009)}]{Chernyshev2009}%
  \BibitemOpen
  \bibfield  {author} {\bibinfo {author} {\bibfnamefont {A.~L.}\ \bibnamefont {Chernyshev}}\ and\ \bibinfo {author} {\bibfnamefont {M.~E.}\ \bibnamefont {Zhitomirsky}},\ }\bibfield  {title} {\bibinfo {title} {Spin waves in a triangular lattice antiferromagnet: Decays, spectrum renormalization, and singularities},\ }\href {https://doi.org/10.1103/PhysRevB.79.144416} {\bibfield  {journal} {\bibinfo  {journal} {Phys. Rev. B}\ }\textbf {\bibinfo {volume} {79}},\ \bibinfo {pages} {144416} (\bibinfo {year} {2009})}\BibitemShut {NoStop}%
\bibitem [{\citenamefont {Mook}\ \emph {et~al.}(2020)\citenamefont {Mook}, \citenamefont {Klinovaja},\ and\ \citenamefont {Loss}}]{Mook2020}%
  \BibitemOpen
  \bibfield  {author} {\bibinfo {author} {\bibfnamefont {A.}~\bibnamefont {Mook}}, \bibinfo {author} {\bibfnamefont {J.}~\bibnamefont {Klinovaja}},\ and\ \bibinfo {author} {\bibfnamefont {D.}~\bibnamefont {Loss}},\ }\bibfield  {title} {\bibinfo {title} {Quantum damping of skyrmion crystal eigenmodes due to spontaneous quasiparticle decay},\ }\href {https://doi.org/10.1103/PhysRevResearch.2.033491} {\bibfield  {journal} {\bibinfo  {journal} {Phys. Rev. Res.}\ }\textbf {\bibinfo {volume} {2}},\ \bibinfo {pages} {033491} (\bibinfo {year} {2020})}\BibitemShut {NoStop}%
\bibitem [{\citenamefont {Koyama}\ and\ \citenamefont {Nasu}(2024)}]{Koyama2024}%
  \BibitemOpen
  \bibfield  {author} {\bibinfo {author} {\bibfnamefont {S.}~\bibnamefont {Koyama}}\ and\ \bibinfo {author} {\bibfnamefont {J.}~\bibnamefont {Nasu}},\ }\bibfield  {title} {\bibinfo {title} {Thermal hall effect incorporating magnon damping in localized spin systems},\ }\href {https://doi.org/10.1103/PhysRevB.109.174442} {\bibfield  {journal} {\bibinfo  {journal} {Phys. Rev. B}\ }\textbf {\bibinfo {volume} {109}},\ \bibinfo {pages} {174442} (\bibinfo {year} {2024})}\BibitemShut {NoStop}%
\bibitem [{\citenamefont {Habel}\ \emph {et~al.}(2024)\citenamefont {Habel}, \citenamefont {Mook}, \citenamefont {Willsher},\ and\ \citenamefont {Knolle}}]{Habel2024}%
  \BibitemOpen
  \bibfield  {author} {\bibinfo {author} {\bibfnamefont {J.}~\bibnamefont {Habel}}, \bibinfo {author} {\bibfnamefont {A.}~\bibnamefont {Mook}}, \bibinfo {author} {\bibfnamefont {J.}~\bibnamefont {Willsher}},\ and\ \bibinfo {author} {\bibfnamefont {J.}~\bibnamefont {Knolle}},\ }\bibfield  {title} {\bibinfo {title} {Breakdown of chiral edge modes in topological magnon insulators},\ }\href {https://doi.org/10.1103/PhysRevB.109.024441} {\bibfield  {journal} {\bibinfo  {journal} {Phys. Rev. B}\ }\textbf {\bibinfo {volume} {109}},\ \bibinfo {pages} {024441} (\bibinfo {year} {2024})}\BibitemShut {NoStop}%
\bibitem [{\citenamefont {Shitade}(2014)}]{Shitade2014}%
  \BibitemOpen
  \bibfield  {author} {\bibinfo {author} {\bibfnamefont {A.}~\bibnamefont {Shitade}},\ }\bibfield  {title} {\bibinfo {title} {Heat transport as torsional responses and keldysh formalism in a curved spacetime},\ }\href {https://doi.org/10.1093/ptep/ptu162} {\bibfield  {journal} {\bibinfo  {journal} {Prog. Theor. Exp. Phys.}\ }\textbf {\bibinfo {volume} {2014}},\ \bibinfo {pages} {123I01} (\bibinfo {year} {2014})}\BibitemShut {NoStop}%
\bibitem [{Note3()}]{Note3}%
  \BibitemOpen
  \bibinfo {note} {The ``Hall'' contributions defined in Eq.~\protect \eqref {eq:kappa_Hall-def} for $\kappa ^{\protect \mathrm {QM}}$ and $\kappa ^{\protect \mathrm {disp}}$ also vanish owing to the $\protect \mathcal C_3^z$ symmetry.}\BibitemShut {Stop}%
\bibitem [{\citenamefont {Owerre}(2018)}]{Owerre2018}%
  \BibitemOpen
  \bibfield  {author} {\bibinfo {author} {\bibfnamefont {S.~A.}\ \bibnamefont {Owerre}},\ }\bibfield  {title} {\bibinfo {title} {{Strain-induced topological magnon phase transitions: applications to kagome-lattice ferromagnets}},\ }\href {https://doi.org/10.1088/1361-648X/aac365} {\bibfield  {journal} {\bibinfo  {journal} {J. Phys.: Condens. Matter}\ }\textbf {\bibinfo {volume} {30}},\ \bibinfo {pages} {245803} (\bibinfo {year} {2018})}\BibitemShut {NoStop}%
\bibitem [{\citenamefont {Kondo}\ and\ \citenamefont {Akagi}(2022)}]{Kondo2022}%
  \BibitemOpen
  \bibfield  {author} {\bibinfo {author} {\bibfnamefont {H.}~\bibnamefont {Kondo}}\ and\ \bibinfo {author} {\bibfnamefont {Y.}~\bibnamefont {Akagi}},\ }\bibfield  {title} {\bibinfo {title} {{Nonlinear magnon spin Nernst effect in antiferromagnets and strain-tunable pure spin current}},\ }\href {https://doi.org/10.1103/PhysRevResearch.4.013186} {\bibfield  {journal} {\bibinfo  {journal} {Phys. Rev. Res.}\ }\textbf {\bibinfo {volume} {4}},\ \bibinfo {pages} {013186} (\bibinfo {year} {2022})}\BibitemShut {NoStop}%
\bibitem [{\citenamefont {Takashima}\ \emph {et~al.}(2018)\citenamefont {Takashima}, \citenamefont {Shiomi},\ and\ \citenamefont {Motome}}]{Takashima2018}%
  \BibitemOpen
  \bibfield  {author} {\bibinfo {author} {\bibfnamefont {R.}~\bibnamefont {Takashima}}, \bibinfo {author} {\bibfnamefont {Y.}~\bibnamefont {Shiomi}},\ and\ \bibinfo {author} {\bibfnamefont {Y.}~\bibnamefont {Motome}},\ }\bibfield  {title} {\bibinfo {title} {Nonreciprocal spin seebeck effect in antiferromagnets},\ }\href {https://doi.org/10.1103/PhysRevB.98.020401} {\bibfield  {journal} {\bibinfo  {journal} {Phys. Rev. B}\ }\textbf {\bibinfo {volume} {98}},\ \bibinfo {pages} {020401} (\bibinfo {year} {2018})}\BibitemShut {NoStop}%
\bibitem [{\citenamefont {Nomura}\ \emph {et~al.}(2019)\citenamefont {Nomura}, \citenamefont {Zhang}, \citenamefont {Zherlitsyn}, \citenamefont {Wosnitza}, \citenamefont {Tokura}, \citenamefont {Nagaosa},\ and\ \citenamefont {Seki}}]{Nomura2019}%
  \BibitemOpen
  \bibfield  {author} {\bibinfo {author} {\bibfnamefont {T.}~\bibnamefont {Nomura}}, \bibinfo {author} {\bibfnamefont {X.-X.}\ \bibnamefont {Zhang}}, \bibinfo {author} {\bibfnamefont {S.}~\bibnamefont {Zherlitsyn}}, \bibinfo {author} {\bibfnamefont {J.}~\bibnamefont {Wosnitza}}, \bibinfo {author} {\bibfnamefont {Y.}~\bibnamefont {Tokura}}, \bibinfo {author} {\bibfnamefont {N.}~\bibnamefont {Nagaosa}},\ and\ \bibinfo {author} {\bibfnamefont {S.}~\bibnamefont {Seki}},\ }\bibfield  {title} {\bibinfo {title} {Phonon magnetochiral effect},\ }\href {https://doi.org/10.1103/PhysRevLett.122.145901} {\bibfield  {journal} {\bibinfo  {journal} {Phys. Rev. Lett.}\ }\textbf {\bibinfo {volume} {122}},\ \bibinfo {pages} {145901} (\bibinfo {year} {2019})}\BibitemShut {NoStop}%
\bibitem [{\citenamefont {Nakai}\ and\ \citenamefont {Nagaosa}(2019)}]{Nakai2019}%
  \BibitemOpen
  \bibfield  {author} {\bibinfo {author} {\bibfnamefont {R.}~\bibnamefont {Nakai}}\ and\ \bibinfo {author} {\bibfnamefont {N.}~\bibnamefont {Nagaosa}},\ }\bibfield  {title} {\bibinfo {title} {Nonreciprocal thermal and thermoelectric transport of electrons in noncentrosymmetric crystals},\ }\href {https://doi.org/10.1103/PhysRevB.99.115201} {\bibfield  {journal} {\bibinfo  {journal} {Phys. Rev. B}\ }\textbf {\bibinfo {volume} {99}},\ \bibinfo {pages} {115201} (\bibinfo {year} {2019})}\BibitemShut {NoStop}%
\bibitem [{\citenamefont {Hirokane}\ \emph {et~al.}(2020)\citenamefont {Hirokane}, \citenamefont {Nii}, \citenamefont {Masuda},\ and\ \citenamefont {Onose}}]{Hirokane2020}%
  \BibitemOpen
  \bibfield  {author} {\bibinfo {author} {\bibfnamefont {Y.}~\bibnamefont {Hirokane}}, \bibinfo {author} {\bibfnamefont {Y.}~\bibnamefont {Nii}}, \bibinfo {author} {\bibfnamefont {H.}~\bibnamefont {Masuda}},\ and\ \bibinfo {author} {\bibfnamefont {Y.}~\bibnamefont {Onose}},\ }\bibfield  {title} {\bibinfo {title} {Nonreciprocal thermal transport in a multiferroic helimagnet},\ }\href {https://doi.org/10.1126/sciadv.abd3703} {\bibfield  {journal} {\bibinfo  {journal} {Sci. Adv.}\ }\textbf {\bibinfo {volume} {6}},\ \bibinfo {pages} {eabd3703} (\bibinfo {year} {2020})}\BibitemShut {NoStop}%
\bibitem [{\citenamefont {Nomura}\ \emph {et~al.}(2023)\citenamefont {Nomura}, \citenamefont {Zhang}, \citenamefont {Takagi}, \citenamefont {Karube}, \citenamefont {Kikkawa}, \citenamefont {Taguchi}, \citenamefont {Tokura}, \citenamefont {Zherlitsyn}, \citenamefont {Kohama},\ and\ \citenamefont {Seki}}]{Nomura2023}%
  \BibitemOpen
  \bibfield  {author} {\bibinfo {author} {\bibfnamefont {T.}~\bibnamefont {Nomura}}, \bibinfo {author} {\bibfnamefont {X.-X.}\ \bibnamefont {Zhang}}, \bibinfo {author} {\bibfnamefont {R.}~\bibnamefont {Takagi}}, \bibinfo {author} {\bibfnamefont {K.}~\bibnamefont {Karube}}, \bibinfo {author} {\bibfnamefont {A.}~\bibnamefont {Kikkawa}}, \bibinfo {author} {\bibfnamefont {Y.}~\bibnamefont {Taguchi}}, \bibinfo {author} {\bibfnamefont {Y.}~\bibnamefont {Tokura}}, \bibinfo {author} {\bibfnamefont {S.}~\bibnamefont {Zherlitsyn}}, \bibinfo {author} {\bibfnamefont {Y.}~\bibnamefont {Kohama}},\ and\ \bibinfo {author} {\bibfnamefont {S.}~\bibnamefont {Seki}},\ }\bibfield  {title} {\bibinfo {title} {Nonreciprocal phonon propagation in a metallic chiral magnet},\ }\href {https://doi.org/10.1103/PhysRevLett.130.176301} {\bibfield  {journal} {\bibinfo  {journal} {Phys. Rev. Lett.}\ }\textbf {\bibinfo {volume} {130}},\ \bibinfo {pages} {176301} (\bibinfo {year} {2023})}\BibitemShut {NoStop}%
\bibitem [{\citenamefont {Sano}\ \emph {et~al.}(2024)\citenamefont {Sano}, \citenamefont {Takikawa}, \citenamefont {Takahashi}, \citenamefont {Yamada}, \citenamefont {Mizushima},\ and\ \citenamefont {Fujimoto}}]{Sano2024}%
  \BibitemOpen
  \bibfield  {author} {\bibinfo {author} {\bibfnamefont {Y.}~\bibnamefont {Sano}}, \bibinfo {author} {\bibfnamefont {D.}~\bibnamefont {Takikawa}}, \bibinfo {author} {\bibfnamefont {M.~O.}\ \bibnamefont {Takahashi}}, \bibinfo {author} {\bibfnamefont {M.~G.}\ \bibnamefont {Yamada}}, \bibinfo {author} {\bibfnamefont {T.}~\bibnamefont {Mizushima}},\ and\ \bibinfo {author} {\bibfnamefont {S.}~\bibnamefont {Fujimoto}},\ }\bibfield  {title} {\bibinfo {title} {Nonreciprocal heat transport in the kitaev chiral spin liquid},\ }\href {https://doi.org/10.1103/PhysRevB.110.214430} {\bibfield  {journal} {\bibinfo  {journal} {Phys. Rev. B}\ }\textbf {\bibinfo {volume} {110}},\ \bibinfo {pages} {214430} (\bibinfo {year} {2024})}\BibitemShut {NoStop}%
\bibitem [{\citenamefont {Terashima}\ \emph {et~al.}(2025)\citenamefont {Terashima}, \citenamefont {Uji}, \citenamefont {Matsuda}, \citenamefont {Shibauchi},\ and\ \citenamefont {Kasahara}}]{Terashima2025}%
  \BibitemOpen
  \bibfield  {author} {\bibinfo {author} {\bibfnamefont {T.}~\bibnamefont {Terashima}}, \bibinfo {author} {\bibfnamefont {S.}~\bibnamefont {Uji}}, \bibinfo {author} {\bibfnamefont {Y.}~\bibnamefont {Matsuda}}, \bibinfo {author} {\bibfnamefont {T.}~\bibnamefont {Shibauchi}},\ and\ \bibinfo {author} {\bibfnamefont {S.}~\bibnamefont {Kasahara}},\ }\bibfield  {title} {\bibinfo {title} {Apparent nonreciprocal transport in fese bulk crystals},\ }\href {https://doi.org/10.1103/PhysRevB.111.054521} {\bibfield  {journal} {\bibinfo  {journal} {Phys. Rev. B}\ }\textbf {\bibinfo {volume} {111}},\ \bibinfo {pages} {054521} (\bibinfo {year} {2025})}\BibitemShut {NoStop}%
\bibitem [{\citenamefont {Zeng}\ \emph {et~al.}(2019)\citenamefont {Zeng}, \citenamefont {Nandy}, \citenamefont {Taraphder},\ and\ \citenamefont {Tewari}}]{Zeng2019}%
  \BibitemOpen
  \bibfield  {author} {\bibinfo {author} {\bibfnamefont {C.}~\bibnamefont {Zeng}}, \bibinfo {author} {\bibfnamefont {S.}~\bibnamefont {Nandy}}, \bibinfo {author} {\bibfnamefont {A.}~\bibnamefont {Taraphder}},\ and\ \bibinfo {author} {\bibfnamefont {S.}~\bibnamefont {Tewari}},\ }\bibfield  {title} {\bibinfo {title} {Nonlinear nernst effect in bilayer ${\mathrm{wte}}_{2}$},\ }\href {https://doi.org/10.1103/PhysRevB.100.245102} {\bibfield  {journal} {\bibinfo  {journal} {Phys. Rev. B}\ }\textbf {\bibinfo {volume} {100}},\ \bibinfo {pages} {245102} (\bibinfo {year} {2019})}\BibitemShut {NoStop}%
\bibitem [{\citenamefont {Zeng}\ \emph {et~al.}(2022)\citenamefont {Zeng}, \citenamefont {Nandy},\ and\ \citenamefont {Tewari}}]{Zeng2022}%
  \BibitemOpen
  \bibfield  {author} {\bibinfo {author} {\bibfnamefont {C.}~\bibnamefont {Zeng}}, \bibinfo {author} {\bibfnamefont {S.}~\bibnamefont {Nandy}},\ and\ \bibinfo {author} {\bibfnamefont {S.}~\bibnamefont {Tewari}},\ }\bibfield  {title} {\bibinfo {title} {Chiral anomaly induced nonlinear nernst and thermal hall effects in weyl semimetals},\ }\href {https://doi.org/10.1103/PhysRevB.105.125131} {\bibfield  {journal} {\bibinfo  {journal} {Phys. Rev. B}\ }\textbf {\bibinfo {volume} {105}},\ \bibinfo {pages} {125131} (\bibinfo {year} {2022})}\BibitemShut {NoStop}%
\bibitem [{\citenamefont {Yamaguchi}\ \emph {et~al.}(2024)\citenamefont {Yamaguchi}, \citenamefont {Nakazawa},\ and\ \citenamefont {Yamakage}}]{Yamaguchi2024}%
  \BibitemOpen
  \bibfield  {author} {\bibinfo {author} {\bibfnamefont {T.}~\bibnamefont {Yamaguchi}}, \bibinfo {author} {\bibfnamefont {K.}~\bibnamefont {Nakazawa}},\ and\ \bibinfo {author} {\bibfnamefont {A.}~\bibnamefont {Yamakage}},\ }\bibfield  {title} {\bibinfo {title} {Microscopic theory of nonlinear hall effect induced by electric field and temperature gradient},\ }\href {https://doi.org/10.1103/PhysRevB.109.205117} {\bibfield  {journal} {\bibinfo  {journal} {Phys. Rev. B}\ }\textbf {\bibinfo {volume} {109}},\ \bibinfo {pages} {205117} (\bibinfo {year} {2024})}\BibitemShut {NoStop}%
\bibitem [{\citenamefont {Nakazawa}\ \emph {et~al.}(2024)\citenamefont {Nakazawa}, \citenamefont {Yamaguchi},\ and\ \citenamefont {Yamakage}}]{Nakazawa2024}%
  \BibitemOpen
  \bibfield  {author} {\bibinfo {author} {\bibfnamefont {K.}~\bibnamefont {Nakazawa}}, \bibinfo {author} {\bibfnamefont {T.}~\bibnamefont {Yamaguchi}},\ and\ \bibinfo {author} {\bibfnamefont {A.}~\bibnamefont {Yamakage}},\ }\bibfield  {title} {\bibinfo {title} {Nonlinear charge transport properties in chiral tellurium},\ }\href {https://doi.org/10.1103/PhysRevMaterials.8.L091601} {\bibfield  {journal} {\bibinfo  {journal} {Phys. Rev. Mater.}\ }\textbf {\bibinfo {volume} {8}},\ \bibinfo {pages} {L091601} (\bibinfo {year} {2024})}\BibitemShut {NoStop}%
\bibitem [{\citenamefont {Nomoto}\ \emph {et~al.}(2025)\citenamefont {Nomoto}, \citenamefont {Kikkawa}, \citenamefont {Nakazawa}, \citenamefont {Yamaguchi},\ and\ \citenamefont {Kagawa}}]{Nomoto2025}%
  \BibitemOpen
  \bibfield  {author} {\bibinfo {author} {\bibfnamefont {T.}~\bibnamefont {Nomoto}}, \bibinfo {author} {\bibfnamefont {A.}~\bibnamefont {Kikkawa}}, \bibinfo {author} {\bibfnamefont {K.}~\bibnamefont {Nakazawa}}, \bibinfo {author} {\bibfnamefont {T.}~\bibnamefont {Yamaguchi}},\ and\ \bibinfo {author} {\bibfnamefont {F.}~\bibnamefont {Kagawa}},\ }\bibfield  {title} {\bibinfo {title} {Observation of the nonlinear chiral thermoelectric hall effect in tellurium},\ }\href {https://doi.org/10.1038/s41567-025-03073-7} {\bibfield  {journal} {\bibinfo  {journal} {Nat. Phys.}\ }\textbf {\bibinfo {volume} {21}},\ \bibinfo {pages} {1920} (\bibinfo {year} {2025})}\BibitemShut {NoStop}%
\bibitem [{\citenamefont {Nakazawa}\ \emph {et~al.}(2025)\citenamefont {Nakazawa}, \citenamefont {Yamaguchi},\ and\ \citenamefont {Yamakage}}]{Nakazawa2025}%
  \BibitemOpen
  \bibfield  {author} {\bibinfo {author} {\bibfnamefont {K.}~\bibnamefont {Nakazawa}}, \bibinfo {author} {\bibfnamefont {T.}~\bibnamefont {Yamaguchi}},\ and\ \bibinfo {author} {\bibfnamefont {A.}~\bibnamefont {Yamakage}},\ }\bibfield  {title} {\bibinfo {title} {Nonlinear charge and thermal transport properties induced by orbital magnetic moment in chiral crystalline cobalt monosilicide},\ }\href {https://doi.org/10.1103/PhysRevB.111.045161} {\bibfield  {journal} {\bibinfo  {journal} {Phys. Rev. B}\ }\textbf {\bibinfo {volume} {111}},\ \bibinfo {pages} {045161} (\bibinfo {year} {2025})}\BibitemShut {NoStop}%
\bibitem [{\citenamefont {Xiao}\ and\ \citenamefont {Niu}(2021)}]{Xiao2021}%
  \BibitemOpen
  \bibfield  {author} {\bibinfo {author} {\bibfnamefont {C.}~\bibnamefont {Xiao}}\ and\ \bibinfo {author} {\bibfnamefont {Q.}~\bibnamefont {Niu}},\ }\bibfield  {title} {\bibinfo {title} {{Conserved current of nonconserved quantities}},\ }\href {https://doi.org/10.1103/PhysRevB.104.L241411} {\bibfield  {journal} {\bibinfo  {journal} {Phys. Rev. B}\ }\textbf {\bibinfo {volume} {104}},\ \bibinfo {pages} {L241411} (\bibinfo {year} {2021})}\BibitemShut {NoStop}%
\bibitem [{\citenamefont {Matsushita}\ \emph {et~al.}(2025)\citenamefont {Matsushita}, \citenamefont {Yanase}, \citenamefont {Mizushima}, \citenamefont {Fujimoto},\ and\ \citenamefont {Vekhter}}]{Matsushita2025}%
  \BibitemOpen
  \bibfield  {author} {\bibinfo {author} {\bibfnamefont {T.}~\bibnamefont {Matsushita}}, \bibinfo {author} {\bibfnamefont {Y.}~\bibnamefont {Yanase}}, \bibinfo {author} {\bibfnamefont {T.}~\bibnamefont {Mizushima}}, \bibinfo {author} {\bibfnamefont {S.}~\bibnamefont {Fujimoto}},\ and\ \bibinfo {author} {\bibfnamefont {I.}~\bibnamefont {Vekhter}},\ }\href@noop {} {\bibinfo {title} {{Intrinsic spin Nernst effect in spin-triplet superconductors}}} (\bibinfo {year} {2025}),\ \Eprint {https://arxiv.org/abs/2512.19971} {arXiv:2512.19971 [cond-mat.supr-con]} \BibitemShut {NoStop}%
\end{thebibliography}%
\end{document}